\documentclass[acmsmall,screen]{acmart}

\usepackage{bm}
\usepackage{graphicx}
\usepackage{subcaption}
\usepackage{enumitem}
\AtBeginDocument{%
  \providecommand\BibTeX{{%
    \normalfont B\kern-0.5em{\scshape i\kern-0.25em b}\kern-0.8em\TeX}}}

\setcopyright{acmcopyright}
\copyrightyear{2024}
\acmYear{2024}
\acmDOI{XXXXXXX.XXXXXXX}

\newcommand{\revise}[1]{\textcolor{black}{#1}}

\renewcommand\footnotetextcopyrightpermission[1]{} % removes footnote with conference information in first column
\setcopyright{none}

\begin{document}

\title{From Capture to Display: A Survey on Volumetric Video}

\author{Yili Jin}
\email{yili.jin@mail.mcgill.ca}
\affiliation{%
  \institution{McGill University}
  \country{Canada}
}

\author{Kaiyuan Hu}
\affiliation{%
  \institution{McGill University}
  \country{Canada}
}

\author{Junhua Liu}
\affiliation{%
  \institution{University of Southern California}
  \country{USA}
}

\author{Fangxin Wang}
\affiliation{%
  \institution{The Chinese University of Hong Kong, Shenzhen}
  \country{China}
}

\author{Xue Liu}
\affiliation{%
  \institution{McGill University}
  \country{Canada}
}


\begin{abstract}
Volumetric video, which offers immersive viewing experiences, is gaining increasing prominence. With its six degrees of freedom, it provides viewers with greater immersion and interactivity compared to traditional videos. Despite their potential, volumetric video services pose significant challenges. This survey conducts a comprehensive review of the existing literature on volumetric video. We firstly provide a general framework of volumetric video services, followed by a discussion on prerequisites for volumetric video, encompassing representations, open datasets, and quality assessment metrics. Then we delve into the current methodologies for each stage of the volumetric video service pipeline, detailing capturing, compression, transmission, rendering, and display techniques. Lastly, we explore various applications enabled by this pioneering technology and we present an array of research challenges and opportunities in the domain of volumetric video services. This survey aspires to provide a holistic understanding of this burgeoning field and shed light on potential future research trajectories, aiming to bring the vision of volumetric video to fruition.
\end{abstract}



%\keywords{Virtual reality, volumetric videos}

\maketitle

\section{Introduction}

In recent years, the landscape of multimedia services over the Internet has undergone significant transformations. Starting from traditional flat videos, it has progressed to panoramic videos (360-degree videos) and now to volumetric videos. Anticipated to reach a business value of 22.5 billion USD by 2024~\cite{DBLP:journals/network/LiuLCWILJ21}, volumetric videos have captured the attention of both researchers and industry players alike.

The concept of volumetric video stems from holograms and 3D virtual environments often portrayed in popular science fiction, such as Star Wars~\cite{lucas1977star} and Blade Runner~\cite{scott1982blade}. These imaginative stories have fueled the desire to replicate reality with incredible detail, transcending the limitations of flat screens. Advancements in computer graphics and information processing have played a crucial role in the evolution from two-dimensional video to three-dimensional volumetric video. Despite over a decade of rapid development, volumetric video technology is still in its infancy, holding immense potential for growth and innovation. Volumetric videos stand apart from traditional videos due to their ability to deliver an unparalleled experience of spatialized immersion and six degrees-of-freedom (DoF) interactivity. This includes three dimensions of watching position $(X, Y, Z)$ and three dimensions of watching orientation $(yaw, pitch, roll)$.

\begin{figure}[t]
\begin{center}
\includegraphics[width=0.7\columnwidth]{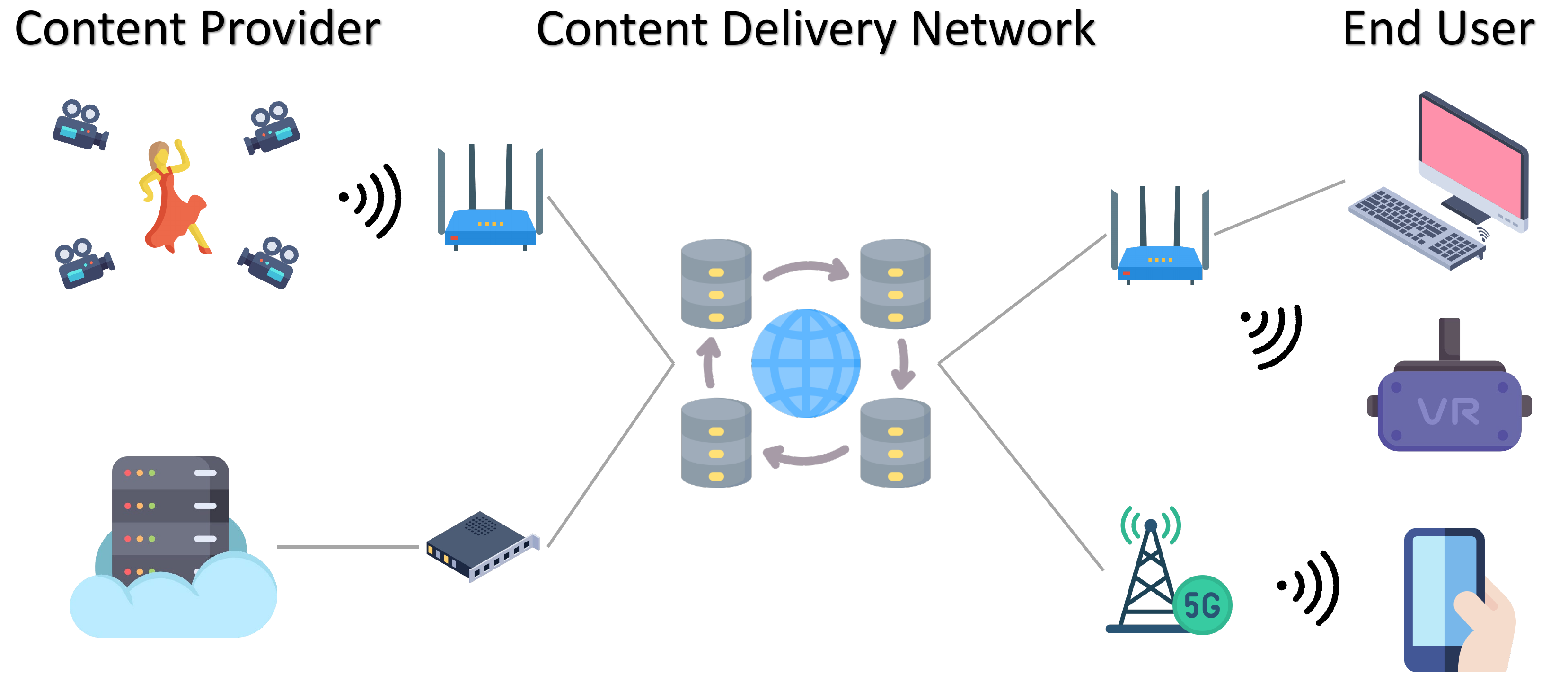}
\end{center}
\vspace{-20pt}
\caption{\label{fig:introsystem}%
Overview of volumetric video delivery systems.}
\vspace{-10pt}
\end{figure}

We provide an overview of volumetric video delivery systems. Fig.~\ref{fig:introsystem} illustrates the high-level architecture of such systems. Volumetric videos can be acquired by cameras or saved video files on cloud servers. These videos are then transmitted through the internet using various access networks, including Ethernet~\cite{DBLP:journals/comsur/SommerGFKMSS10}, WiFi~\cite{DBLP:conf/iwcmc/ZhongKCFA15}, or cellular networks~\cite{DBLP:journals/comsur/AsadiWM14}. \revise{After transmission, volumetric videos can be displayed across a range of devices such as desktops, mobile devices, and Head-Mounted Displays (HMDs). HMDs, such as the Apple Vision Pro~\cite{visionpro}, HTC VIVE~\cite{vive}, and Sony PlayStation VR~\cite{psvr}, provide a more immersive viewing experience compared to traditional flat screens.} Volumetric video will revolutionize the way we consume and experience video content. It allows us to feel like we are truly present in the environment, providing a much more engaging and captivating experience for viewers.

\begin{table*}[!t]\centering
\caption{Terms and synonyms related to volumetric video.}
\vspace{-8pt}
\begin{tabular}{|p{2.7cm}|p{6.95cm}|p{3.2cm}|}
\hline
\textbf{Term} & \textbf{Definition} & \textbf{Synonym} \\ \hline
Volumetric video  & Volumetric video captures objects and environments in full 3D. It can be viewed with 6 DoF. &     \\ \hline
360° video \cite{DBLP:journals/csur/FanLPH19}            & 360° video captures lights from all directions to a camera. It can be viewed with 3 DoF.               & Panoramic video; Omnidirectional video            \\ \hline
\revise{3D video \cite{DBLP:journals/tce/MerkleMW10} }           & \revise{3D video provides depth perception of its contents, encompassing volumetric video and other formats such as light field video.}              &            \\ \hline
Virtual reality (VR) \cite{DBLP:journals/annals/DzardanovaK23,DBLP:journals/cga/SerafinGENN18} & VR is a simulated experience that gives the user an immersive feel of a virtual world. &  \\ \hline
Mixed reality (MR) \cite{DBLP:conf/chi/SpeicherHN19,DBLP:conf/chi/RatcliffeSBTF21} & MR combines virtual objects with the real environment in which users are currently situated. & \\ \hline
Metaverse \cite{DBLP:journals/comsur/WangSZXLLS23} & Metaverse is a new internet paradigm creating a virtual shared space for immersive social interaction, entertainment, work, and commerce. & \\ \hline
Degrees of freedom (DoF) & DoF describe ways an object can move in 3D space. There are six DoF: three rotational and three translational movements along the x, y, and z axes. & \\ \hline
Head-mounted display (HMD) & HMD is a display device, worn on the head, for an immersive viewing experience. & VR headset \\ \hline
Viewport & A portion of videos that are visible to a volumetric or 360° video viewer. & Field of View (FoV), Region of Interest (RoI) \\ \hline
\end{tabular}
\label{table:terms}
\vspace{-12pt}
\end{table*}

Volumetric video services and their underlying technologies have a huge potential in revolutionizing multimedia applications for the future. However, there is still a gap in the existing literature when it comes to providing a comprehensive overview of the current state of volumetric video services, including their architecture, opportunities, and challenges. This survey paper aims to fill this gap by offering a detailed understanding of the entire process of volumetric video services, from capture to display, and presenting the latest research on volumetric video services. Furthermore, we discuss the open challenges and opportunities faced by volumetric videos from various angles, providing valuable insights into future research directions. As an emerging field, the terminology used in volumetric video studies is inconsistent. Table~\ref{table:terms} defines some of the terms used and their synonyms, if any. To ensure accuracy, when presenting research works in the rest of this article, we may modify the terminology used by those works in the literature.

\revise{In this survey, we conduct an extensive search of relevant literature on volumetric video across multiple platforms. To ensure comprehensive coverage, we include works from a wide range of related fields, such as computer vision, multimedia systems, and telecommunications. However, given the vast number of publications in these areas, not all existing works could be included. Therefore, we prioritized works that (1) introduced novel methods, (2) addressed key challenges in volumetric video, (3) were frequently cited in recent literature, or (4) offered comprehensive datasets or benchmarks.
This selection process ensures that the survey covers key contributions while acknowledging that some works may not be included due to the breadth of the field.}

\subsection{Related Surveys}
This article concentrates on the burgeoning field of volumetric video service, a topic that, to the best of our knowledge, has not been thoroughly surveyed in the existing literature. The most relevant works are a tutorial by Hooft et al.~\cite{DBLP:journals/comsur/HooftAVSSST23} and a chapter by Eisert et al.~\cite{IVT11}. The former offers an introduction to the creation, streaming, and evaluation of immersive videos. In contrast to our study, the authors cover a broader range of immersive video formats, including 360° video. They outline the technological progression from traditional video to 3 DoF video, and eventually to 6 DoF video, comparing the various video formats along the way. For readers interested in a wider scope of immersive video, we recommend referring to their work. Eisert et al.~\cite{IVT11} concentrate on the topic of virtual humans, beginning with an overview of current methods for capturing 3D human models, including image pre-processing and 3D mesh processing. They then discuss techniques for animating the body and face of virtual humans to enable them to respond to user behavior. Finally, the authors address the topic of streaming captured virtual humans. For readers interested in a finer scope of virtual humans, we recommend referring to their work.

Fan et al.~\cite{DBLP:journals/csur/FanLPH19} present a comprehensive survey on 360° video streaming. It includes video and viewer datasets for simulations, and detailed discussions of optimization tools. Although volumetric video has a completely different representation from 360° video, it serves as a precursor to volumetric video. Many concepts in volumetric video systems are inspired by 360° video, such as the tile-based viewport adaptive streaming framework. For those interested in learning more about 360° video streaming, Fan et al.~\cite{DBLP:journals/csur/FanLPH19} can provide valuable insights, which may also facilitate a better understanding of volumetric video approaches.

Volumetric video has the potential to be employed in immersive computing applications. Apostolopoulos et al.~\cite{DBLP:journals/pieee/ApostolopoulosCCKTW12} and Han~\cite{DBLP:journals/cm/Han19} explore immersive computing from the perspectives of communication systems and mobile systems, respectively. Moreover, volumetric video can be applied to VR, MR, and Metaverse applications. In VR, entire 3D scenes are created using computer graphics, while MR combines synthesized content with real environments. The Metaverse, an emerging paradigm for the next-generation Internet, aims to establish a fully immersive and self-sustaining virtual shared space. Readers interested in those applications can refer to the respective sources: VR~\cite{DBLP:journals/annals/DzardanovaK23,DBLP:journals/cga/SerafinGENN18}, MR~\cite{DBLP:conf/chi/SpeicherHN19,DBLP:conf/chi/RatcliffeSBTF21}, and Metaverse~\cite{DBLP:journals/comsur/WangSZXLLS23}.

\begin{figure}[t]
\begin{center}
\includegraphics[width=0.85\columnwidth]{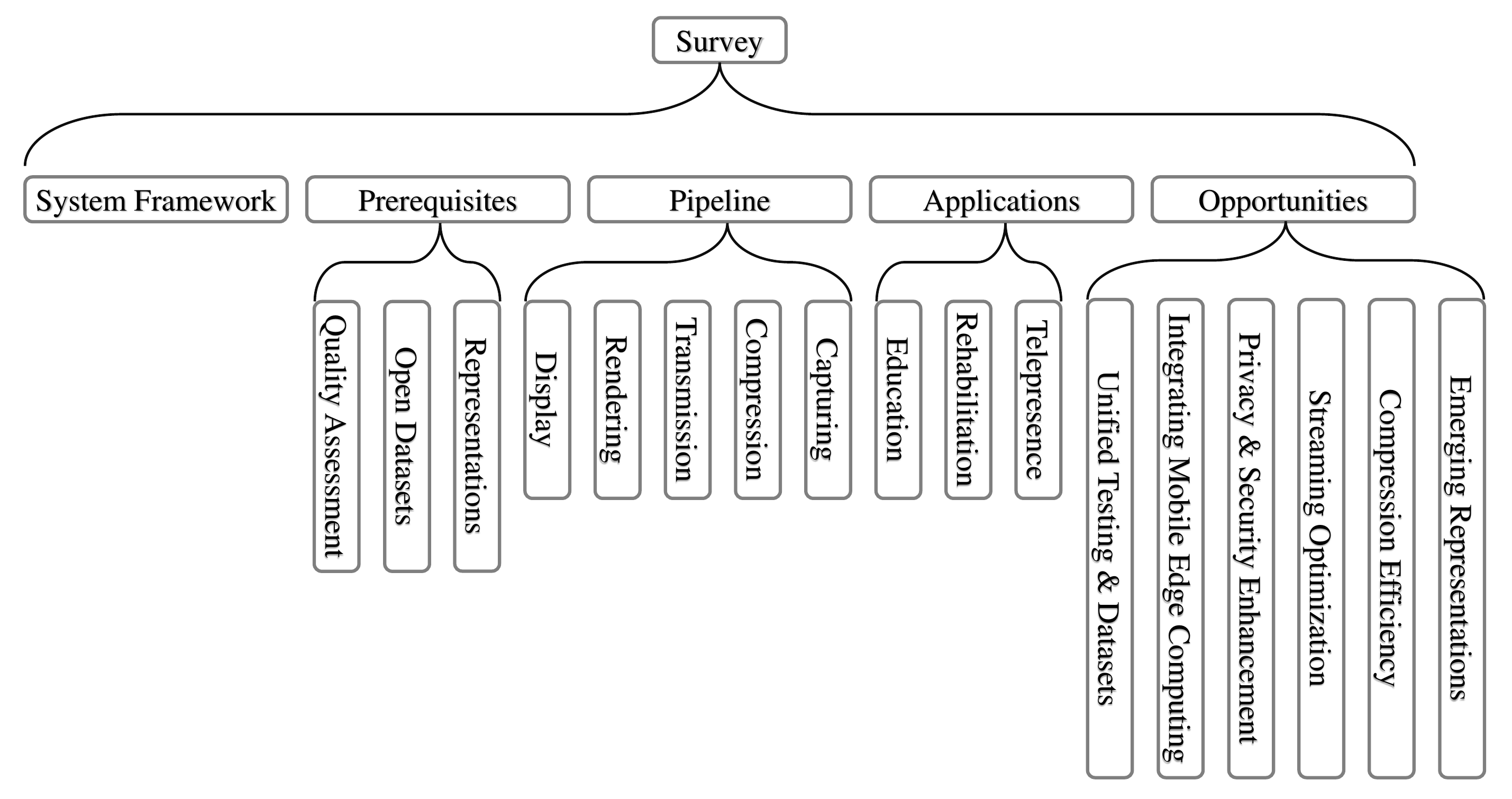}
\end{center}
\vspace{-25pt}
\caption{\label{fig:organizations}%
\revise{Organization of this survey.}}
\end{figure}

\subsection{Organization}
Fig.~\ref{fig:organizations} provides a comprehensive overview of our survey's structure, which is organized into five primary categories: System Framework, Prerequisites, Pipeline, Applications, and Opportunities.

\begin{itemize}[leftmargin=15pt]
    \item \emph{System Framework (Section~\ref{sec:system}):} This section introduces a general framework for volumetric video service, detailing its core components and their interactions.

    \item \emph{Prerequisites (Section~\ref{sec:prerequisite}):} To lay the foundation for understanding volumetric video systems, this section outlines essential prerequisites. These include various 3D representations, relevant open datasets, and quality assessments.

    \item \emph{Pipeline (Section~\ref{sec:pipeline}):} This section delves into the end-to-end pipeline of volumetric video services, examining related works in each stage. The stages covered include Capturing, Compression, Transmission, Rendering, and Display, offering a thorough discussion of the processes involved.

    \item \emph{Applications (Section~\ref{sec:application}):} This section explores emerging applications of volumetric videos, emphasizing their growing impact across various domains. This section highlights the potential of volumetric videos in revolutionizing these fields.

    \item \emph{Opportunities (Section~\ref{sec:opportunity}):} This section discusses the various research challenges and opportunities in the field of volumetric video services.
\end{itemize}

\section{System Framework}
\label{sec:system}

\begin{figure}[t]
\begin{center}
\includegraphics[width=1\columnwidth]{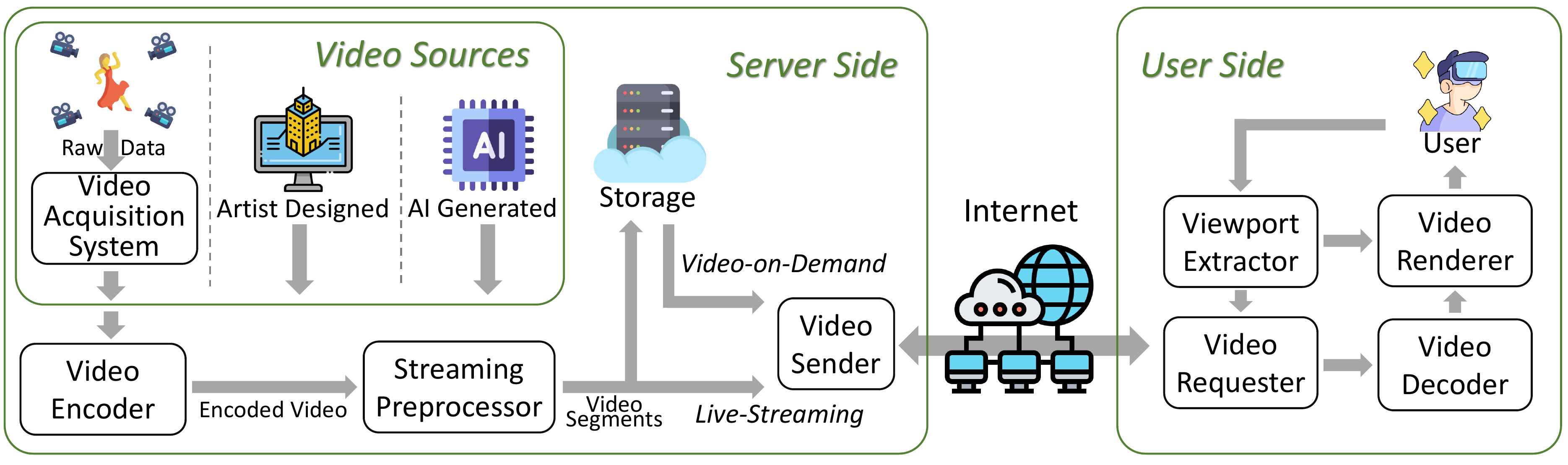}
\end{center}
\vspace{-15pt}
\caption{\label{fig:system}%
A general framework for volumetric video service.}
\vspace{-5pt}
\end{figure}

This section presents a general framework for volumetric video service, as shown in Fig.~\ref{fig:system}. \revise{The framework focuses on the essential components that handle various aspects of volumetric video processing. These components were selected based on their fundamental roles in managing the complexity of volumetric data, which involves capturing, encoding, preprocessing, and rendering 3D content for interactive applications.}

\begin{itemize}[leftmargin=15pt]
\item \emph{\revise{Video Acquisition System}:} It captures volumetric videos using a variety of input data, such as RGB data, depth data, and LiDAR data.

\item \emph{Video Encoder:} It encodes the captured volumetric videos. It may also support tiling for partial streaming and rendering, which can help reduce bandwidth consumption. The algorithm employed varies depending on the 3D representations used.

\item \emph{Streaming Preprocessor:} \revise{It preprocesses raw video data into a format suitable for streaming. For example, it can segment the video into temporal chunks, divide it into spatial tiles for partial streaming, and adjust video quality for adaptive bit-rate streaming. Not all preprocessing steps are necessary; their application depends on the streaming method employed.}

\item \emph{Video Sender:} It is responsible for transmitting the requested content from the server to the user.

\item \emph{Video Requester:} It generates requests for video segments with varying bit-rates, timestamps, or locations. It is typically the core decision-maker for optimizing the streaming system.

\item \emph{Video Decoder:} It decodes the received videos, which is the opposite of the encoder.

\item \emph{Video Renderer:} \revise{It converts 3D content into a format suitable for display, adapting based on the type of display device used. For 2D screens or HMDs, the renderer converts 3D content into 2D projections based on the user's viewpoint. For holographic or light field displays, it enables direct interaction with the 3D content without conversion.}

\item \emph{Viewport Extractor:} It receives viewport information and predicts future viewport trajectory. This information assists the Video Requester in making decisions and enables the Video Renderer to accurately render the 3D content.

\end{itemize}

\section{Prerequisites}
\label{sec:prerequisite}

This section outlines essential prerequisites for volumetric video services, which include 3D representations, relevant open datasets, and quality assessment issues.

\subsection{Representation}

Over the course of several decades, traditional video has reached a relatively mature form of representation. However, its counterpart, 3D volumetric video is still in its early stages, with a plethora of representation formats. Depending on the type of volumetric representation that is transmitted and rendered, a range of streaming strategies and techniques are developed. \revise{These representations can be categorized into two types: explicit and implicit, based on how the 3D data is structured and represented. Explicit representations define 3D content using clearly defined geometric elements, where the shape and position of objects are directly represented. In contrast, implicit representations use mathematical functions or neural networks to represent the 3D content indirectly, storing the content in a more abstract form.} Most existing works mainly focus on utilizing explicit representations such as 3D mesh and point cloud.  These are generally preferred due to their ease of implementation and optimization. The prevalent representation formats used in volumetric videos are summarized below. Table~\ref{table:repre} shows comparisons and examples of different representations. \revise{In this table, for \textit{Visual Quality}, a Low level indicates limited detail, lower resolution, and less realistic rendering, while a High level offers high detail, fine resolution, and photorealistic capabilities. For \textit{Computing Resources}, a Low level can be managed on consumer-grade hardware with minimal load, while a High level demands significant resources, often requiring high-end or cloud-based processing. For \textit{Editability}, an Easy level allows for simple modifications with common tools, making it suitable for frequent updates or real-time changes, while a Hard level is challenging to modify due to its complex representation.}

\begin{table*}[!t]\centering
%\resizebox{1.0\linewidth}{!}{
\centering
\caption{Comparisons and examples of different representations.}
\vspace{-8pt}
\begin{tabular}{|p{2.5cm}|l|p{1.8cm}|p{1.8cm}|p{1.5cm}|l|}
\hline
\textbf{Representation} & \textbf{Size} & \textbf{Visual Quality} & \textbf{Computing Resources} & \textbf{Editability} &\textbf{Example} \\ \hline
Point Cloud \cite{DBLP:conf/icra/RusuC11,DBLP:journals/pami/GuoWHLLB21}  & Large & Low & Low & Easy & \raisebox{-.5\height}{\includegraphics[width=0.1\columnwidth]{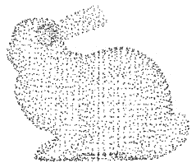}}\\ \hline
Mesh \cite{DBLP:conf/imr/Owen98, DBLP:journals/cgf/BommesLPPSTZ13} & Medium & Medium & Medium & Medium & \raisebox{-.5\height}{\includegraphics[width=0.1\columnwidth]{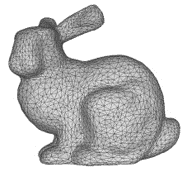}}\\ \hline
Voxel \cite{xu2021voxel,DBLP:journals/sensors/AleksandrovZH21} & Medium & Low & Low & Easy & \raisebox{-.5\height}{\includegraphics[width=0.1\columnwidth]{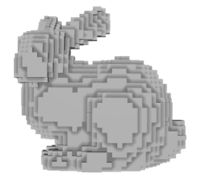}}\\ \hline
Plenoptic Point Cloud \cite{DBLP:journals/tcsv/ChanNGCS05,DBLP:journals/tip/SandriQC19}  & Huge & High & High & Medium &\raisebox{-.5\height}{\includegraphics[width=0.1\columnwidth]{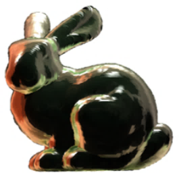}}\\ \hline
%V-PCC  & Small & Low & No & \raisebox{-.5\height}{\includegraphics[width=0.1\columnwidth]{Figures/repre/VPCC.png}}\\ \hline
Implicit surfaces \cite{DBLP:journals/csur/AraujoLJJW15,DBLP:journals/tvcg/Cani-GascuelD97} & Medium & Medium & Medium & Hard & \raisebox{-.5\height}{\includegraphics[width=0.15\columnwidth]{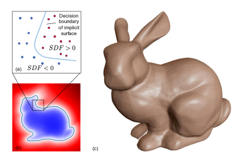}}\\ \hline
Neural Radiance Fields \cite{DBLP:journals/cacm/MildenhallSTBRN22,DBLP:journals/cgf/XieTSLYKTTSS22} & Medium & Very high & Very high & Hard & \raisebox{-.5\height}{\includegraphics[width=0.12\columnwidth]{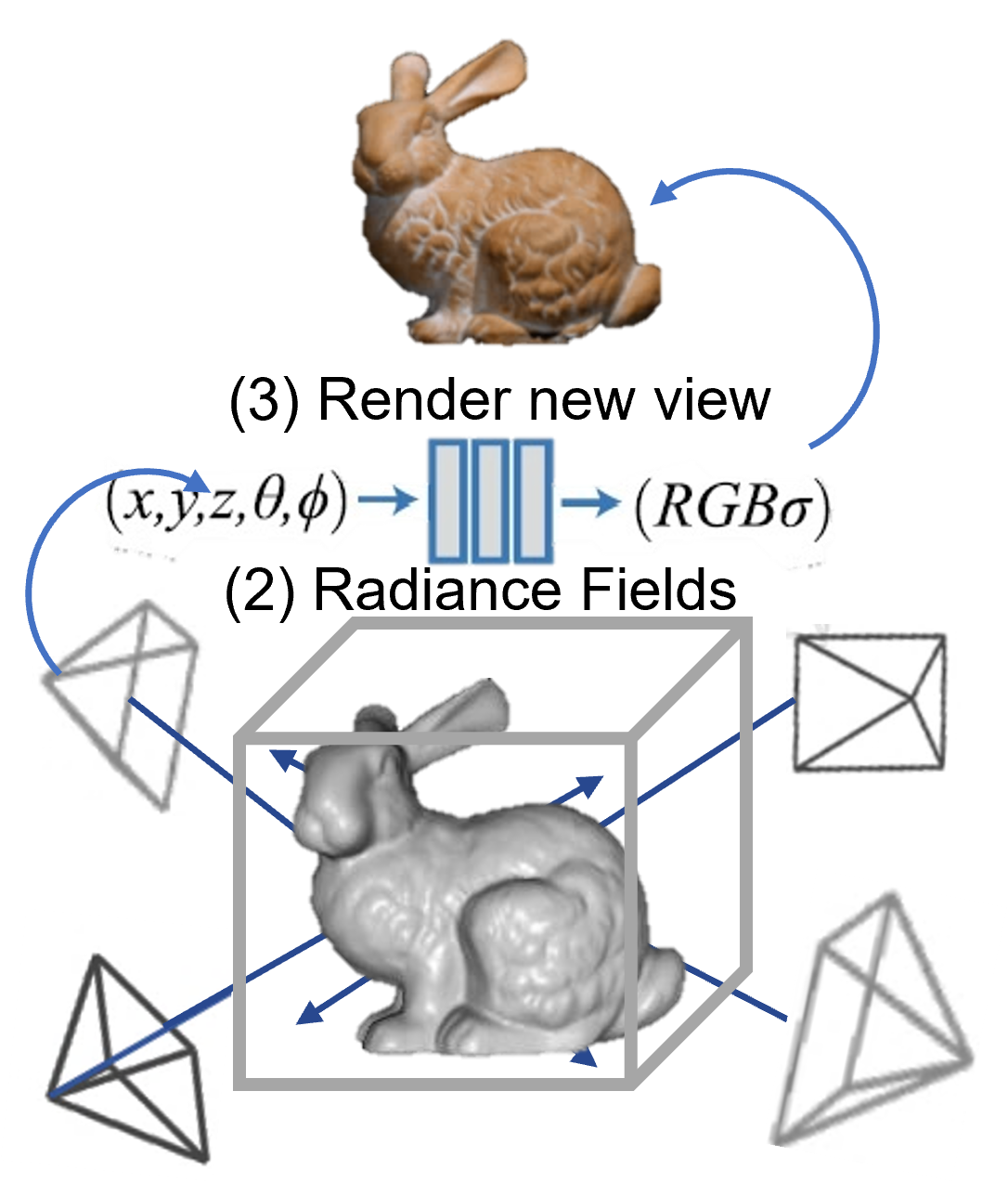}}\\ \hline
\end{tabular}
%}
\label{table:repre}
\vspace{-12pt}
\end{table*}

\begin{itemize}[leftmargin=15pt]
    \item \emph{Point cloud (PtCl):} This format employs a large group of individual data points in space to represent a 3D object. Each point contains spatial coordinates and additional attributes (e.g., RGB color). PtCl is the raw form collected from LiDAR and RGB-D cameras. It is relatively simple and flexible to handle on client devices, allowing easy manipulation and enabling live streaming. PtCls can accurately capture the geometry and shape of objects as they directly represent the surface points. However, because of the discrete nature of the PtCl, it requires a huge bandwidth, and its defects in detail expression and limited resolution hinder its ability to achieve a photorealistic perspective~\revise{\cite{DBLP:conf/qomex/ZermanOGS20}}.
    \item \emph{Mesh:} A polygon mesh consists of vertices, edges, and faces that define the shape of a polyhedral object. The 3D mesh format is a collection of meshes that represent the spatial surface, color, and texture of the object. Compared with PtCl, the mesh is suitable for representing complex geometry with smooth surfaces. Mesh enables the capture of intricate details and textures on object surfaces, facilitating accurate modeling of surface properties and reflectance. Consequently, they enhance the production of visually compelling volumetric videos. However, real-time capture and manipulation of mesh can be challenging, as changes to the topology or connectivity between vertices often require extensive computational effort and more memory compared to other representations~\revise{\cite{DBLP:conf/nips/LiLMKW0K20}}, which impedes live volumetric video streaming to be deployed on common devices.
    \item \emph{Voxel:} The concept of a voxel is derived from the pixel, where the 2D pixel is extended to a 3D voxel. The difference is that a voxel represents the value of a regular cube in three-dimensional space. However, voxel grids are difficult to capture fine geometric details and can consume significant memory~\revise{\cite{DBLP:journals/cga/KaufmanB88}}, especially for high-resolution or large-scale datasets.
    \item \emph{Plenoptic Point Cloud:} It represents both point cloud and light field information that captures both the geometric and photometric properties of the scene~\cite{DBLP:journals/tip/SandriQC19}. The color appears different depending on the viewing direction, which enables novel view synthesis and free viewpoint rendering~\cite{riegler2020free}. However, it is highly memory-intensive and storage-intensive due to capturing both geometry and radiance information. It also requires complex algorithms for reconstruction and rendering~\revise{\cite{DBLP:journals/tip/SandriQC19}}.
    \item \emph{Implicit surfaces:} Implicit surfaces represent 3D objects as the zero-level set of a function, allowing for intuitive handling of complex topology. The function takes the 3D coordinates as input and outputs a signed distance value, indicating whether the point is inside or outside the object. Implicit surfaces have the advantage of being smooth and continuous, which makes them useful for rendering and shape reconstruction. As they do not require training a neural network, they may struggle with complex shapes and detailed structures. Moreover, they require solving complex equations to determine the surface properties, which can be computationally expensive and challenging to obtain using classical methods.
    
    \item \emph{Neural Radiance Fields (NeRF):} It is a recent technique that uses neural networks to model the volumetric primitive. The captured scene is optimized using multiple 2D views into a neural radiance field model, a 6D function $\Phi$ that generates 2D views (represented by volume density value $\sigma$ and color $c$) from different perspectives related to time $t$ and view direction $(x, y, z)$. i.e. $ \Phi(x, y, z, \theta, \phi)=\sigma, c.$ Compared to other representations, NeRF can represent higher-resolution geometry and appearance to render photorealistic novel views of complex geometry and appearance. However, it requires a large amount of training data and computational resources and additional time for training and inference. The requirements of real-time inference for volumetric video streaming also pose great challenges.
\end{itemize}

Our survey has undertaken an exploration of the diverse representations applicable to volumetric video. Each method manifests its own unique strengths and weaknesses. Ultimately, the choice of representation depends on the specific application requirements and priorities. \revise{It should be noted that some frameworks~\cite{DBLP:journals/tmm/CarballeiraCDBC22,DBLP:journals/tcsv/MielochGMJJRS22,DBLP:journals/esticas/LingGGC19} transmit RGB and depth information directly from the camera to the user, enabling the synthesis of new viewing angles without relying on 3D representations.}

\subsection{Open Datasets}

\begin{table}[!t]\centering
\centering
\caption{Basic information of different datasets.}
\vspace{-8pt}
\resizebox{1.0\textwidth}{!}{
\begin{tabular}{|l|l|l|p{2.45cm}|p{3.5cm}|}
\hline
\textbf{Dataset} & \textbf{Format} & \textbf{Size} & \textbf{Content} & \textbf{Resolution (per frame)} \\ \hline
Owlii \cite{Owlii} & Mesh & 4$\times$20s$\times$30fps & Full human body & \textasciitilde40k triangles with 2048$\times$2048 texture map\\ \hline
8iVFB \cite{8iVFB} & Point cloud & 4$\times$10s$\times$30fps & Full human body & 1024$\times$1024$\times$1024 points\\ \hline
8iVSLF \cite{8iVSLF}  & Voxel & 1$\times$10s$\times$30fps & Full human body & 4096$\times$4096$\times$4096 points\\ \hline
Pag{\'e}s et al. \cite{Volograms} & Mesh & 3$\times$5s$\times$30fps & Full human body & \textasciitilde40k polygons with 4096$\times$4096 texture map\\ \hline
MVUB \cite{MVUB} & Point cloud & 5$\times$(7\textasciitilde10)s$\times$30fps & Upper human body & 4096$\times$4096$\times$4096 points\\ \hline
CWIPC-SXR \cite{cwipc} & Point cloud & 45$\times$(20\textasciitilde50)s$\times$30fps & Full human body & \textasciitilde80k points\\ \hline
Sun et al. \cite{DBLP:conf/mmsys/SunHSOHH23} & Point cloud & 27$\times$600frames & Shape, Full human body, Textile &\textasciitilde120k points \\ \hline
FSVVD \cite{fsvvd} & Point cloud & 26$\times$(4\textasciitilde73)s$\times$30fps& Full human body with full scenes& 700k\textasciitilde1500k points\\ \hline
\end{tabular}
}
\label{table:dataset}
\vspace{-15pt}
\end{table}

Datasets play a vital role in enabling researchers and developers to explore novel ideas and carry out reproducible analyses, ensuring fair comparisons among different solutions. In this section, we present an overview of the existing volumetric video datasets. It is important to note that our focus is solely on volumetric video, and as such, we do not include datasets that feature static content, such as ModelNet~\cite{DBLP:conf/cvpr/WuSKYZTX15}. We make a brief overview of these datasets, and summarize the basic information, which is illustrated in Table ~\ref{table:dataset}.

The Owlii dataset~\cite{Owlii} consists of four dynamic textured human mesh sequences: basketball player, dancer, exercise, and model. Each sequence is captured at 30 frames per second over a 20-second period, containing around 40,000 triangles.

The 8iVFB dataset~\cite{8iVFB} includes four voxelized point cloud sequences: longdress, loot, redandblack, and soldier. Each sequence captures the full body of a human subject using 42 RGB cameras configured in 14 clusters, with each cluster acting as a logical RGBD camera. The sequences are captured at 30 frames per second over a 10-second period.

The 8iVSLF dataset~\cite{8iVSLF} features one 300-frame sequence and six single-frame point clouds, capturing the full body of a human subject using 39 synchronized RGB cameras at 30 frames per second. Each cluster of cameras captured RGB and computed depth-from-stereo.

Pag\'es et al.~\cite{Volograms} provide another volumetric sequence dataset, which comprises three sequences featuring three distinct characters. Each sequence is captured using 12 HD cameras for different purposes and applications, with varying characteristics in terms of texture and movement.

The MVUB dataset~\cite{MVUB} includes five subjects: Andrew, David, Phil, Ricardo, and Sara. The upper bodies of these subjects are captured using four frontal RGBD cameras at 30 frames per second over a 7-10 second period for each sequence.

One of the biggest challenges with the previously mentioned datasets is their relatively small size, as they contain only a few videos. The CWIPC-SXR dataset~\cite{cwipc} offers a much larger selection of 45 unique sequences, designed for various use cases in social scenarios, including "Education and Training," "Healthcare," "Communication and Social Interactions," and "Performance and Sports."

While the above datasets are limited to a single human body, recent developments have led to the creation of more diverse datasets. Sun et al.~\cite{DBLP:conf/mmsys/SunHSOHH23} have collected a dataset containing nine objects in three categories (shape, human body, and textile) with different animation patterns. Their dataset features synthetically generated objects with pre-determined motion patterns, enabling the generation of motion vectors for the points.

Another notable dataset is the FSVVD~\cite{fsvvd}, which depicts human interactions with objects and full related scenes. This dataset offers over 30 different daily scenarios and aims to provide a universal dataset for evaluation and research on the application of volumetric representation in real-life scenarios.

\subsection{Quality Assessment}
\revise{While processes such as compression, transmission, and rendering can introduce distortions that degrade content quality, other steps, like pre- and post-processing, can enhance the visual experience. Quality assessment is crucial in ensuring that the processed content remains true to its intended form. For a deeper understanding of how quality can be preserved and improved throughout the processing chain, readers may refer to relevant works by Qualinet~\cite{Qualinet1,Qualinet2}.}
As a result, it is crucial to develop mechanisms capable of quantifying these distortions to create effective compression, transmission, or rendering methods. For example, to evaluate the effectiveness of a compression model, two primary metrics are typically employed: the compression rate and the distortion level. Therefore, it is essential to have a mechanism to quantify them. Similarly, when training a neural-based model, a mechanism to quantify distortion is also indispensable to use as a loss function.

Quality assessment has been extensively studied for traditional video~\cite{DBLP:journals/tbc/ChikkerurSRK11,DBLP:journals/comsur/ChenWZ15}, with decades of research leading to standardized test methodologies and evaluation procedures. \revise{However, applying traditional methodologies and algorithms to volumetric videos is not straightforward. Unlike traditional 2D video, which is constrained to a regular grid of pixels, volumetric video is represented in a more complex, 3D format. Moreover, because observers are free to navigate and explore the content from different viewpoints, traditional objective quality metrics (and even subjective methodologies) must be redesigned to account for this added level of interaction and immersion.}

Quality assessment approaches can be broadly categorized into two types: subjective and objective. \revise{Subjective quality assessment involves the direct evaluation of video quality by a large number of observers. Typically, this approach requires these observers to evaluate the quality of the videos, and the final quality score is calculated by averaging or analyzing the differences in the scores provided.} While subjective quality assessment is essential, it is not practical for widespread use due to the significant amount of manpower, time, and financial resources required. Therefore, many researchers have focused on developing reliable and effective objective quality assessment methods. \revise{Objective quality assessment includes the vision modeling approach, which simulates the human visual system, and the engineering approach, which analyzes specific features or artifacts from video compression or transmission~\cite{winkler2006perceptual}.} This approach eliminates the need for subjective evaluation and provides a more efficient and cost-effective way to assess video quality.

In the following subsections, we will describe related works in subjective and objective quality assessment methods. Alexiou et al.~\cite{IVT18} conduct a survey that focused on point cloud and mesh quality assessment. In contrast, we will provide a more general perspective on volumetric video.

\subsubsection{Subjective Quality Assessment}
In the case of traditional videos, \revise{subjective methods are created based on recommendations from established standardization organizations, such as the International Telecommunication Union (ITU)~\cite{ITU}, or expert panels brought together by researchers, like the Society of Motion Picture and Television Engineers (SMPTE)~\cite{SMPTE}.} A novel approach for assessing 360° video quality has recently been standardized~\cite{DBLP:journals/tmm/GutierrezPOSCMV22}. Meanwhile, the process of establishing standards for volumetric videos is still in progress. As of now, there are no specific guidelines or recommendations in place for the emerging field of volumetric video.

It is impossible to view the entirety of the volumetric video at once. To obtain accurate subjective quality scores for the entire volumetric video, the experimenter must ensure that the video is inspected thoroughly by the participants of the subjective experiment. This can be achieved in two primary ways: either by allowing viewers to interact with the volumetric video themselves, or by presenting a representative stimulus that does not allow for viewer interaction, such as a sequence of images from predetermined viewpoints. While the former method more closely mimics real-life volumetric video consumption, the latter method provides a consistent experience across all subjects, ensuring reproducibility.

For non-interactive ways, a significant portion of research concentrates on static content~\cite{DBLP:conf/qomex/CruzDAPDPPE19,DBLP:journals/tmm/PanCB05}, which is less complex compared to dynamic videos, as there is no need to consider potential interactions between camera movement and video actions. Schwarz et al.~\cite{DBLP:journals/esticas/SchwarzPBBCCCKL19} analyze both static and dynamic colored point cloud models across various encoding types, configurations, and bit-rates. Hooft et al.~\cite{DBLP:conf/qomex/HooftVTBTS20} investigate the subjective quality assessment of dynamic, colored point clouds within an adaptive streaming context. V\'asa and Skala~\cite{DBLP:journals/tvcg/VasaS11} and Torkhani et al.~\cite{DBLP:journals/spic/TorkhaniWC15}, suggest quality assessment experiments that involve dynamic meshes, incorporating different noise and compression distortions. The evaluated stimuli consist of mesh sequence videos, rendered from fixed perspectives. The methodologies employed are single stimulus rating and multiple stimulus rating, respectively. Zerman et al.~\cite{DBLP:conf/qomex/ZermanOGS20} employ the absolute category rating with hidden reference method to juxtapose dynamic textured meshes and colored point clouds in a compression setting.

For interactive ways, Subramanyam et al.~\cite{DBLP:conf/vr/SubramanyamLVC20} conduct an experiment to evaluate the quality of digital humans represented as dynamic point clouds, in both 3 DoF and 6 DoF conditions. The models were displayed using fixed-sized quads in a virtual scene, and participants assessed them using an ACR-HR protocol. In the 6 DoF scenario, users were able to navigate using physical movements, while in the 3 DoF counterpart, they remained seated. The researchers also extend their work to 2DTV in subsequent study~\cite{DBLP:conf/vr/SubramanyamLVC202}. \revise{Paudyal et al.~\cite{DBLP:journals/tbc/PaudyalBCGC21} examine the impact of visualization techniques on the quality of light field images, recommending a visualization technique for subjective quality assessment. Their study also explored the perceptual visual impact of compression and noise artifacts, and analyzed the performance of 2D image quality measures applied to light field images.}

\revise{In summary, subjective quality assessment methods, though essential for accurately capturing human perception of video quality, are resource-intensive and often impractical for large-scale or real-time systems. While they provide valuable ground truth for validating objective metrics, their reliance on human testers limits their scalability. Therefore, subjective assessment is best suited for benchmarking new quality assessment methods, particularly in controlled environments such as laboratory settings. However, for practical deployment in volumetric video streaming, it must be complemented by objective methods to ensure real-time performance and scalability.}

\subsubsection{Objective Quality Assessment}
When evaluating the quality of volumetric videos, it may seem reasonable to incorporate methods used in traditional video quality assessment~\cite{DBLP:conf/icpr/HoreZ10}, including structural similarity index measure (SSIM) and peak signal-to-noise ratio (PSNR). However, it is important to note that simple geometric or color distances between 3D models are not strongly correlated with human perception due to the lack of consideration for perceptual characteristics of the human visual system~\cite{DBLP:journals/cgf/Lavoue11}.

There are two main types of volumetric quality assessment approaches: model-based and image-based. Model-based approaches involve comparing the 3D representation itself directly, while image-based approaches compare the projection image that the viewer sees in their viewport. \revise{For image-based quality assessment, after projection, quality assessment methods for traditional 2D images~\cite{DBLP:journals/chinaf/ZhaiM20} can be employed. But they may not always perform well in the presence of 3D-specific distortions, such as view synthesis artifacts. These distortions can degrade the perceived quality in ways that are not adequately captured by conventional 2D metrics. As a result, alternative or extended metrics that account for 3D-specific issues may be necessary to ensure accurate quality assessment in volumetric video.}

The earliest model-based quality assessment for volumetric video utilized simple distances between attributes of matched points to measure local errors~\cite{DBLP:conf/icip/TianOFCV17}. However, these point-to-point metrics do not account for perceptual characteristics of the human visual system. To address this limitation, an alternative was proposed, which used distances that are more perceptually relevant, known as the point-to-plane metric~\cite{DBLP:conf/icip/TianOFCV17}.
Recent proposals have expanded beyond surface properties extracted from point samples, incorporating statistics to capture relationships between points in the same local neighborhood. For instance, PC-MSDM~\cite{DBLP:conf/qomex/MeynetDL19} was proposed to use the relative difference between local curvature statistics and PCQM~\cite{DBLP:conf/qomex/MeynetNDL20} leverage a weighting function to regularize feature contributions in the final quality prediction. The PointSSIM~\cite{DBLP:conf/icmcs/AlexiouE20} captures perceptual degradations based on the relative difference of statistical dispersion estimators applied on local populations of location, normal, curvature, and luminance data. VQA-CPC~\cite{hua2020vqa} relies on statistics of geometric and color quantities.
More recently, GraphSIM~\cite{DBLP:journals/pami/YangMXLS22} denotes a graph signal processing-based approach, which evaluates statistical moments of color gradients computed over graphs. A multi-scale version of this metric, known as MS-GraphSIM~\cite{DBLP:conf/mm/ZhangYX21}, was presented as an extension. Xu et al.~\cite{DBLP:journals/tbc/XuYYH22} presented the EPES, a metric based on potential energy. In the work of Diniz et al.~\cite{DBLP:conf/qomex/DinizFF20}, local binary patterns on the luminance channel are applied in local neighborhoods. This work was later extended~\cite{DBLP:conf/icip/DinizFF20} to consider the point-to-plane distance and the point-to-point distance between corresponding feature maps in the quality prediction. Another proposed descriptor~\cite{DBLP:conf/mmsp/DinizFF20}, known as local luminance patterns, introduces a voxelization stage in the metric's pipeline to alleviate its sensitivity to different voxelization parameters. \revise{Ling et al.~\cite{DBLP:journals/esticas/LingGGC19} examine the impact of hypothetical rendering trajectories on perceived quality.}

\revise{Objective quality assessment methods are essential for evaluating volumetric video in real-time applications where subjective assessment is impractical. Broadly, these methods can be classified into model-based and image-based approaches, each suited to different scenarios. Model-based approaches focus on geometric and color differences in point clouds and meshes. They are effective for applications requiring high accuracy in geometry, such as 3D reconstructions, but may not fully capture the perceptual quality experienced by viewers. These methods are best applied when geometry preservation is the primary concern. Image-based methods evaluate 2D projections of 3D content, making them more aligned with human visual perception. These metrics are particularly useful for real-time applications, where rendering occurs in 2D for displays. They offer a balance between computational efficiency and visual quality, making them suitable for dynamic, interactive environments. In summary, for real-time streaming, image-based metrics are recommended for their balance between quality and speed. For applications requiring precise geometry, model-based metrics are ideal.}

\begin{figure}[t]
\begin{center}
\includegraphics[width=0.8\columnwidth]{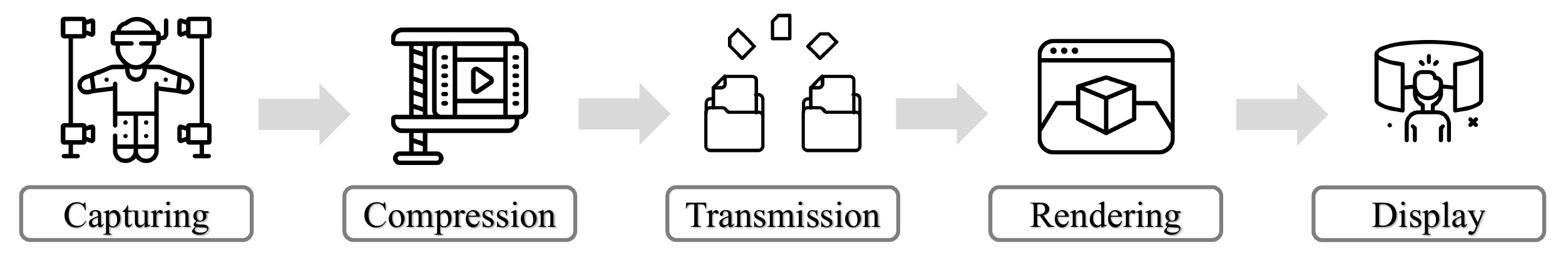}
\end{center}
\vspace{-10pt}
\caption{\label{fig:pipeline}%
End-to-end pipeline of volumetric video services.}
\end{figure}

\section{Pipeline}

\label{sec:pipeline}
This section delves into the end-to-end pipeline of volumetric video services, as shown in Fig.~\ref{fig:pipeline}, examining related works in each stage. The stages covered include Capturing, Compression, Transmission, Rendering, and Display.

\subsection{Capturing}
Video capturing is the first step to producing volumetric videos. Since volumetric video is represented by 3D content, the capture process is quite different from traditional flat videos consisting of arrays of pixels, which involves more sophisticated devices and requires additional post-processing steps. In this section, we cover the different techniques for capturing volumetric videos.

%\noindent\textbf{Post-preprocessing:}
\subsubsection{Capture Setup}
Volumetric video capture involves intricate setups and multiple post-processing steps to produce reconstructed 3D scenes. We classify current volumetric video capture setups into three categories: \textit{Calibrated Camera Array}, \textit{Monocular Camera}, and \textit{Advanced Capture Techniques}.

\textit{Calibrated Camera Array:}
Volumetric videos are typically captured using depth camera arrays, such as the Microsoft Azure Kinect~\cite{Azure_Kinect} and Intel RealSense~\cite{realsense}. These camera arrays are positioned around the target region, facing inwards. Since each camera captures data from a different angle, it is necessary to merge the data into the same coordinate system using camera calibration parameters, which allows for the construction of a complete 3D scene. \revise{The calibration process normally generates two sets of parameters: \textit{intrinsic parameters}, which include characteristics of the cameras, such as focal length and principal points~\cite{principal_points}, describing the characteristics of the cameras, and \textit{extrinsic parameters}, which define the camera's relative positions and orientations.}

\begin{figure}[t]
    \centering
        \begin{subfigure}[t]{0.3\textwidth}
        \centering
        \includegraphics[width=\linewidth, trim=90 0 90 0]{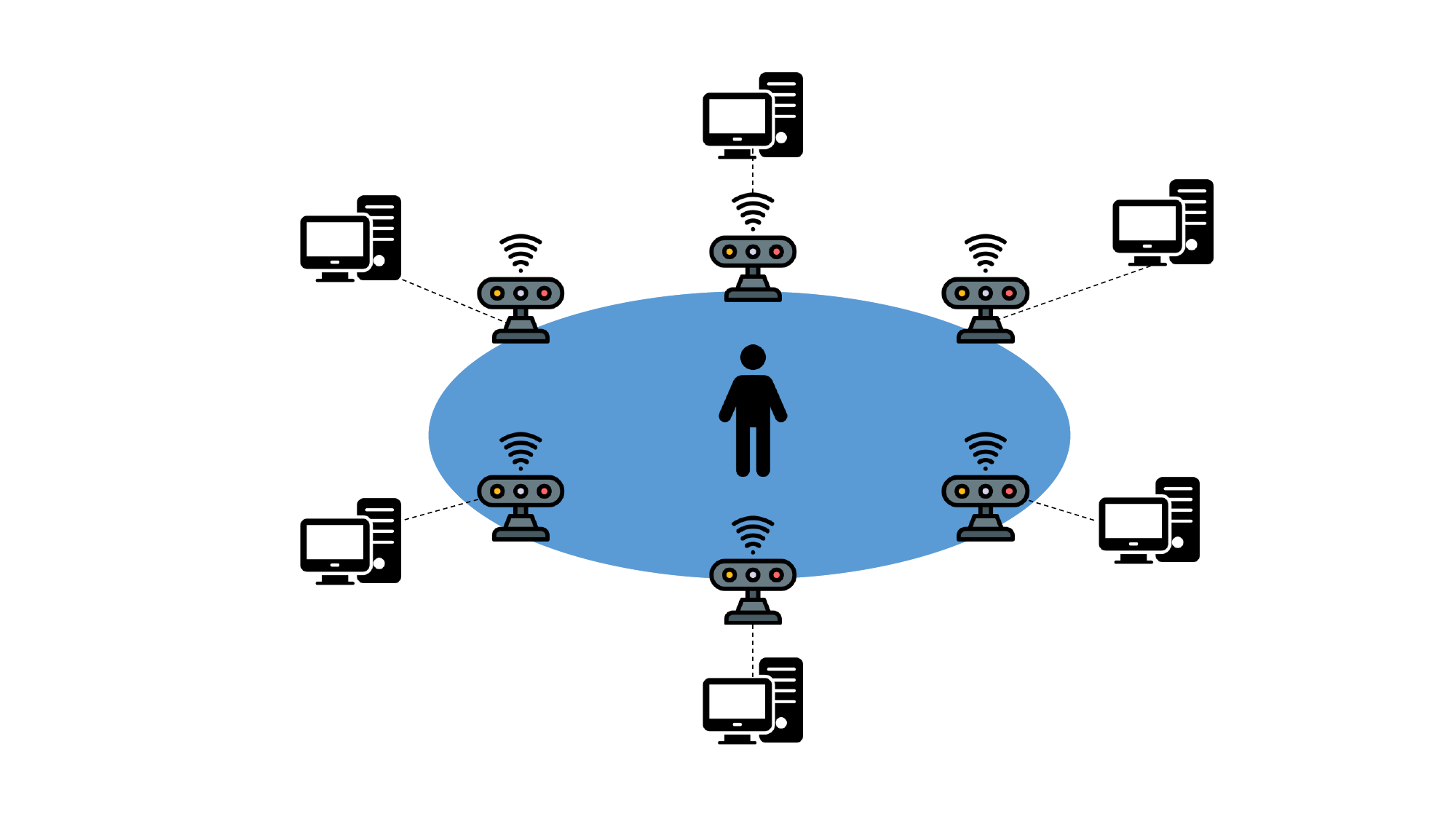}
        \caption{Camera Array}
        \label{fig: camera_array}
    \end{subfigure}
    \hfill
        \begin{subfigure}[t]{0.3\textwidth}  
        \centering
        \includegraphics[width=\linewidth]{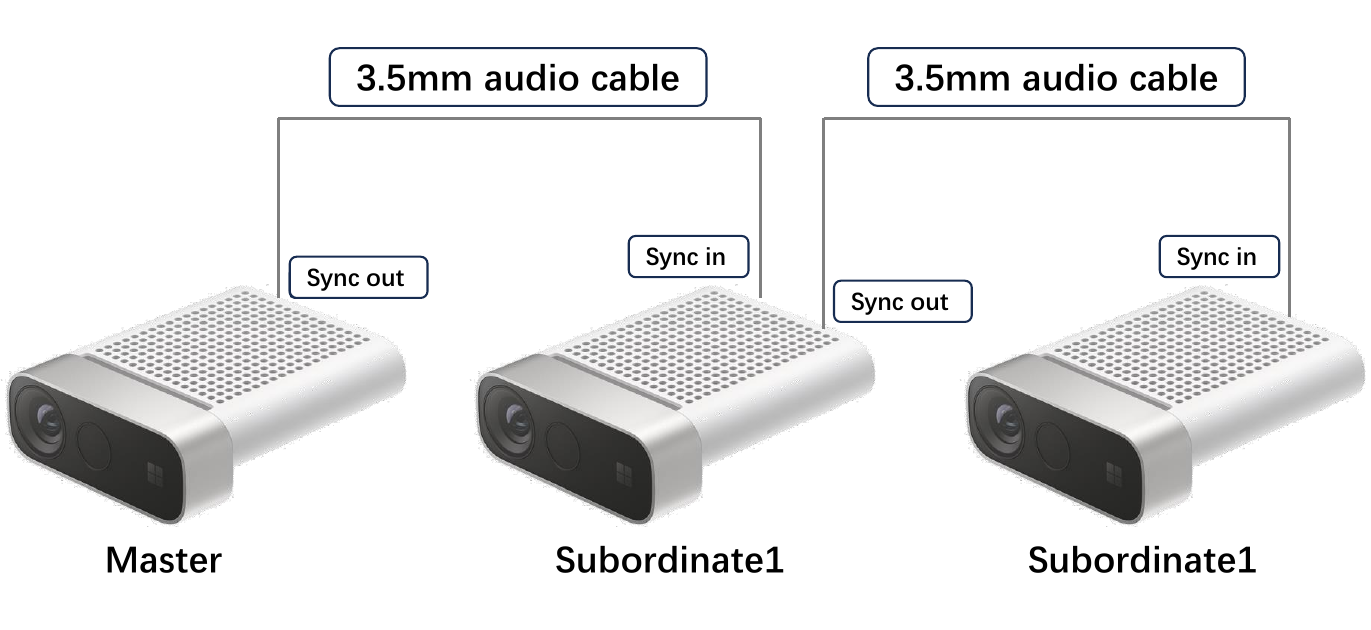}
        \caption{Daisy-chain Topology}
        \label{fig: Daisy-chain}
    \end{subfigure}
    \hfill
    \begin{subfigure}[t]{0.3\textwidth}    
        \centering
        \includegraphics[width=\linewidth]{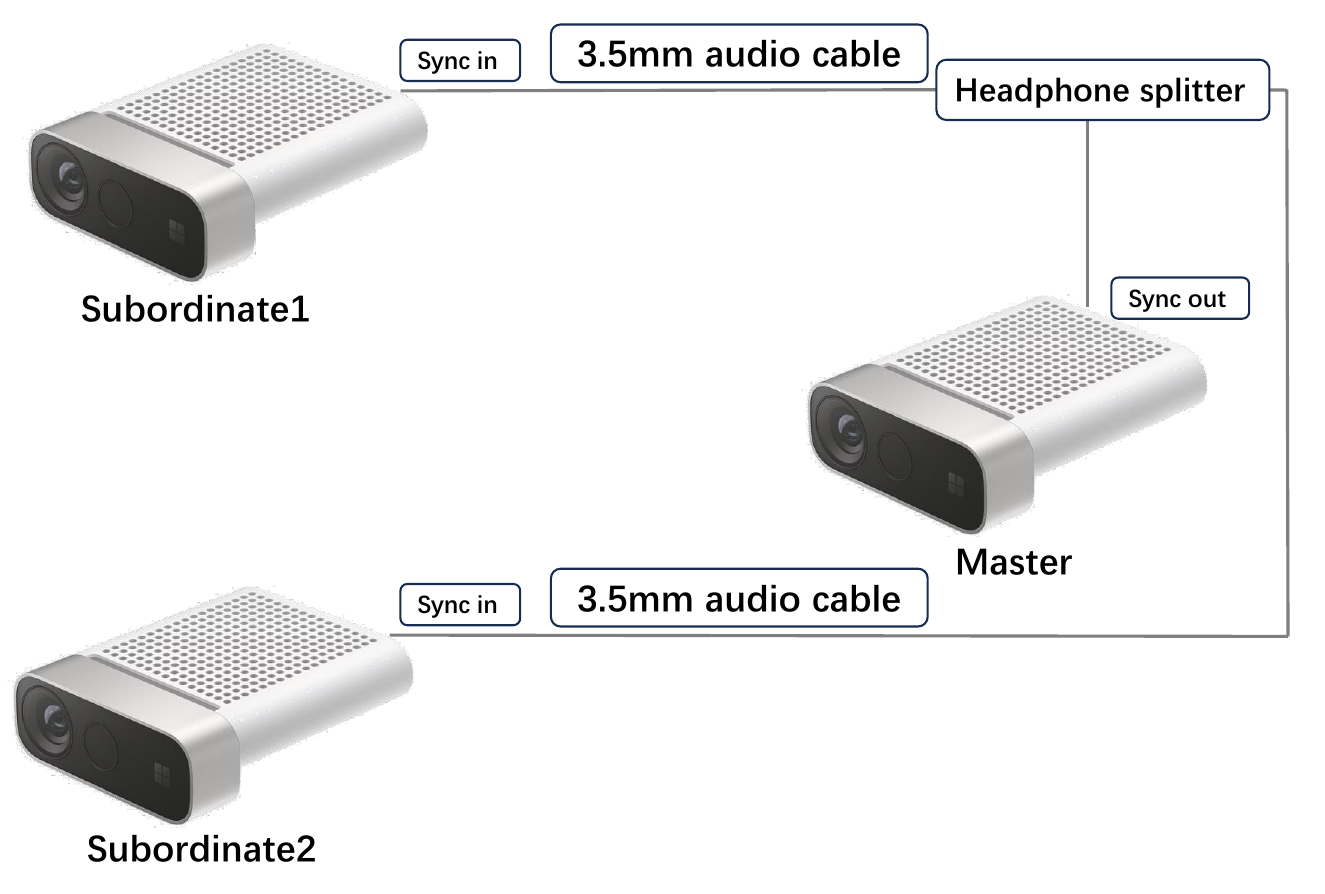}
        \caption{Star Topology}
        \label{fig: Star}
    \end{subfigure}
    \vspace{-5pt}
    \caption{Illustrations for camera array setup.}
    \vspace{-5pt}
    \label{fig:all-three-figures}
\end{figure}

Common camera array setups are illustrated in Fig.~\ref{fig: camera_array}, where cameras are placed around the target region, each attached to a processing unit. Before the capture begins, \emph{multi-camera calibration} and \emph{temporal synchronization} are conducted for the convenience of subsequent processing.

Multi-camera calibration is traditionally achieved through marker-based methods, where camera views are aligned using markers that are then registered among themselves~\cite{camera_calibration_ETH,camera_calibration_ZZY}. Alternatively, structure-based methods can be used, where physical objects such as a stack of specific-sized boxes are placed in the center of the capture region for calibration \cite{DBLP:conf/icra/ZeislP16}. Data-driven correspondence establishment is used to initially match images, followed by global optimization to estimate a solution with respect to the coordinate system of the structure. 
\revise{Recently, Artificial intelligence (AI) techniques have emerged as effective tools for camera calibration~\cite{liao2023deep}. These methods use deep learning models to automate the calibration process by learning spatial correspondences and camera parameters from large datasets. AI-based techniques can eliminate the need for physical markers or structures, offering greater flexibility and adaptability, especially in dynamic environments. For example, Convolutional Neural Networks (CNNs)~\cite{9412653} can be trained to detect keypoints and match them across different camera views, improving calibration accuracy and reducing manual effort. AI-driven calibration also enables real-time adjustments to camera positions and orientations~\cite{hu2024teleor}, accommodating setups where cameras are frequently moved or adjusted, which would otherwise require re-calibration using traditional methods.}

Temporal synchronization is crucial in capturing volumetric video using a camera array. Both hardware and software synchronization~\cite{DBLP:conf/mm/ShresthaBW07,DBLP:conf/iccp/AnsariWGC19} are necessary to ensure that every camera captures the scene simultaneously from different angles. Hardware synchronization involves physically connecting the cameras using cables in a specific topology, such as a daisy chain or star, with one device as the 'master device' and others as 'subordinate devices'. The master device triggers the subordinate devices to capture the scene simultaneously. Fig.~\ref{fig: Daisy-chain} and Fig.~\ref{fig: Star} illustrate the daisy-chain and star topologies, respectively. However, when capturing with a server and multiple host PCs attached to each device, the streams of each device may not be in sync with each other or with the server. Therefore, software-level synchronization is necessary to synchronize the clocks of each sensor's processing unit and the server. The Precision Time Protocol (PTP) \cite{DBLP:conf/ict/KannistoVHH04} is commonly used in practice to align the clocks of every sensor processing unit and the server to a single global timeline. This ensures that every frame captured by each device is synchronized and can be merged seamlessly into a single volumetric video.

\textit{Monocular Camera:}
Recent developments in deep learning and computer vision have made it possible to capture volumetric video using just a single RGB camera~\cite{Wimbauer_2023_CVPR,Geng_2023_CVPR,xu2018monoperfcap,yang2024depth}. This simplifies the process and makes volumetric video capture more accessible to a wider range of creators and industries.
There are several methods for capturing volumetric video using a monocular camera, but two of the most popular techniques are Structure from Motion (SfM)~\cite{DBLP:conf/cvpr/SchonbergerF16,vijayanarasimhan2017sfm} and Single-View Depth Estimation~\cite{yang2024depth,DBLP:journals/ijon/MingMFY21,garg2016unsupervised}.

SfM involves capturing multiple images of a subject from different viewpoints and using algorithms to estimate the 3D structure of the scene \cite{wei2020deepsfm,DBLP:conf/cvpr/SchonbergerF16}. The process of SfM typically involves several steps, including feature extraction \cite{brachmann2018learning}, feature matching, camera pose estimation, triangulation, and bundle adjustment. Feature extraction involves identifying distinctive features in each image, such as corners, edges, or blobs. Feature matching involves determining which features in different images correspond to the same 3D point. Camera pose estimation involves estimating the position and orientation of the camera for each image. Triangulation involves computing the 3D position of each feature point by intersecting the rays emanating from the camera centers. Finally, bundle adjustment involves refining the camera parameters and feature positions to minimize the reprojection error, which measures the difference between the observed and predicted image locations of the feature points.

Single-View Depth Estimation, on the other hand, involves estimating the depth of a scene from a single image~\cite{mertan2022single,laina2016deeper}. To capture volumetric video using this technique, multiple images of the subject are captured from different viewpoints, and the depth of each image is estimated using single-view depth estimation. The estimated depths are then combined to create a 3D point cloud, which is further processed by surface reconstruction algorithms to create a 3D mesh of the scene. Finally, texture mapping techniques are applied to map the captured images onto the 3D mesh to create a textured 3D model of the subject.

\revise{SfM and Single-View Depth Estimation each offer unique advantages and drawbacks, necessitating a thoughtful selection based on project-specific criteria. SfM facilitates comprehensive scene reconstruction, enabling immersive exploration but with lengthier processing times. In contrast, Single-View Depth Estimation excels in speedy object reconstruction but struggles to capture entire scenes in a single shot, potentially compromising scene integrity. The choice of method depends on processing speed, scene complexity, and desired detail, highlighting the need to align the technique with project requirements for optimal results.}

\revise{\textit{Advanced Capture Techniques:}
In addition to camera arrays and monocular setups, other advanced techniques are gaining prominence in volumetric video capture:}

\revise{Light Field Cameras capture both the intensity and direction of light rays in a scene, enabling post-capture perspective changes and offering more freedom of movement in the viewing experience~\cite{zhou2021review}. This technology captures 4D light fields, which can be rendered as realistic volumetric scenes with accurate depth perception.}

\revise{Holographic Capture Systems such as Looking Glass Factory's holographic displays use specialized sensors to capture and display volumetric content in 3D without the need for VR headsets. These systems record and reconstruct light waves from the scene to create fully immersive 3D holograms~\cite{wang2020holographic}.}

\subsubsection{Data Post-processing}

To generate a continuous complete volumetric video sequence, the captured data often need to undergo a series of post-processing procedures. Typically, the captured raw data contains color image sequences along with corresponding depth or pose information. %Since temporal synchronization is usually conducted prior to the capturing process, the resulting data can be saved in the same set of frames. 
We introduce several data processing procedures required for generating the complete volumetric scene from the raw data.

%\noindent\textbf{Noise Removal}
%make some survey about PtCl noise removal
 
\noindent\textit{Data Alignment:}
Depth camera arrays capture both texture (color) and geometry (depth) information, but aligning these two types of data is necessary to reconstruct the original 3D scene accurately. First, the raw color and depth data are processed to eliminate noise~\cite{6756961} and correct any distortions~\cite{clarkson2013distortion}. Next, the RGB image and depth image are aligned so that the corresponding pixels in each image occupy the same position in the RGB-D image. This alignment process is accomplished using the calibration data obtained during the Multi-camera Calibration step, which provides information about the intrinsic and extrinsic parameters of the camera. The depth map is transformed into the coordinate system of the color image, and the depth values are assigned to corresponding pixels in the color image, resulting in aligned RGB-D images where each pixel contains both color and depth information~\cite{DBLP:conf/ismar/NewcombeIHMKDKSHF11}.

\noindent\textit{Merging}
Once RGB-D images have been obtained, they can be used to produce a reconstructed scene composed of 3D representations. However, due to the limited field of view of depth sensors, each reconstructed scene only covers a limited area of the target scene. Therefore, it is imperative to merge these sub-scenes in order to compose a complete scene. Using the calibration parameters obtained during the Multi-camera Calibration process, all of the sub-scenes can be projected onto the same coordinate system. The sub-sections are then merged together to construct the complete volumetric scene. It is worth noting that the calibration process must be precise to avoid defects at the edges of the scenes.
    
%\noindent\textit{Training of NeRF}
%For data captured with monocular cameras, a training process should be conducted to acquire the NeRF network to represent the 3D scene. 
%RGB+Depth-->RGBD-->Mesh/PtCl
%multi-camera synchronization, merge, alignment, and noise removal. Several softwares provide such functions, such as VCL3D\footnote{https://vcl3d.github.io/VolumetricCapture/}. 

% \subsubsection{Capture Device}
% Different from traditional flat video that only requires a single RGB camera, capturing of volumetric video usually requires multiple devices such as depth cameras or a specialized combination of Li-DAR and RGB cameras to form an array to capture the scene from different angles. Currently, the most commonly used volumetric capture devices are commodity Microsoft Azure Kinect \cite{Azure_Kinect} and Intel RealSense, both using integrated RGB and depth cameras to capture RGB-D data. 
% %需不需要分开介绍

% We take Azure Kinect as an example, that contains a depth sensor, a spatial microphone array with a video camera, and an orientation sensor. In a volumetric video capture setup, multiple (3\textasciitilde12) Azure Kinects are deployed around the center of the scene, as shown in Figure ~\ref{fig: camera_array}. The devices are interconnected via audio cable to form an array, with one serving as the master device and others as slave devices to achieve time synchronization. During capturing, the master device sends a trigger signal to invoke other slave devices to take a shot simultaneously.

\subsubsection{Open Software}
Currently, only a few open-source volumetric data capturing systems are available. One such system is VCL3D~\cite{sterzentsenko2018low}, which is an open-source software that requires a host PC and several client PCs attached to capturing devices for data acquisition and processing. Each system uses commodity capture devices as input sensors, and after further processing, the volumetric data is represented in either .ply or .pgm file format. To visualize the volumetric data, a 3D visualization program supporting .ply file format, such as Meshlab~\cite{DBLP:conf/egItaly/CignoniCCDGR08}, can be employed.

\subsection{Compression}
Compression is a crucial aspect of volumetric video services because the raw data captured is often large and has redundant information. In order to reduce the data size, efficient compression techniques are necessary. While traditional 2D videos have been extensively researched~\cite{DBLP:journals/imst/Clarke99} and standardized, such as H.265~\cite{DBLP:journals/tcsv/PastuszakA16}, compressing volumetric video is still a relatively new and challenging area. Existing compression techniques for 2D video cannot be directly applied to volumetric video because the data structure and characteristics are fundamentally different. Therefore, the compression of volumetric video remains a new and challenging area. Since the compression algorithm employed varies depending on the 3D representation used, we discuss them categorized by point cloud compression, mesh compression, and NeRF compression.

\subsubsection{Point Cloud Compression}
\revise{Traditional point cloud compression methods are often categorized into two types: transform-based and predictive coding. However, these approaches are not mutually exclusive and can be combined in hybrid methods. Transform-based methods, such as Octree-based~\cite{DBLP:conf/spbg/SchnabelK06} and Wavelet-based methods~\cite{DBLP:journals/tip/NadenauRK03}, apply mathematical transforms to the point cloud data, followed by quantization and encoding. Octree methods achieve high compression ratios but can introduce geometric distortions, while wavelet methods offer better rate-distortion trade-offs at the cost of higher computational requirements. Predictive coding methods, like Delta coding~\cite{DBLP:conf/visualization/DevillersG00} and Context-based methods~\cite{DBLP:conf/icip/GarciaQ17}, predict points based on previously encoded ones and encode the residuals. Their performance depends on point order, providing moderate compression ratios. In practice, hybrid methods combine transform-based techniques with predictive coding to enhance compression efficiency.}

The Moving Picture Experts Group (MPEG)~\cite{MPEG} has standardized point cloud compression through MPEG-PCC~\cite{MPEG-PCC}, which encompasses three distinct technologies targeting specific categories of point cloud data: LIDAR point cloud compression (L-PCC) for dynamically acquired data, surface point cloud compression (S-PCC) for static data, and video-based point cloud compression (V-PCC) for dynamic content. Finalized in early 2020, the MPEG-PCC standard features two classes of solutions~\cite{DBLP:journals/esticas/SchwarzPBBCCCKL19}: the video-based class, represented by V-PCC, suitable for point sets with a relatively uniform distribution of points, and the geometry-based class (G-PCC), which combines L-PCC and S-PCC, making it better suited for sparser distributions. \revise{The core algorithm of V-PCC projects 3D point cloud data onto a 2D plane using an efficient segmentation and directional projection method, followed by compression encoding with the well-established 2D image compression tool HEVC. Compared with previous compression technologies, V-PCC offers high compression efficiency by utilizing the established HEVC tool, which allows easy integration into existing media pipelines for real-time services. However, due to the inherent nature of loss during 3D-2D projection and visual artifacts like patch discontinuities, V-PCC may struggle with scenarios where point cloud data is sparse or contains high detail, compared with geometry-based approaches like G-PCC.}

In recent years, learning-based approaches have become popular due to their high effectiveness. These methods use machine learning algorithms to either learn efficient representations or predict missing points. An initial attempt was made to propose a simple yet effective architecture, consisting only of convolution layers, which achieved promising results~\cite{DBLP:conf/icip/QuachVD19}. This was followed by the introduction of several parameters, including a hyper-prior model, deeper transforms, fine-tuning of the loss function, and adaptive threshold~\cite{DBLP:conf/mmsp/QuachVD20}. The experiments revealed that these additions significantly improved the performance of the network. Another study was conducted using a small number of convolution layers~\cite{DBLP:conf/pcs/GuardaRP19,DBLP:conf/euvip/GuardaRP19}, and interestingly, the performance evaluation results demonstrated that a larger number of filters per layer only contributed to better results at larger bit-rates. Autoencoder-based methods have been shown to learn a compact representation of the point cloud data and achieve high compression ratios~\cite{DBLP:conf/icml/AchlioptasDMG18}, but the quality of the reconstructed point cloud may be compromised. Other methods, such as Generative Adversarial Networks (GANs)~\cite{DBLP:conf/dcc/XuFGMJZW21} and Transformers~\cite{DBLP:conf/mir/LiangL22}, have also been used for point cloud compression. While these methods can generate high-quality point clouds, they often require large amounts of training data.

\subsubsection{Mesh Compression}
The earliest and simplest approach to mesh compression is based on quantization and entropy coding~\cite{DBLP:conf/siggraph/Deering95}. In this method, vertex coordinates are quantized, then compressed using entropy coding. The index data structure is then compressed separately~\cite{DBLP:conf/graphicsinterface/ToumaG98}. More advanced techniques involve exploiting the connectivity information of the mesh. The Edgebreaker algorithm~\cite{DBLP:journals/tvcg/Rossignac99} and the Topological Surgery algorithm~\cite{DBLP:conf/siggraph/GumholdS98} are two pioneering methods in this domain. They both operate by traversing the mesh in a specific order and recording the operations needed to reconstruct it~\cite{DBLP:journals/cgf/AlliezD01}. Predictive coding is another approach that is based on the idea of predicting a vertex's position based on its neighbors~\cite{DBLP:conf/siggraph/KarniG00}. The parallelogram prediction scheme~\cite{DBLP:conf/smi/Rossignac01} is a common technique used in this method. The spectral methods, such as the Laplacian spectral approach~\cite{DBLP:conf/siggraph/KarniG00}, exploit the spectral properties of the mesh to achieve compression. \revise{DRACO~\cite{DRACO}, developed by Google, is another powerful compression algorithm. By combining vertex quantization, connectivity encoding, and entropy encoding, DRACO achieves a high compression rate while maintaining visual fidelity, supporting progressive transmission for real-time applications}. These methods perform well with smooth meshes but may not be the best choice for models with sharp features~\cite{DBLP:conf/sgp/SorkineCLARS04}.

The strength of traditional methods lies in their simplicity and efficiency. However, they often fail to leverage spatial coherency and global structures in the mesh, which can lead to suboptimal compression rates. In recent years, machine-learning approaches have been explored for mesh compression. These include variational autoencoders (VAEs)~\cite{DBLP:conf/cvpr/Tan0LX18} and CNNs~\cite{DBLP:journals/tip/HanLVLBHC18}. These methods leverage the ability of neural networks to learn compact and expressive representations of data. The choice of compression technique depends on the specific requirements of the application, including the acceptable loss of quality, the storage capacity, and the computational resources available.
\revise{We conclude the scenarios where these compression methods are most effective. Traditional methods are ideal for basic mesh compression tasks where simplicity and computational efficiency are essential. They work well for objects with smooth surfaces and simple geometries, offering a balance between compression ratio and visual quality. Their low computational demands make them compatible with a wide range of hardware, including older or lower-powered devices. In contrast, Draco excels in web-based applications where fast decoding and low computational overhead are crucial, particularly for devices with limited processing power, such as smartphones and tablets. For high-detail models or near-lossless compression, more robust hardware or machine-learning-based methods may be required. They are more suitable for applications that demand higher compression efficiency and are capable of leveraging modern hardware.}

\subsubsection{NeRF Compression}
The primary obstacle in compressing NeRF is to maintain the high-quality rendering of 3D scenes while significantly reducing the model size. It is also crucial to ensure that the compressed model can support efficient inference. Despite NeRF's growing popularity, few studies have concentrated on compressing NeRF. Since NeRF is represented by neural network models, most current approaches are inspired by model compression techniques~\cite{DBLP:journals/corr/HanMD15}.

There are four key approaches to compressing models: (1) model pruning~\cite{DBLP:conf/atait/LiM22}, which involves removing redundant connections or layers; (2) weight quantization~\cite{DBLP:conf/cvpr/JacobKCZTHAK18}, which reduces the model size by converting full precision float numbers to lower bit representations; (3) low-rank approximation~\cite{DBLP:conf/bmvc/JaderbergVZ14}, which involves decomposing high-rank matrices into smaller counterparts; and (4) knowledge distillation~\cite{DBLP:journals/ijcv/GouYMT21}, which uses a well-trained large network to guide the training of a smaller network. These techniques are mostly independent and can be combined for better results. Some NeRF research have already adopted these techniques. PlenOctrees~\cite{DBLP:conf/iccv/YuLT0NK21} and Re:NeRF~\cite{DBLP:conf/wacv/DengT23} use weight quantization, while Plenoxels~\cite{DBLP:conf/cvpr/Fridovich-KeilY22} employ a similar mechanism to weight pruning. CCNeRF~\cite{DBLP:conf/nips/TangC0Z22} and TensoRF~\cite{DBLP:conf/eccv/ChenXGYS22} use low-rank approximation to decompose full-size tensors. More recently, Li et al.~\cite{li2023compressing} introduced VQRF, an end-to-end compression framework for NeRF. Their approach uses an adaptive voxel pruning mechanism, a learnable vector quantization, and a weight quantization method.

\revise{NeRF compression is particularly advantageous when high-quality rendering of complex 3D scenes is required but data size and transmission bandwidth are limited. This technique is well-suited for applications where photorealism and detailed scene representation are critical. However, due to its computational intensity, NeRF compression is best applied in scenarios where offline processing or cloud-based rendering is feasible, rather than in real-time applications. It is ideal for use cases that prioritize visual fidelity over latency, such as cinematic rendering, virtual tourism, and architectural visualization, where pre-rendering can be leveraged to minimize delays in real-time playback. For real-time applications, alternative compression methods that are less resource-intensive may be more appropriate, depending on the system's hardware capabilities and latency requirements.}

\subsection{Transmission}

\begin{figure}[t]
    \centering    
    \begin{subfigure}[t]{0.320\textwidth}    
        \centering
        \includegraphics[width=\linewidth]{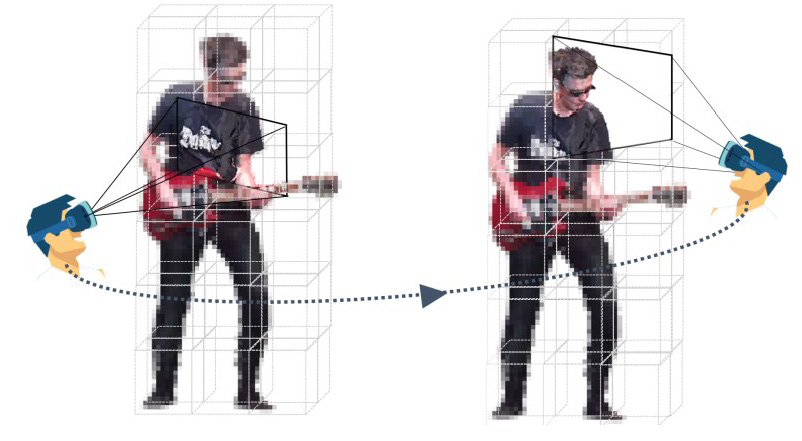}
        \caption{Tile-based}
        \label{fig:Tile-based}
    \end{subfigure}
    \hfill
    \begin{subfigure}[t]{0.320\textwidth}    
        \centering
        \includegraphics[width=\linewidth]{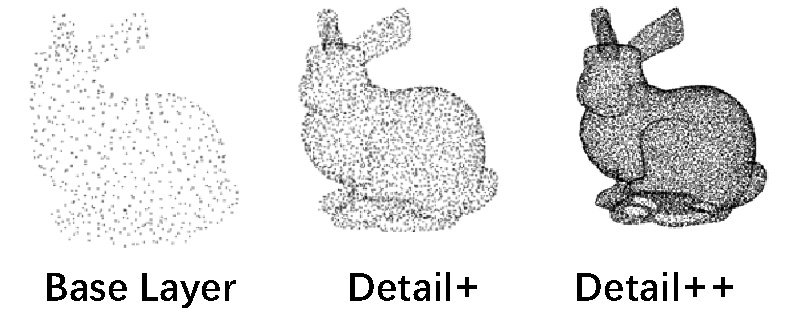}
        \caption{Layered}
        \label{fig:Layered}
    \end{subfigure}
    \hfill
    \begin{subfigure}[t]{0.320\textwidth}    
        \centering
        \raisebox{8pt}{
        \includegraphics[width=\linewidth]{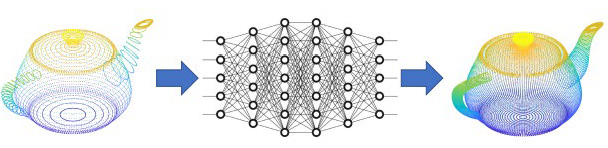}}
        \caption{SR-based}
        \label{fig:SR-based}
    \end{subfigure}
    \hfill
    \vspace{-8pt}
    \caption{Illustrations for methods of transmission.}
        \vspace{-5pt}
    \label{fig:transmission}
\end{figure}

Transmission is a crucial step in delivering volumetric content from a server to end users. However, the large data size of volumetric content presents significant challenges to the transmission process. To address this issue, various methods have been proposed to optimize the transmission cost of volumetric videos. \revise{These methods can be classified into three categories: tile-based, layered, and super-resolution (SR)-based.} Fig.~\ref{fig:transmission} provides straightforward illustrations of these methods. \revise{The content shown in the figures is sourced from the following datasets: CWIPC-SXR~\cite{cwipc}, Stanford Bunny~\cite{DBLP:conf/siggraph/TurkL94}, and Utah Teapot~\cite{DBLP:conf/siggraph/Torrence06a} respectively.}

In the following subsections, we will describe related works in transmission methods. \revise{Viola et al.~\cite{IVT15} conduct a survey that focused on point cloud and mesh streaming. In contrast, we will provide a more general perspective on volumetric video.}

\subsubsection{Tile-based Transmission}
Users can freely move their heads in 6 DOF while watching volumetric videos. However, due to the limited viewport of users (approximately 120°~\cite{knapp1938introduction}), only a portion of the volumetric scene falls within the users' viewing frustum~\cite{viewing_frustum} at any given time, rendering the remaining part redundant. Such a feature provides insights to achieve tile-based transmission for volumetric videos.

By predicting the user's future viewports, streaming systems can reduce bandwidth by prioritizing only the tiles falling into the user's viewports. A pioneering work called ViVo~\cite{DBLP:conf/mobicom/HanLQ20} proposed by Han et al. introduced the concept of visibility-aware optimization to reduce mobile data usage and decoding overhead for volumetric video streaming. By predicting the users' future viewports, the system can predict which part of the volumetric scene will fall into the users' viewing frustum, allowing it to reduce the quality of the remaining part. \revise{Further, Liu et al.~\cite{DBLP:conf/vr/LiuZWJZXC23, 10634203} enhance this concept with a caching mechanism that predicts viewports using a Long-short Term Sequential Prediction Model, integrating gaze and attention inference. Prioritized tiles are cached based on predicted viewing patterns, dynamically adapting to user movements to optimize cache utilization and reduce data transmission. To address the challenge of significant motion within the users' viewport, GROOT~\cite{DBLP:conf/mobicom/LeeYLCK20} introduce a fast tiling scheme that utilizes the hierarchical structure of Parallel Decodable Tree. It organizes tiles into a hierarchical structure, streaming only those intersecting the user’s viewport, thus reducing bandwidth consumption while ensuring high responsiveness and performance in dynamic environments.}

\revise{In addition to viewport prediction techniques, tiling schemes play a critical role in improving transmission efficiency.
Li et al.~\cite{DBLP:journals/tmm/LiZLHH23} propose a novel hybrid visual saliency and hierarchical clustering empowered 3D tiling scheme that can better match the user's viewport. The scheme is accompanied by a joint computational and communication resource allocation mechanism that achieves a trade-off between communication and computational resources to maximize the quality of experience (QoE).} \revise{Park et al.~\cite{park2019rate} propose to leverage 3D tiles and a window-based buffer, allowing faster insertions near the head rather than at the tail, to respond quickly to user actions. To maximize tile utility, they developed a greedy yet optimal algorithm that adjusts the tile requests within rate constraints, selecting the best set at each transmission opportunity.}

\revise{Semantic information can further enhance tile-based transmission. Existing object detection and scene understanding techniques \cite{DBLP:journals/mta/LiangXGBZC21,DBLP:journals/mta/YuanAL21} can identify key objects or regions of interest within user's viewport, allowing the system to prioritize the tiles for higher-quality transmission. Furthermore, context awareness can adapt tile selection based on user interaction and the environment, ensuring that the transmission strategy focuses on the most relevant content. By integrating semantic information with tile-based transmission, bandwidth efficiency can be further optimized.}

\revise{In addition to the aforementioned viewport-based tiling transmission, several works have explored bit-rate adaptation to optimize video-on-demand streaming. Hosseini and Timmerer~\cite{hosseini2018dynamic} propose a dynamic adaptive streaming solution for point cloud data, focusing on adjusting the bit-rate dynamically based on network conditions. Van der Hooft et al.~\cite{DBLP:conf/mm/HooftWTTH19} further develop a system that dynamically adjusts both quality and bit-rate based on the user’s viewport and current network conditions, ensuring a seamless and optimized streaming experience.}

\subsubsection{Layered Transmission}
Layered transmission has been a fundamental technique employed in conventional video delivery~\cite{DBLP:journals/jsac/RejaieHE00}. The underlying concept involves encoding the video at various levels of quality, with the video chunk possessing the lowest quality referred to as the "base layer." This base layer contains the most crucial information of each frame and is given the highest priority throughout the transmission process. In situations where network conditions are favorable, additional detail can be incorporated into the base layer to enhance the visual quality. However, when it comes to transmitting volumetric video, there are additional factors to consider in order to optimize layered transmission.

Shi et al.~\cite{DBLP:conf/mmsys/ShiVDO23} have explored the utilization of redundant information present in point clouds to extend the bit-rate range of MPEG's V-PCC compression standard. \revise{They achieved this by simplifying the point clouds through down-sampling and down-scaling techniques, resulting in a collection of point cloud data across various density levels, facilitating layer transmission as well.} They achieved up to a 48.5\% reduction in bit-rate while maintaining the same quality. \revise{Low-latency DASH has also been a key focus for improving real-time volumetric video delivery. Jansen et al.~\cite{jansen2020pipeline} propose a multiparty conferencing system that leverages point cloud compression and low-latency DASH to support real-time interactions over the network, making it highly suitable for applications like virtual conferencing.}

%\revised{Furthermore, in conjunction with the aforementioned intuitive techniques, Zhang et al.~\cite{DBLP:conf/hotnets/ZhangHPW21, DBLP:conf/sensys/ZhangZ0P22} have tackled the complexities of multi-user broadcasts, including scenarios like classroom education and collaborative design. Their studies underscore the viability of exploiting viewport-similarity to enhance network resource efficiency via multicast over mmWave communication. By capitalizing on this similarity, they suggest transmitting shared content among users to minimize motion-to-photon latency. Nebula~\cite{DBLP:conf/wmcsa/QianHPG19} has further implemented layered transmission on mobile devices with the assistance of edge computing to reduce overall latency and adapt to fluctuating network conditions. Li et al.~\cite{10129903} have employed a rolling optimization to predict user's future viewport in a short rolling window to reduce prediction error and a deep reinforcement learning-based approach to optimize tile selection.}

\revise{In addition to simply adjusting the quality of volumetric content corresponding to the currently available bit-rate, Liu et al. introduce V2RA~\cite{gurel2024v2ra}, a grid-based rate adaptation logic for volumetric video streaming, which enhances layered transmission by prioritizing the streaming of key components like geometry and texture based on the user's viewport. The V2RA method adapts the video bit-rate dynamically by using a quality ladder for each viewport, optimizing the trade-off between bandwidth usage and visual quality. By leveraging the combination of geometry and texture layers, V2RA minimizes the loss of perceptual quality while achieving substantial bandwidth savings.} 

\revise{Furthermore, Nebula~\cite{DBLP:conf/wmcsa/QianHPG19} extends the concept of layered transmission by utilizing edge computing to handle the decoding and rendering of volumetric videos, particularly for mobile devices. By organizing content into layers, Nebula allows for incremental upgrades based on network conditions and device capability. It adapts to bandwidth fluctuations through rate adaptation algorithms and optimizes content delivery via viewport adaptation, balancing high QoE with efficient bandwidth usage. This layered approach reduces the computational load on mobile devices while maintaining a seamless streaming experience.}

\subsubsection{Super-resolution-based Transmission}
Super-resolution techniques have been utilized in 2D video transmission~\cite{DBLP:conf/nossdav/WangLH0CQ22} to enhance the visual quality of low-resolution videos. This technique allows the original video to be initially transmitted to the end-user at a lower quality. However, with the aid of local computing power, the video can be pre-processed using pre-trained super-resolution models. As a result, users can experience video with higher quality even when network resources are limited. When it comes to 3D content, research in this area is still relatively nascent.

Zhang et al. are the pioneers in proposing a volumetric video streaming system that utilizes 3D super-resolution techniques~\cite{DBLP:conf/wmcsa/ZhangW0021}. Building upon this work, they introduced YuZu~\cite{DBLP:conf/nsdi/ZhangW0022}. In their research, the SR process was divided into two steps: \textit{intra-frame SR} and \textit{inter-frame SR}. For \textit{intra-frame SR}, they strategically employed off-the-shelf 3D SR models such as PU-GAN~\cite{PU_GAN} and MPU~\cite{MPU}. They accelerated the up-sampling approach through various techniques, including model optimization, reduction of input data, and improved patch generation. In the case of \textit{inter-frame SR}, they expedited the up-sampling process by caching and reusing previous SR results across consecutive frames. Furthermore, their method is also applicable to mobile devices.

There are several other intriguing ideas related to SR-based transmission. Firstly, one approach is to apply SR processing specifically to the texture information while keeping the geometry information unchanged. Given that texture information significantly contributes to perceived visual quality, performing SR processing on texture data alone can be more cost-effective compared to processing the entire dataset. Another idea involves directly leveraging well-established 2D SR methods and applying SR processing to the rendered 2D frames before they are displayed to the user. However, the feasibility of this approach has yet to be proven and requires further investigation and validation.

\subsection{Rendering}
%Render algorithm of different representation, Cube Local render, remote render, render cost, quality of different rendering algorithm, 

Rendering refers to the intricate process of creating a realistic visual depiction of a 3D model or scene. It encompasses the transformation of geometric data and material properties into visually appealing images or videos. As a result, the rendering performance, including factors like quality, frame rate, and resource consumption, significantly impacts the overall QoE. We discuss related works from the perspectives of representation and system, respectively.

\subsubsection{Representation Perspective}
In point cloud-based volumetric video systems, rendering often involves treating each point as an individual pixel. This simplistic approach employs straightforward rendering algorithms such as visibility splatting~\cite{pfister2000surfels}. \revise{However, rendering tools often allow for different representations of point clouds, where each point can be displayed as spheres, cubes, or other primitives, depending on the application and desired visual effect.} On the other hand, rendering a mesh entails connecting adjacent points using geometric primitives to form triangles, which are then rasterized onto a 2D display surface. However, these methods are not specifically designed to provide an immersive viewing experience, leading to a sustained lower quality.

Rendering implicit surfaces involves evaluating the implicit function at each pixel on the display. One commonly used approach is ray marching, where a ray is cast from the camera position, and the implicit function is evaluated along the ray until a surface intersection is found. However, NeRF~\cite{DBLP:journals/cacm/MildenhallSTBRN22} has emerged as a superior representation in terms of interactivity and photorealism. In NeRF rendering, a ray is cast from the camera position, and neural networks are used to evaluate the function's value and gradients at each point along the ray. This enables high-quality rendering with realistic lighting and reflections, which greatly enhances the viewing experience.

However, there are still several challenges to overcome. Firstly, the use of ray casting-based neural models requires evaluating a large multi-layer perceptron (MLP) at numerous sample positions along the ray for every pixel. This demands significant computational resources and time for model training, making it a resource-intensive approach. Secondly, the current volume rendering process is excessively slow for interactive visualization, necessitating the use of specialized rendering algorithms that don't align well with commonly available hardware.

The pursuit of high quality at a reduced cost has generated considerable interest in alternative neural-based approaches. While NeRFs possess the capability to accurately depict 3D scenes for image rendering, it is important to acknowledge that meshes continue to serve as the primary scene representation. Consequently, recent advancements have focused on leveraging the concept of NeRF to improve mesh-based representations in two significant ways: (1) Recent works~\cite{rakotosaona2023nerfmeshing,chen2023mobilenerf} have introduced methods to distill the volumetric 3D representation obtained from training NeRF into an approximation network. This network can then extract the 3D mesh and its appearance, resulting in a physically accurate representation. The final 3D mesh can be rendered in real time on readily available devices, offering practicality and efficiency. (2) Some works~\cite{guo2023vmesh} propose a hybrid representation that combines mesh and NeRF. This approach retains the advantages of mesh-based assets while incorporating the ability to represent subtle geometric structures provided by NeRF, resulting in more versatile and detailed representations.

Researchers have also shown interest in integrating neural network-based approaches into point cloud rendering~\cite{huang2022ponder,dai2020neural,xu2022point}. These approaches enhance points with neural features and employ CNNs to render them, resulting in improved visual quality. However, this quality-focused approach often comes at the expense of other factors. The rendering algorithms used in these methods are typically time-consuming, requiring a significant amount of time to render each frame, especially on high-throughput and computationally intensive devices. Moreover, these operations often require additional per-scene training, which is not suitable for volumetric video streaming tasks. To provide a satisfactory user experience, a system should be capable of rendering videos at a minimum rate of 30 frames per second, a goal not currently attainable with existing neural rendering models. Looking ahead, future advancements could aim to develop a neural point cloud renderer capable of rendering at an interactive rate on commonly available hardware, without the need for per-scene training, while still maintaining satisfactory quality. This could potentially be achieved by leveraging natively supported point types in graphics APIs or implementing parallel software rasterization on the GPU.

\subsubsection{System Perspective}

Rendering processes can be categorized as local rendering and remote rendering. Local rendering refers to rendering performed on the user's own device, such as a computer or a mobile device. It offers several advantages, including real-time interaction, user control, and the ability to handle sensitive or private data without leaving the user's machine. However, local rendering faces scalability issues, particularly when dealing with complex scenes or high-resolution output. The computational resources required for rendering may exceed the capabilities of the user's device, leading to slow performance or even crashes~\cite{DBLP:journals/tmm/LiZLHH23}. 

On the other hand, remote rendering involves offloading the rendering process to a remote server or cloud infrastructure. It overcomes the scalability limitations of local rendering by utilizing the computational power and resources available in the cloud~\cite{DBLP:conf/mobicom/LiuHQNZ22}. Remote rendering can handle large scenes and compute-intensive rendering techniques more efficiently, resulting in faster and more realistic visualizations. Furthermore, it provides the flexibility to render on various devices, including low-powered devices like smartphones or tablets. In summary, while local rendering offers real-time interaction and control, remote rendering addresses scalability issues and enables efficient rendering of complex scenes. The choice between local and remote rendering depends on factors such as scene complexity, computational resources, and desired output quality.

\subsection{Display}
The final step in displaying volumetric contents is crucial. While 2D displays~\cite{DBLP:journals/cg/MasiaWDG13} can provide various visual cues such as shading, occlusion, relative size, and perspectives, they lack certain elements that are exclusive to volumetric displays. One such cue is binocular disparity~\cite{DBLP:conf/apgv/HeldB08}, also known as stereopsis, which is only present in binocular vision. This cue results from the formation of two slightly different images of the same scene in each eye, due to the differing viewpoints of each eye. When an object is closer, the difference between the left and right eye's images is greater, and as the object moves further away, the difference decreases. Inaccurate binocular disparity can lead to distortions in the perceived depth of the scene. Another binocular cue is vergence~\cite{palmer1999vision}, which is an oculomotor cue where the optical axes of the two eyes rotate and converge toward the object in focus. The kinaesthetic sensations from the extraocular muscles provide information for depth perception, as the angle of vergence is inversely proportional to the depth of the object. The combination of binocular disparity and vergence is referred to as stereo cues.

The main design of volumetric displays is based on delivering stereo cues by presenting each eye with a separate planar image. Two main approaches to HMDs' volumetric displays are varifocal displays and multifocal displays, both of which we will describe.

One way to enhance standard head-mounted stereo displays is to incorporate varifocal displays, which actively adjust the focal distance of the image plane seen by each eye using active optics, such as liquid lenses~\cite{DBLP:journals/tog/AksitLKSL17,DBLP:journals/tvcg/DunnTTKADMLF17}. This adjustment is based on the observer's gaze, producing a varying depth of field effect. However, these displays can introduce lens distortions that are unwanted due to the use of active optics like deformable membrane mirrors~\cite{DBLP:journals/tvcg/DunnTTKADMLF17}. Additionally, accurate synchronization between the optics and the 2D image source generation (e.g., digital micromirror device~\cite{sampsell1994digital}) with the 3D gaze location is necessary. Any inaccuracies between the optics and the observer's gaze can result in errors in the reproduced focal plane. Varifocal displays also require the defocus blur to be synthesized in rendering~\cite{DBLP:journals/tog/XiaoKFCL18}, instead of being optically reproduced, since they only allow for a uniform focal depth throughout the scene for a fixed gaze. This mechanism can be limiting and may not always provide the most realistic simulation of natural vision.

Multifocal displays are a type of volumetric display that has a fixed viewing position. This type of display renders a stack of images for each eye at a fixed number of focal planes located at various distances. Each plane adds a particular amount of light, allowing the viewer to accommodate appropriately at the desired depth. These focal planes can consist of superimposed image planes with beam-splitters~\cite{DBLP:journals/tog/AkeleyWGB04} or time-multiplexed image slices~\cite{DBLP:journals/tog/ChangKS18,Love:09} that sweep a 3D volume with high-speed switchable lenses. Compared to varifocal displays, multifocal displays do not require strict synchronization of the optics and rendering with the gaze location. However, they still maintain high resolution and contrast, as they can adopt well-established 2D display techniques~\cite{DBLP:journals/tog/ZhongJYHWM21}. Architectures with fixed focal planes also prevent optical aberrations. However, the accuracy of the eye position is crucial for the quality of a multifocal display, as a slight misalignment in the focal cues can immediately break sharp edges and realism. Differences in eye positions of individual observers can be compensated for with a homography correction~\cite{DBLP:journals/tog/MercierSMZHNL17}. The integration of a high dynamic range (HDR) with a multifocal display has been shown to achieve a level of realism that transcends any existing 3D display technique, confusing naive observers between a physical object and its virtual 3D reproduction~\revise{\cite{DBLP:journals/tog/ZhongJYHWM21}}.

\section{Applications}
\label{sec:application}

This section presents an overview of the three most promising applications of volumetric video technology: telepresence, rehabilitation, and education.

\subsection{Telepresence}
Volumetric video can enhance telepresence by providing a more realistic and immersive representation of remote participants. One of the key benefits of volumetric video in telepresence is its ability to capture and transmit a more realistic representation of a remote participant's body language, gestures, and facial expressions. Traditional video conferencing systems~\cite{DBLP:conf/cscw/Egido88} often struggle to convey these nonverbal cues, which are critical to effective communication and collaboration~\cite{troje2023zoom}. With volumetric video, remote participants can be captured and rendered in 3D, allowing the receiving party to see and interact with them as if they were in the same room. This can significantly improve communication and collaboration in remote teams, particularly for tasks that require a high degree of visual and spatial understanding.

\revise{A prominent example of this is Holoportation~\cite{orts2016holoportation}. It is a real-time 3D teleportation system that enables remote users to interact with each other as if they were physically present in the same space. By capturing 3D volumetric video and transmitting it in real time, Holoportation allows users to see and engage with full-body representations of remote participants, improving the sense of immersion and realism in remote collaboration.}

Another advantage of volumetric video in telepresence is its ability to provide a more immersive experience. With traditional video conferencing systems, participants are typically limited to a 2D view of the remote location~\cite{tcsvt,HeadsetOff}. This can make it challenging to get a sense of the space and environment, which can limit collaboration and problem-solving. Volumetric video, on the other hand, can capture and render a 3D representation of the remote location, allowing participants to explore and interact with the space as if they were physically present. This can be particularly useful for remote inspections, virtual site visits, and remote training sessions.

% In conclusion, volumetric video has the potential to revolutionize telepresence by providing a more realistic and immersive representation of remote participants. With its ability to capture and render 3D video, volumetric video can enhance communication and collaboration in remote teams, and provide a more immersive experience. While there are still challenges to be addressed, the potential benefits of volumetric video in telepresence are significant, and it is likely to play an increasingly important role in remote collaboration and communication in the years to come.

\subsection{Rehabilitation}
Volumetric video has the potential to revolutionize the field of rehabilitation by providing a more immersive and engaging experience for patients, allowing them to interact with their environment and practice real-world scenarios~\cite{ijcai2024p920}.

One of the key benefits of volumetric video in rehabilitation is its ability to provide patients with an immersive environment in which to practice their skills. For example, a patient who has suffered a stroke~\cite{laver2017virtual} may have difficulty with balance and coordination, making it challenging to perform everyday tasks such as walking or reaching for objects. Using volumetric video, the patient can be placed in a virtual environment that simulates real-world situations, such as walking on uneven terrain or reaching for objects on a high shelf. This allows the patient to practice their skills in a safe and controlled environment, increasing their confidence and reducing their risk of injury.

Volumetric video can also be used to create operational room simulations for rehabilitation purposes~\cite{hu2024teleor}. These virtual operating environments allow medical professionals, such as surgeons and nurses, to practice and refine their skills in realistic, high-pressure settings without the risk of patient harm. By enabling patients to practice dexterity, precision, and coordination through hands-on tasks within the virtual operational room, volumetric video provides a valuable tool for motor skill recovery and professional development in healthcare.

In addition, volumetric video can be used to monitor the patient's progress and provide feedback in real time~\cite{DBLP:journals/jsac/PostolacheHAGGK21}. By capturing data on the patient's movements and performance, therapists can track their progress over time and adjust their rehabilitation program as needed. This can help to ensure that the patient is making steady progress toward their goals and can also provide motivation and encouragement to continue with their therapy.

% In conclusion, volumetric video has the potential to revolutionize the field of rehabilitation by providing a more immersive and engaging experience for patients. By creating realistic and personalized virtual environments, therapists can help patients to practice their skills in a safe and controlled environment, monitor their progress in real-time, and provide feedback and encouragement to help them achieve their goals. As this technology continues to evolve, it is likely that we will see even more applications of volumetric video in healthcare and rehabilitation.

\subsection{Education}
Another area where volumetric video has the potential to make a significant impact is education. Volumetric video has the ability to create immersive and interactive experiences, which can help learners to better understand complex concepts. For example, in medical education~\cite{DBLP:conf/siggrapha/PapagiannakisLK18}, volumetric video can be used to create 3D models of the human body, allowing medical students to explore the body in a way that was not possible before. This can help them to better understand the anatomy and physiology of the human body, as well as the various medical conditions that can affect it.

In engineering education~\cite{DBLP:conf/coginfocom/Horvath16b}, volumetric video can be used to create 3D models of complex machinery and equipment. This can help students to better understand how these machines work and how they can be maintained and repaired. In addition, volumetric video can be used to create simulations of real-world scenarios, allowing students to practice their problem-solving skills in a safe and controlled environment.

Volumetric video can also be used to create virtual field trips~\cite{makransky2022benefits}, allowing students to explore different parts of the world without ever leaving the classroom. For example, a history class could use volumetric videos to take students on a virtual tour of ancient ruins, allowing them to explore and learn about different cultures and civilizations.

Furthermore, volumetric video can be used to create personalized learning experiences~\cite{horvath2021analysis}. By creating 3D models of individual students, educators can tailor the learning experience to the individual needs and preferences of each student. For example, a student who is struggling with a particular concept could be presented with a more detailed and interactive 3D model, while a student who is more advanced could be presented with a more challenging model.

% In conclusion, volumetric video has the potential to revolutionize education by providing immersive, interactive, and personalized learning experiences. By creating 3D models of real-life objects and scenarios, volumetric video can help students to better understand complex concepts and develop their problem-solving skills. As this technology continues to evolve, it is likely that we will see even more innovative applications in the field of education.

\revise{\textbf{In conclusion}, volumetric video holds immense potential to revolutionize various industries by offering immersive and interactive experiences that go beyond the limitations of traditional media. In telepresence, volumetric video enhances remote communication by providing lifelike 3D representations of participants, improving engagement and non-verbal communication. However, realizing true real-time interaction in telepresence will require advanced transmission protocols that surpass those currently discussed. In healthcare, volumetric video can provide a more engaging and effective environment for patient rehabilitation, offering personalized and immersive simulations that help patients practice real-world skills safely. Similarly, in education, volumetric video creates interactive, 3D learning environments that deepen students’ understanding of complex concepts and provide experiences like virtual field trips and personalized learning paths. Each of these sectors stands to benefit greatly from the adoption of volumetric video, but the full realization of its potential hinges on continued advancements in compression, rendering, and transmission technologies. As these underlying technologies evolve, volumetric video could reshape how we communicate, learn, and engage with digital content across various fields.}

\section{Opportunities}
\label{sec:opportunity}

In this section, we delve into the various research challenges and opportunities in the field of volumetric video services.

\subsection{Emerging Representations}
Despite numerous attempts to explore different types of representations, mesh~\cite{DBLP:conf/imr/Owen98, DBLP:journals/cgf/BommesLPPSTZ13} and point cloud~\cite{DBLP:conf/icra/RusuC11,DBLP:journals/pami/GuoWHLLB21} remain the most commonly used methods in volumetric video transmission due to their straightforwardness. However, the substantial data size and limited representation accuracy associated with these methods present a persistent challenge, necessitating the development of more advanced techniques. A comparison of various representation methods is presented in Table~\ref{table:repre}.

The emergence of implicit representation techniques like NeRF~\cite{DBLP:journals/cacm/MildenhallSTBRN22} has offered a solution to the limitations of traditional discrete 3D representations. However, utilizing NeRF as a volumetric video representation is not a straightforward task and presents several obstacles. Firstly, the ray casting-based neural model used by NeRF evaluates a large MLP at numerous sample positions along the ray for each pixel, which necessitates significant resources and training time. Secondly, the volume rendering process is too slow for real-time visualization and requires specialized rendering algorithms that are not easily compatible with commonly available hardware, thereby impeding its widespread adoption. Finally, the baseline NeRF fails to accurately represent and reconstruct non-static or dynamic scenes, posing a significant challenge.

As each representation has its own strengths and limitations, it is intriguing to explore the possibility of hybridizing them for volumetric video. For instance, the NeRF performs well in representing static scenes but faces difficulties with dynamic content. Thus, we could use NeRF to represent static scenes and mesh or point cloud to depict dynamic content. The majority of current research concentrates on a single representation. Hybridizing different representations presents both challenges and opportunities, including the need to determine how to effectively combine multiple pipelines and how to decide when to utilize each representation.

\subsection{Compression Efficiency}
The compression system has two main components: intra-frame compression and inter-frame compression. While much research has been devoted to developing and improving the 3D representations of intra-frame compression, inter-frame compression has received relatively little attention and thus presents numerous opportunities for further exploration~\cite{DBLP:journals/tvcg/LiangLDW24}. Specifically, there is a noticeable research gap in addressing the temporal redundancy of volumetric data, which is an area that warrants further investigation.

While compression algorithms are designed to minimize quality loss as perceived by the human visual system, they currently lack the ability to take into account the semantics of video content or identify which parts of a video are most important to viewers. Instead, they operate solely at the pixel level, such as points within a point cloud or vertices within a 3D mesh~\cite{DBLP:conf/qomex/ZermanOGS20}. However, advances in 3D vision have given machines the capability to extract semantic information from video content~\cite{PointNet++}. By leveraging this information, it becomes possible to code most of the content at a higher level, resulting in more efficient compression and improved quality retention.

High-level representations can be leveraged for compression beyond the pixel level. In the case of volumetric video conferencing, the primary content transmitted is the human body and its facial expressions. Instead of coding at the pixel level, the motion of the human can be captured and used to reconstruct the current frame based on the reference frame's 3D motion and the human body's position. The motion of a human can be accurately represented using fewer than 100 parameters~\cite{10.1109/TPAMI.2023.3330935}, which is significantly smaller than a 3D motion vector, making it an ideal choice for transmitting each frame. By utilizing these techniques, the efficiency of inter-frame compression for 3D content can be significantly improved.

\subsection{Streaming Optimization} 
Inspired by the concept of 360° video streaming~\cite{DBLP:journals/comsur/YaqoobBM20}, visibility-aware video streaming aims to transmit only the video content within a viewer's field of view, optimizing the video streaming experience. However, this approach presents significant challenges that must be overcome to achieve its goals.

Firstly, selecting the appropriate bit-rate is challenging due to the dynamic nature of network conditions, individual users' behavioral patterns, and the complexity of volumetric videos. In particular, volumetric videos pose a significant challenge as different viewports may encompass varying amounts of video objects with different data sizes, resulting in an uncertain and cascading effect on bit-rate adaptation.

Secondly, the unique 3D characteristics of objects in volumetric videos and their complex spatial relationships create a challenge in allocating bit-rate in a precise and granular manner. Traditional approaches to unified bit-rate assignment are inadequate in volumetric videos with 3D scenes and new data formats. Achieving a balance between maximizing QoE and minimizing bandwidth usage through fine-grained bit-rate allocation that takes into account spatial features is a significant challenge that must be addressed.

Lastly, handling rapid viewport changes is critical. With head-mounted displays, users can quickly change their viewport by turning their heads, resulting in sudden, frequent viewport switches. The streaming system needs to be able to react and adapt to rapid viewport changes without much latency or buffering. Fast viewport prediction and flexible segment fetching are required to provide a smooth viewing experience.

\subsection{Privacy \& Security Enhancement}
Volumetric videos offer an immersive experience that can transport viewers to another world. However, this technology also raises a host of security concerns. One major issue is the potential for volumetric data to include highly sensitive biometric information, such as facial contours and gait patterns, \revise{which could be used for identification purposes~\cite{miller2020personal}}. This information could be easily obtained if someone's volumetric representation is available.

Another privacy concern comes from the viewer's side, as head motion data can reveal a lot about a person's psychological state. Researchers have found that head motion data can be linked to medical conditions like autism~\cite{jarrold2013social} and post-traumatic stress disorder (PTSD)~\cite{loucks2019you}. Moreover, there is mounting evidence that tracking data can be used to diagnose dementia~\cite{werner2009use, tarnanas2013ecological}. Overall, while volumetric videos offer a cutting-edge experience, their potential privacy and security risks must be carefully considered and addressed, unfortunately, few studies have focused on this issue.

One direct approach to preserving privacy is to use data perturbation~\cite{DBLP:conf/etra/LiuXDBHJ19,10660498}, which involves adding a moderate amount of random noise to specific regions of sensitive data. This technique can help to mask biometric data or head motion information that could be used to infer a person's psychological state or medical condition. However, this method is not foolproof and may not be effective against sophisticated attacks. Therefore, additional security measures such as encryption~\cite{DBLP:journals/pieee/SmidB88} and anonymization~\cite{DBLP:conf/bigdatasec/MurthyBRR19} may also be necessary to ensure confidentiality. It is worth noting that implementing privacy protection measures may require modifying the original data or adding complex modules to the system, which could affect performance. Thus, finding a balance between privacy and performance is an important consideration.

\subsection{Integrating Mobile Edge Computing}

\revise{After analyzing the characteristics of volumetric video content~\cite{fsvvd} and viewer behaviors~\cite{DBLP:conf/mm/HuYJLCZ023,DBLP:conf/mm/JinL0C22} under various scenarios, including static versus dynamic movement and single versus multiple characters, we find that a large portion of the video content is viewed repeatedly from slightly different angles, even over extended periods. This behavior is unsurprising, as users are free to move, while the majority of scenes and background objects remain static.}

Mobile edge computing (MEC)~\cite{DBLP:journals/comsur/MaoYZHL17} has created numerous opportunities by using geo-distributed edge servers like base stations to cache frequently accessed content. This caching significantly reduces network latency and bandwidth consumption, thereby enhancing users' QoE while saving on service costs~\cite{DBLP:journals/iotj/JinLWC23}. However, designing an edge caching system is a complex task that comes with several challenges. One major challenge in developing an edge caching system is resource allocation. It is crucial to allocate resources dynamically to ensure fairness in QoE while optimizing resource utilization. Storage, bandwidth, and processing power are resources that need to be allocated, and a mechanism must be developed to allocate them based on user demand, network conditions, and system load. Another significant challenge in designing an edge caching system is user mobility. As users move from one location to another, their proximity to edge nodes changes, affecting their QoE. To address this challenge, the system must be adaptive to user mobility patterns. The system should predict user movements and pre-cache content to ensure the content is available when the user moves to a new location. The placement of content in the edge caching system is another significant challenge. The system must decide which content to cache at which edge node, taking into account the popularity of the content, the frequency of access, and the resources available at each edge node. The system should also consider data privacy and security requirements when deciding where to place the content. In summary, while mobile edge computing presents significant opportunities for enhancing users' QoE and saving on service costs, designing an edge caching system comes with challenges like resource allocation, user mobility, and content placement.

\subsection{Unified Testing \& Datasets}
Although there are many existing works on volumetric video services, unlike the AI domain, most of those works are tested in disparate setups and datasets. The lack of standardized testing procedures and datasets for volumetric video presents a significant challenge and opportunity for researchers. Due to the complex and diverse nature of volumetric video, developing a universal benchmarking framework that accurately evaluates different methods is challenging. Moreover, the lack of a common dataset inhibits researchers' ability to compare and validate results across studies.

Current datasets have several drawbacks: 1) Most only contain video content, without additional data like user behaviors. 2) Existing video content datasets only have a single representation format, preventing the comparison of methods using different representations. 3) Existing datasets are relatively small, sufficient for testing but insufficient for training machine learning models. A unified, large-scale, multimodal dataset would greatly benefit the research community. Such a dataset should incorporate diverse video representations and other data like user interactions. By unifying data formats, researchers could seamlessly apply and compare methods.

\revise{Efforts by organizations such as MPEG~\cite{MPEG}, ITU~\cite{ITU}, and Video Quality Experts Group (VQEG)~\cite{VQEG} in developing standardized testing procedures and datasets for traditional video content serve as valuable precedents. Leveraging similar principles in the volumetric video space could drive the creation of a common benchmark for testing algorithms and methods. A standardized framework for volumetric video would allow for accurate comparison of different techniques, ultimately advancing the state-of-the-art in this domain.}

\section{Conclusion}
\revise{In conclusion, this survey paper has offered a comprehensive and in-depth examination of volumetric video, an emerging technology poised to transform various industries. It provided a thorough system overview, covering representations, datasets, and quality assessment, followed by a detailed exploration of the entire pipeline from capturing to display. It delved into the various applications and future opportunities that volumetric video presents.}

\revise{As technology continues to evolve, volumetric video is poised to play a crucial role in advancing fields such as telepresence, healthcare, and education. The continued development of underlying technologies like compression, rendering, and transmission will be key to realizing the full potential of volumetric video, setting the stage for its broader adoption and impact across industries.}


%\begin{acks}
%To Robert, for the bagels and explaining CMYK and color spaces.
%\end{acks}

\bibliographystyle{ACM-Reference-Format}
\bibliography{sample-base}


\begin{thebibliography}{223}

%%% the Web address.

\ifx \showCODEN    \undefined \def \showCODEN     #1{\unskip}     \fi
\ifx \showDOI      \undefined \def \showDOI       #1{#1}\fi
\ifx \showISBNx    \undefined \def \showISBNx     #1{\unskip}     \fi
\ifx \showISBNxiii \undefined \def \showISBNxiii  #1{\unskip}     \fi
\ifx \showISSN     \undefined \def \showISSN      #1{\unskip}     \fi
\ifx \showLCCN     \undefined \def \showLCCN      #1{\unskip}     \fi
\ifx \shownote     \undefined \def \shownote      #1{#1}          \fi
\ifx \showarticletitle \undefined \def \showarticletitle #1{#1}   \fi
\ifx \showURL      \undefined \def \showURL       {\relax}        \fi
% invisible to TeX
\providecommand\bibfield[2]{#2}
\providecommand\bibinfo[2]{#2}
\providecommand\natexlab[1]{#1}
\providecommand\showeprint[2][]{arXiv:#2}

\bibitem[Achlioptas et~al\mbox{.}(2018)]%
        {DBLP:conf/icml/AchlioptasDMG18}
\bibfield{author}{\bibinfo{person}{Panos Achlioptas}, \bibinfo{person}{Olga
  Diamanti}, \bibinfo{person}{Ioannis Mitliagkas}, {and}
  \bibinfo{person}{Leonidas~J. Guibas}.} \bibinfo{year}{2018}\natexlab{}.
\newblock \showarticletitle{Learning Representations and Generative Models for
  3D Point Clouds}. In \bibinfo{booktitle}{\emph{Proceedings of the 35th
  International Conference on Machine Learning, {ICML} 2018}}
  \emph{(\bibinfo{series}{Proceedings of Machine Learning Research},
  Vol.~\bibinfo{volume}{80})}. \bibinfo{publisher}{{PMLR}},
  \bibinfo{pages}{40--49}.
\newblock


\bibitem[Akeley et~al\mbox{.}(2004)]%
        {DBLP:journals/tog/AkeleyWGB04}
\bibfield{author}{\bibinfo{person}{Kurt Akeley}, \bibinfo{person}{Simon~J.
  Watt}, \bibinfo{person}{Ahna~Reza Girshick}, {and} \bibinfo{person}{Martin~S.
  Banks}.} \bibinfo{year}{2004}\natexlab{}.
\newblock \showarticletitle{A Stereo Display Prototype with Multiple Focal
  Distances}.
\newblock \bibinfo{journal}{\emph{{ACM} Trans. Graph.}} \bibinfo{volume}{23},
  \bibinfo{number}{3} (\bibinfo{year}{2004}), \bibinfo{pages}{804--813}.
\newblock


\bibitem[Aksit et~al\mbox{.}(2017)]%
        {DBLP:journals/tog/AksitLKSL17}
\bibfield{author}{\bibinfo{person}{Kaan Aksit}, \bibinfo{person}{Ward Lopes},
  \bibinfo{person}{Jonghyun Kim}, \bibinfo{person}{Peter Shirley}, {and}
  \bibinfo{person}{David Luebke}.} \bibinfo{year}{2017}\natexlab{}.
\newblock \showarticletitle{Near-eye Varifocal Augmented Reality Display Using
  See-through Screens}.
\newblock \bibinfo{journal}{\emph{{ACM} Trans. Graph.}} \bibinfo{volume}{36},
  \bibinfo{number}{6} (\bibinfo{year}{2017}), \bibinfo{pages}{189:1--189:13}.
\newblock


\bibitem[Aleksandrov et~al\mbox{.}(2021)]%
        {DBLP:journals/sensors/AleksandrovZH21}
\bibfield{author}{\bibinfo{person}{Mitko Aleksandrov}, \bibinfo{person}{Sisi
  Zlatanova}, {and} \bibinfo{person}{David~J. Heslop}.}
  \bibinfo{year}{2021}\natexlab{}.
\newblock \showarticletitle{Voxelisation Algorithms and Data Structures: {A}
  Review}.
\newblock \bibinfo{journal}{\emph{Sensors}} \bibinfo{volume}{21},
  \bibinfo{number}{24} (\bibinfo{year}{2021}), \bibinfo{pages}{8241}.
\newblock


\bibitem[Alexiou and Ebrahimi(2020)]%
        {DBLP:conf/icmcs/AlexiouE20}
\bibfield{author}{\bibinfo{person}{Evangelos Alexiou} {and}
  \bibinfo{person}{Touradj Ebrahimi}.} \bibinfo{year}{2020}\natexlab{}.
\newblock \showarticletitle{Towards a Point Cloud Structural Similarity
  Metric}. In \bibinfo{booktitle}{\emph{2020 {IEEE} International Conference on
  Multimedia {\&} Expo Workshops, {ICME} Workshops 2020}}.
  \bibinfo{publisher}{{IEEE}}, \bibinfo{pages}{1--6}.
\newblock


\bibitem[Alexiou et~al\mbox{.}(2023)]%
        {IVT18}
\bibfield{author}{\bibinfo{person}{Evangelos Alexiou}, \bibinfo{person}{Yana
  Nehm\'e}, \bibinfo{person}{Emin Zerman}, \bibinfo{person}{Irene Viola},
  \bibinfo{person}{Guillaume Lavou\'e}, \bibinfo{person}{Ali Ak},
  \bibinfo{person}{Aljosa Smolic}, \bibinfo{person}{Patrick~Le Callet}, {and}
  \bibinfo{person}{Pablo Cesar}.} \bibinfo{year}{2023}\natexlab{}.
\newblock \showarticletitle{Subjective and Objective Quality Assessment for
  Volumetric Video}.
\newblock In \bibinfo{booktitle}{\emph{Immersive Video Technologies}}.
  \bibinfo{publisher}{Elsevier}, \bibinfo{pages}{501--552}.
\newblock


\bibitem[Alliez and Desbrun(2001)]%
        {DBLP:journals/cgf/AlliezD01}
\bibfield{author}{\bibinfo{person}{Pierre Alliez} {and}
  \bibinfo{person}{Mathieu Desbrun}.} \bibinfo{year}{2001}\natexlab{}.
\newblock \showarticletitle{Valence-Driven Connectivity Encoding for 3D
  Meshes}.
\newblock \bibinfo{journal}{\emph{Comput. Graph. Forum}} \bibinfo{volume}{20},
  \bibinfo{number}{3} (\bibinfo{year}{2001}), \bibinfo{pages}{480--489}.
\newblock


\bibitem[Ansari et~al\mbox{.}(2019)]%
        {DBLP:conf/iccp/AnsariWGC19}
\bibfield{author}{\bibinfo{person}{Sameer Ansari}, \bibinfo{person}{Neal
  Wadhwa}, \bibinfo{person}{Rahul Garg}, {and} \bibinfo{person}{Jiawen Chen}.}
  \bibinfo{year}{2019}\natexlab{}.
\newblock \showarticletitle{Wireless Software Synchronization of Multiple
  Distributed Cameras}. In \bibinfo{booktitle}{\emph{{IEEE} International
  Conference on Computational Photography, {ICCP} 2019}}.
  \bibinfo{publisher}{{IEEE}}, \bibinfo{pages}{1--9}.
\newblock


\bibitem[Apostolopoulos et~al\mbox{.}(2012)]%
        {DBLP:journals/pieee/ApostolopoulosCCKTW12}
\bibfield{author}{\bibinfo{person}{John~G. Apostolopoulos},
  \bibinfo{person}{Philip~A. Chou}, \bibinfo{person}{W.~Bruce Culbertson},
  \bibinfo{person}{Ton Kalker}, \bibinfo{person}{Mitchell~D. Trott}, {and}
  \bibinfo{person}{Susie~J. Wee}.} \bibinfo{year}{2012}\natexlab{}.
\newblock \showarticletitle{The Road to Immersive Communication}.
\newblock \bibinfo{journal}{\emph{Proc. {IEEE}}} \bibinfo{volume}{100},
  \bibinfo{number}{4} (\bibinfo{year}{2012}), \bibinfo{pages}{974--990}.
\newblock


\bibitem[Ara{\'{u}}jo et~al\mbox{.}(2015)]%
        {DBLP:journals/csur/AraujoLJJW15}
\bibfield{author}{\bibinfo{person}{Bruno Rodrigues~De Ara{\'{u}}jo},
  \bibinfo{person}{Daniel~S. Lopes}, \bibinfo{person}{Pauline Jepp},
  \bibinfo{person}{Joaquim~A. Jorge}, {and} \bibinfo{person}{Brian Wyvill}.}
  \bibinfo{year}{2015}\natexlab{}.
\newblock \showarticletitle{A Survey on Implicit Surface Polygonization}.
\newblock \bibinfo{journal}{\emph{{ACM} Comput. Surv.}} \bibinfo{volume}{47},
  \bibinfo{number}{4} (\bibinfo{year}{2015}), \bibinfo{pages}{60:1--60:39}.
\newblock


\bibitem[Asadi et~al\mbox{.}(2014)]%
        {DBLP:journals/comsur/AsadiWM14}
\bibfield{author}{\bibinfo{person}{Arash Asadi}, \bibinfo{person}{Qing Wang},
  {and} \bibinfo{person}{Vincenzo Mancuso}.} \bibinfo{year}{2014}\natexlab{}.
\newblock \showarticletitle{A Survey on Device-to-Device Communication in
  Cellular Networks}.
\newblock \bibinfo{journal}{\emph{{IEEE} Commun. Surv. Tutorials}}
  \bibinfo{volume}{16}, \bibinfo{number}{4} (\bibinfo{year}{2014}),
  \bibinfo{pages}{1801--1819}.
\newblock


\bibitem[Assarsson and M{\"{o}}ller(2000)]%
        {viewing_frustum}
\bibfield{author}{\bibinfo{person}{Ulf Assarsson} {and} \bibinfo{person}{Tomas
  M{\"{o}}ller}.} \bibinfo{year}{2000}\natexlab{}.
\newblock \showarticletitle{Optimized View Frustum Culling Algorithms for
  Bounding Boxes}.
\newblock \bibinfo{journal}{\emph{J. Graphics, GPU, {\&} Game Tools}}
  \bibinfo{volume}{5}, \bibinfo{number}{1} (\bibinfo{year}{2000}),
  \bibinfo{pages}{9--22}.
\newblock


\bibitem[Bommes et~al\mbox{.}(2013)]%
        {DBLP:journals/cgf/BommesLPPSTZ13}
\bibfield{author}{\bibinfo{person}{David Bommes}, \bibinfo{person}{Bruno
  L{\'{e}}vy}, \bibinfo{person}{Nico Pietroni}, \bibinfo{person}{Enrico Puppo},
  \bibinfo{person}{Cl{\'{a}}udio~T. Silva}, \bibinfo{person}{Marco Tarini},
  {and} \bibinfo{person}{Denis Zorin}.} \bibinfo{year}{2013}\natexlab{}.
\newblock \showarticletitle{Quad-Mesh Generation and Processing: {A} Survey}.
\newblock \bibinfo{journal}{\emph{Comput. Graph. Forum}} \bibinfo{volume}{32},
  \bibinfo{number}{6} (\bibinfo{year}{2013}), \bibinfo{pages}{51--76}.
\newblock


\bibitem[Brachmann and Rother(2018)]%
        {brachmann2018learning}
\bibfield{author}{\bibinfo{person}{Eric Brachmann} {and}
  \bibinfo{person}{Carsten Rother}.} \bibinfo{year}{2018}\natexlab{}.
\newblock \showarticletitle{Learning Less is More-6d Camera Localization via 3d
  Surface Regression}. In \bibinfo{booktitle}{\emph{Proceedings of the IEEE
  conference on computer vision and pattern recognition}}.
  \bibinfo{pages}{4654--4662}.
\newblock


\bibitem[Callet et~al\mbox{.}(2013)]%
        {Qualinet2}
\bibfield{author}{\bibinfo{person}{Patrick~Le Callet},
  \bibinfo{person}{Sebastian M{\"o}ller}, \bibinfo{person}{Andrew Perkis},
  \bibinfo{person}{Kjell Brunnstr{\"o}m}, \bibinfo{person}{Sergio Beker},
  \bibinfo{person}{Katrien~De Moor}, \bibinfo{person}{Ann Dooms},
  \bibinfo{person}{Sebastian Egger}, \bibinfo{person}{Marie-Neige Garcia},
  \bibinfo{person}{Tobias Ho{\ss}feld}, \bibinfo{person}{Satu
  Jumisko-Pyykk{\"o}}, \bibinfo{person}{Christian Keimel},
  \bibinfo{person}{Chaker Larabi}, \bibinfo{person}{Bob Lawlor},
  \bibinfo{person}{Patrick Le~Callet}, \bibinfo{person}{Sebastian M{\"o}ller},
  \bibinfo{person}{Fernando Pereira}, \bibinfo{person}{Manuela Pereira},
  \bibinfo{person}{Andrew Perkis}, \bibinfo{person}{Jesenka Pibernik},
  \bibinfo{person}{Ant{\'o}nio Pinheiro}, \bibinfo{person}{Alexander Raake},
  \bibinfo{person}{Peter Reichl}, \bibinfo{person}{Ulrich Reiter},
  \bibinfo{person}{Raimund Schatz}, \bibinfo{person}{Peter Schelkens},
  \bibinfo{person}{Lea Skorin-Kapov}, \bibinfo{person}{Dominik Strohmeier},
  \bibinfo{person}{Christian Timmerer}, \bibinfo{person}{Martin Varela},
  \bibinfo{person}{Ina Wechsung}, \bibinfo{person}{Junyong You}, {and}
  \bibinfo{person}{Andrej Zgank}.} \bibinfo{year}{2013}\natexlab{}.
\newblock \bibinfo{booktitle}{\emph{{Qualinet White Paper on Definitions of
  Quality of Experience}}}.
\newblock \bibinfo{type}{{T}echnical {R}eport}. \bibinfo{institution}{{Qualinet
  (www.qualinet.eu)}}.
\newblock


\bibitem[Cani and Desbrun(1997)]%
        {DBLP:journals/tvcg/Cani-GascuelD97}
\bibfield{author}{\bibinfo{person}{Marie{-}Paule Cani} {and}
  \bibinfo{person}{Mathieu Desbrun}.} \bibinfo{year}{1997}\natexlab{}.
\newblock \showarticletitle{Animation of Deformable Models Using Implicit
  Surfaces}.
\newblock \bibinfo{journal}{\emph{{IEEE} Trans. Vis. Comput. Graph.}}
  \bibinfo{volume}{3}, \bibinfo{number}{1} (\bibinfo{year}{1997}),
  \bibinfo{pages}{39--50}.
\newblock


\bibitem[Carballeira et~al\mbox{.}(2022)]%
        {DBLP:journals/tmm/CarballeiraCDBC22}
\bibfield{author}{\bibinfo{person}{Pablo Carballeira}, \bibinfo{person}{Carlos
  Carmona}, \bibinfo{person}{C{\'{e}}sar D{\'{\i}}az}, \bibinfo{person}{Daniel
  Berj{\'{o}}n}, \bibinfo{person}{Daniel Corregidor},
  \bibinfo{person}{Juli{\'{a}}n Cabrera}, \bibinfo{person}{Francisco
  Mor{\'{a}}n}, \bibinfo{person}{Carmen Doblado}, \bibinfo{person}{Sergio
  Arnaldo}, \bibinfo{person}{Mar{\'{\i}}a del Mar~Mart{\'{\i}}n}, {and}
  \bibinfo{person}{Narciso Garc{\'{\i}}a}.} \bibinfo{year}{2022}\natexlab{}.
\newblock \showarticletitle{{FVV} Live: {A} Real-Time Free-Viewpoint Video
  System With Consumer Electronics Hardware}.
\newblock \bibinfo{journal}{\emph{{IEEE} Trans. Multim.}}  \bibinfo{volume}{24}
  (\bibinfo{year}{2022}), \bibinfo{pages}{2378--2391}.
\newblock


\bibitem[Chan et~al\mbox{.}(2005)]%
        {DBLP:journals/tcsv/ChanNGCS05}
\bibfield{author}{\bibinfo{person}{Shing{-}Chow Chan}, \bibinfo{person}{King~To
  Ng}, \bibinfo{person}{Zhi{-}Feng Gan}, \bibinfo{person}{Kin{-}Lok Chan},
  {and} \bibinfo{person}{Heung{-}Yeung Shum}.} \bibinfo{year}{2005}\natexlab{}.
\newblock \showarticletitle{The Plenoptic Video}.
\newblock \bibinfo{journal}{\emph{{IEEE} Trans. Circuits Syst. Video Technol.}}
  \bibinfo{volume}{15}, \bibinfo{number}{12} (\bibinfo{year}{2005}),
  \bibinfo{pages}{1650--1659}.
\newblock


\bibitem[Chang et~al\mbox{.}(2018)]%
        {DBLP:journals/tog/ChangKS18}
\bibfield{author}{\bibinfo{person}{Jen{-}Hao~Rick Chang},
  \bibinfo{person}{B.~V. K.~Vijaya Kumar}, {and} \bibinfo{person}{Aswin~C.
  Sankaranarayanan}.} \bibinfo{year}{2018}\natexlab{}.
\newblock \showarticletitle{Towards Multifocal Displays with Dense Focal
  Stacks}.
\newblock \bibinfo{journal}{\emph{{ACM} Trans. Graph.}} \bibinfo{volume}{37},
  \bibinfo{number}{6} (\bibinfo{year}{2018}), \bibinfo{pages}{198}.
\newblock


\bibitem[Chen et~al\mbox{.}(2022)]%
        {DBLP:conf/eccv/ChenXGYS22}
\bibfield{author}{\bibinfo{person}{Anpei Chen}, \bibinfo{person}{Zexiang Xu},
  \bibinfo{person}{Andreas Geiger}, \bibinfo{person}{Jingyi Yu}, {and}
  \bibinfo{person}{Hao Su}.} \bibinfo{year}{2022}\natexlab{}.
\newblock \showarticletitle{TensoRF: Tensorial Radiance Fields}. In
  \bibinfo{booktitle}{\emph{Computer Vision - {ECCV} 2022 - 17th European
  Conference, Proceedings, Part {XXXII}}} \emph{(\bibinfo{series}{Lecture Notes
  in Computer Science}, Vol.~\bibinfo{volume}{13692})}.
  \bibinfo{publisher}{Springer}, \bibinfo{pages}{333--350}.
\newblock


\bibitem[Chen et~al\mbox{.}(2015)]%
        {DBLP:journals/comsur/ChenWZ15}
\bibfield{author}{\bibinfo{person}{Yanjiao Chen}, \bibinfo{person}{Kaishun Wu},
  {and} \bibinfo{person}{Qian Zhang}.} \bibinfo{year}{2015}\natexlab{}.
\newblock \showarticletitle{From QoS to QoE: {A} Tutorial on Video Quality
  Assessment}.
\newblock \bibinfo{journal}{\emph{{IEEE} Commun. Surv. Tutorials}}
  \bibinfo{volume}{17}, \bibinfo{number}{2} (\bibinfo{year}{2015}),
  \bibinfo{pages}{1126--1165}.
\newblock


\bibitem[Chen et~al\mbox{.}(2023)]%
        {chen2023mobilenerf}
\bibfield{author}{\bibinfo{person}{Zhiqin Chen}, \bibinfo{person}{Thomas
  Funkhouser}, \bibinfo{person}{Peter Hedman}, {and} \bibinfo{person}{Andrea
  Tagliasacchi}.} \bibinfo{year}{2023}\natexlab{}.
\newblock \showarticletitle{Mobilenerf: Exploiting the Polygon Rasterization
  Pipeline for Efficient Neural Field Rendering on Mobile Architectures}. In
  \bibinfo{booktitle}{\emph{Proceedings of the IEEE/CVF Conference on Computer
  Vision and Pattern Recognition}}. \bibinfo{pages}{16569--16578}.
\newblock


\bibitem[Chikkerur et~al\mbox{.}(2011)]%
        {DBLP:journals/tbc/ChikkerurSRK11}
\bibfield{author}{\bibinfo{person}{Shyamprasad Chikkerur},
  \bibinfo{person}{Vijay Sundaram}, \bibinfo{person}{Martin Reisslein}, {and}
  \bibinfo{person}{Lina~J. Karam}.} \bibinfo{year}{2011}\natexlab{}.
\newblock \showarticletitle{Objective Video Quality Assessment Methods: {A}
  Classification, Review, and Performance Comparison}.
\newblock \bibinfo{journal}{\emph{{IEEE} Trans. Broadcast.}}
  \bibinfo{volume}{57}, \bibinfo{number}{2} (\bibinfo{year}{2011}),
  \bibinfo{pages}{165--182}.
\newblock


\bibitem[Cignoni et~al\mbox{.}(2008)]%
        {DBLP:conf/egItaly/CignoniCCDGR08}
\bibfield{author}{\bibinfo{person}{Paolo Cignoni}, \bibinfo{person}{Marco
  Callieri}, \bibinfo{person}{Massimiliano Corsini}, \bibinfo{person}{Matteo
  Dellepiane}, \bibinfo{person}{Fabio Ganovelli}, {and} \bibinfo{person}{Guido
  Ranzuglia}.} \bibinfo{year}{2008}\natexlab{}.
\newblock \showarticletitle{MeshLab: An Open-Source Mesh Processing Tool}. In
  \bibinfo{booktitle}{\emph{Eurographics Italian Chapter Conference 2008}}.
  \bibinfo{publisher}{Eurographics}, \bibinfo{pages}{129--136}.
\newblock


\bibitem[Clarke(1999)]%
        {DBLP:journals/imst/Clarke99}
\bibfield{author}{\bibinfo{person}{Roger~J. Clarke}.}
  \bibinfo{year}{1999}\natexlab{}.
\newblock \showarticletitle{Image and Video Compression: {A} Survey}.
\newblock \bibinfo{journal}{\emph{Int. J. Imaging Syst. Technol.}}
  \bibinfo{volume}{10}, \bibinfo{number}{1} (\bibinfo{year}{1999}),
  \bibinfo{pages}{20--32}.
\newblock


\bibitem[Clarkson et~al\mbox{.}(2013)]%
        {clarkson2013distortion}
\bibfield{author}{\bibinfo{person}{Sean Clarkson}, \bibinfo{person}{Jonathan
  Wheat}, \bibinfo{person}{Ben Heller}, \bibinfo{person}{James Webster}, {and}
  \bibinfo{person}{Simon Choppin}.} \bibinfo{year}{2013}\natexlab{}.
\newblock \showarticletitle{Distortion Correction of Depth Data from Consumer
  Depth Cameras}.
\newblock \bibinfo{journal}{\emph{3D Body Scanning Technologies, Long Beach,
  California, Hometrica Consulting}} (\bibinfo{year}{2013}),
  \bibinfo{pages}{426--437}.
\newblock


\bibitem[Corporation(2023a)]%
        {vive}
\bibfield{author}{\bibinfo{person}{HTC Corporation}.}
  \bibinfo{year}{2023}\natexlab{a}.
\newblock \bibinfo{title}{VIVE: Discover Virtual Reality Beyond Imagination.}
\newblock
\newblock
\urldef\tempurl%
\url{https://www.vive.com/}
\showURL{%
Retrieved June 14, 2023 from \tempurl}


\bibitem[Corporation(2023b)]%
        {realsense}
\bibfield{author}{\bibinfo{person}{Intel Corporation}.}
  \bibinfo{year}{2023}\natexlab{b}.
\newblock \bibinfo{title}{Intel RealSense}.
\newblock
\newblock
\urldef\tempurl%
\url{https://www.anthropic.com/index/introducing-claude}
\showURL{%
Retrieved June 27, 2023 from \tempurl}


\bibitem[da~Silva~Cruz et~al\mbox{.}(2019)]%
        {DBLP:conf/qomex/CruzDAPDPPE19}
\bibfield{author}{\bibinfo{person}{Lu{\'{\i}}s~Alberto da Silva~Cruz},
  \bibinfo{person}{Emil Dumic}, \bibinfo{person}{Evangelos Alexiou},
  \bibinfo{person}{Jo{\~{a}}o Prazeres}, \bibinfo{person}{Carlos~Rafael
  Duarte}, \bibinfo{person}{Manuela Pereira}, \bibinfo{person}{Ant{\'{o}}nio
  M.~G. Pinheiro}, {and} \bibinfo{person}{Touradj Ebrahimi}.}
  \bibinfo{year}{2019}\natexlab{}.
\newblock \showarticletitle{Point Cloud Quality Evaluation: Towards a
  Definition for Test Conditions}. In \bibinfo{booktitle}{\emph{11th
  International Conference on Quality of Multimedia Experience QoMEX 2019}}.
  \bibinfo{publisher}{{IEEE}}, \bibinfo{pages}{1--6}.
\newblock


\bibitem[Dai et~al\mbox{.}(2020)]%
        {dai2020neural}
\bibfield{author}{\bibinfo{person}{Peng Dai}, \bibinfo{person}{Yinda Zhang},
  \bibinfo{person}{Zhuwen Li}, \bibinfo{person}{Shuaicheng Liu}, {and}
  \bibinfo{person}{Bing Zeng}.} \bibinfo{year}{2020}\natexlab{}.
\newblock \showarticletitle{Neural Point Cloud Rendering via Multi-plane
  Projection}. In \bibinfo{booktitle}{\emph{Proceedings of the IEEE/CVF
  Conference on Computer Vision and Pattern Recognition}}.
  \bibinfo{pages}{7830--7839}.
\newblock


\bibitem[Deering(1995)]%
        {DBLP:conf/siggraph/Deering95}
\bibfield{author}{\bibinfo{person}{Michael Deering}.}
  \bibinfo{year}{1995}\natexlab{}.
\newblock \showarticletitle{Geometry compression}. In
  \bibinfo{booktitle}{\emph{Proceedings of the 22nd Annual Conference on
  Computer Graphics and Interactive Techniques, {SIGGRAPH} 1995}}.
  \bibinfo{publisher}{{ACM}}, \bibinfo{pages}{13--20}.
\newblock


\bibitem[Deng and Tartaglione(2023)]%
        {DBLP:conf/wacv/DengT23}
\bibfield{author}{\bibinfo{person}{Chenxi~Lola Deng} {and}
  \bibinfo{person}{Enzo Tartaglione}.} \bibinfo{year}{2023}\natexlab{}.
\newblock \showarticletitle{Compressing Explicit Voxel Grid Representations:
  fast NeRFs become also small}. In \bibinfo{booktitle}{\emph{{IEEE/CVF} Winter
  Conference on Applications of Computer Vision, {WACV} 2023}}.
  \bibinfo{publisher}{{IEEE}}, \bibinfo{pages}{1236--1245}.
\newblock


\bibitem[d'Eon et~al\mbox{.}(2017)]%
        {8iVFB}
\bibfield{author}{\bibinfo{person}{Eugene d'Eon}, \bibinfo{person}{Bob
  Harrison}, \bibinfo{person}{Taos Myers}, {and} \bibinfo{person}{Philip~A.
  Chou}.} \bibinfo{year}{2017}\natexlab{}.
\newblock \showarticletitle{8i Voxelized Full Bodies - A Voxelized Point Cloud
  Dataset}.
\newblock \bibinfo{journal}{\emph{ISO/IEC JTC1/SC29 Joint WG11/WG1 (MPEG/JPEG)
  input document WG11M40059/WG1M74006}} (\bibinfo{year}{2017}).
\newblock


\bibitem[Devillers and Gandoin(2000)]%
        {DBLP:conf/visualization/DevillersG00}
\bibfield{author}{\bibinfo{person}{Olivier Devillers} {and}
  \bibinfo{person}{Pierre{-}Marie Gandoin}.} \bibinfo{year}{2000}\natexlab{}.
\newblock \showarticletitle{Geometric Compression for Interactive
  Transmission}. In \bibinfo{booktitle}{\emph{11th {IEEE} Visualization
  Conference, {IEEE} Vis 2000, Proceedings}}. \bibinfo{publisher}{{IEEE}
  Computer Society and {ACM}}, \bibinfo{pages}{319--326}.
\newblock


\bibitem[Diniz et~al\mbox{.}(2020a)]%
        {DBLP:conf/mmsp/DinizFF20}
\bibfield{author}{\bibinfo{person}{Rafael Diniz}, \bibinfo{person}{Pedro~Garcia
  Freitas}, {and} \bibinfo{person}{Myl{\`{e}}ne C.~Q. Farias}.}
  \bibinfo{year}{2020}\natexlab{a}.
\newblock \showarticletitle{Local Luminance Patterns for Point Cloud Quality
  Assessment}. In \bibinfo{booktitle}{\emph{22nd {IEEE} International Workshop
  on Multimedia Signal Processing, {MMSP} 2020}}. \bibinfo{publisher}{{IEEE}},
  \bibinfo{pages}{1--6}.
\newblock


\bibitem[Diniz et~al\mbox{.}(2020b)]%
        {DBLP:conf/icip/DinizFF20}
\bibfield{author}{\bibinfo{person}{Rafael Diniz}, \bibinfo{person}{Pedro~Garcia
  Freitas}, {and} \bibinfo{person}{Myl{\`{e}}ne C.~Q. Farias}.}
  \bibinfo{year}{2020}\natexlab{b}.
\newblock \showarticletitle{Multi-Distance Point Cloud Quality Assessment}. In
  \bibinfo{booktitle}{\emph{{IEEE} International Conference on Image
  Processing, {ICIP} 2020}}. \bibinfo{publisher}{{IEEE}},
  \bibinfo{pages}{3443--3447}.
\newblock


\bibitem[Diniz et~al\mbox{.}(2020c)]%
        {DBLP:conf/qomex/DinizFF20}
\bibfield{author}{\bibinfo{person}{Rafael Diniz}, \bibinfo{person}{Pedro~Garcia
  Freitas}, {and} \bibinfo{person}{Myl{\`{e}}ne C.~Q. Farias}.}
  \bibinfo{year}{2020}\natexlab{c}.
\newblock \showarticletitle{Towards a Point Cloud Quality Assessment Model
  using Local Binary Patterns}. In \bibinfo{booktitle}{\emph{Twelfth
  International Conference on Quality of Multimedia Experience, QoMEX 2020}}.
  \bibinfo{publisher}{{IEEE}}, \bibinfo{pages}{1--6}.
\newblock


\bibitem[Dunn et~al\mbox{.}(2017)]%
        {DBLP:journals/tvcg/DunnTTKADMLF17}
\bibfield{author}{\bibinfo{person}{David Dunn}, \bibinfo{person}{Cary Tippets},
  \bibinfo{person}{Kent Torell}, \bibinfo{person}{Petr Kellnhofer},
  \bibinfo{person}{Kaan Aksit}, \bibinfo{person}{Piotr Didyk},
  \bibinfo{person}{Karol Myszkowski}, \bibinfo{person}{David Luebke}, {and}
  \bibinfo{person}{Henry Fuchs}.} \bibinfo{year}{2017}\natexlab{}.
\newblock \showarticletitle{Wide Field Of View Varifocal Near-Eye Display Using
  See-Through Deformable Membrane Mirrors}.
\newblock \bibinfo{journal}{\emph{{IEEE} Trans. Vis. Comput. Graph.}}
  \bibinfo{volume}{23}, \bibinfo{number}{4} (\bibinfo{year}{2017}),
  \bibinfo{pages}{1322--1331}.
\newblock


\bibitem[Dzardanova and Kasapakis(2023)]%
        {DBLP:journals/annals/DzardanovaK23}
\bibfield{author}{\bibinfo{person}{Elena Dzardanova} {and}
  \bibinfo{person}{Vlasios Kasapakis}.} \bibinfo{year}{2023}\natexlab{}.
\newblock \showarticletitle{Virtual Reality: {A} Journey From Vision to
  Commodity}.
\newblock \bibinfo{journal}{\emph{{IEEE} Ann. Hist. Comput.}}
  \bibinfo{volume}{45}, \bibinfo{number}{1} (\bibinfo{year}{2023}),
  \bibinfo{pages}{18--30}.
\newblock


\bibitem[Egido(1988)]%
        {DBLP:conf/cscw/Egido88}
\bibfield{author}{\bibinfo{person}{Carmen Egido}.}
  \bibinfo{year}{1988}\natexlab{}.
\newblock \showarticletitle{Video Conferencing as a Technology to Support Group
  Work: {A} Review of its Failures}. In \bibinfo{booktitle}{\emph{{CSCW} '88,
  Proceedings of the Conference on Computer-Supported Cooperative Work}}.
  \bibinfo{publisher}{{ACM}}, \bibinfo{pages}{13--24}.
\newblock


\bibitem[Eisert et~al\mbox{.}(2023)]%
        {IVT11}
\bibfield{author}{\bibinfo{person}{Peter Eisert}, \bibinfo{person}{Oliver
  Schreer}, \bibinfo{person}{Ingo Feldmann}, \bibinfo{person}{Cornelius
  Hellge}, {and} \bibinfo{person}{Anna Hilsmann}.}
  \bibinfo{year}{2023}\natexlab{}.
\newblock \showarticletitle{Volumetric Video--Acquisition, Interaction,
  Streaming and Rendering}.
\newblock In \bibinfo{booktitle}{\emph{Immersive Video Technologies}}.
  \bibinfo{publisher}{Elsevier}, \bibinfo{pages}{289--326}.
\newblock


\bibitem[Fan et~al\mbox{.}(2019)]%
        {DBLP:journals/csur/FanLPH19}
\bibfield{author}{\bibinfo{person}{Ching{-}Ling Fan},
  \bibinfo{person}{Wen{-}Chih Lo}, \bibinfo{person}{Yu{-}Tung Pai}, {and}
  \bibinfo{person}{Cheng{-}Hsin Hsu}.} \bibinfo{year}{2019}\natexlab{}.
\newblock \showarticletitle{A Survey on 360{\textdegree} Video Streaming:
  Acquisition, Transmission, and Display}.
\newblock \bibinfo{journal}{\emph{{ACM} Comput. Surv.}} \bibinfo{volume}{52},
  \bibinfo{number}{4} (\bibinfo{year}{2019}), \bibinfo{pages}{71:1--71:36}.
\newblock


\bibitem[Fridovich{-}Keil et~al\mbox{.}(2022)]%
        {DBLP:conf/cvpr/Fridovich-KeilY22}
\bibfield{author}{\bibinfo{person}{Sara Fridovich{-}Keil},
  \bibinfo{person}{Alex Yu}, \bibinfo{person}{Matthew Tancik},
  \bibinfo{person}{Qinhong Chen}, \bibinfo{person}{Benjamin Recht}, {and}
  \bibinfo{person}{Angjoo Kanazawa}.} \bibinfo{year}{2022}\natexlab{}.
\newblock \showarticletitle{Plenoxels: Radiance Fields without Neural
  Networks}. In \bibinfo{booktitle}{\emph{{IEEE/CVF} Conference on Computer
  Vision and Pattern Recognition, {CVPR} 2022}}. \bibinfo{publisher}{{IEEE}},
  \bibinfo{pages}{5491--5500}.
\newblock


\bibitem[Garcia and de~Queiroz(2017)]%
        {DBLP:conf/icip/GarciaQ17}
\bibfield{author}{\bibinfo{person}{Diogo~C. Garcia} {and}
  \bibinfo{person}{Ricardo~L. de Queiroz}.} \bibinfo{year}{2017}\natexlab{}.
\newblock \showarticletitle{Context-based Octree Coding for Point-cloud Video}.
  In \bibinfo{booktitle}{\emph{2017 {IEEE} International Conference on Image
  Processing, {ICIP} 2017}}. \bibinfo{publisher}{{IEEE}},
  \bibinfo{pages}{1412--1416}.
\newblock


\bibitem[Garg et~al\mbox{.}(2016)]%
        {garg2016unsupervised}
\bibfield{author}{\bibinfo{person}{Ravi Garg}, \bibinfo{person}{Vijay~Kumar
  Bg}, \bibinfo{person}{Gustavo Carneiro}, {and} \bibinfo{person}{Ian Reid}.}
  \bibinfo{year}{2016}\natexlab{}.
\newblock \showarticletitle{Unsupervised {CNN} for Single View Depth
  Estimation: Geometry to the Rescue}. In \bibinfo{booktitle}{\emph{Computer
  Vision--ECCV 2016: 14th European Conference, Amsterdam, The Netherlands,
  October 11-14, 2016, Proceedings, Part VIII 14}}. Springer,
  \bibinfo{pages}{740--756}.
\newblock


\bibitem[Geng et~al\mbox{.}(2023)]%
        {Geng_2023_CVPR}
\bibfield{author}{\bibinfo{person}{Chen Geng}, \bibinfo{person}{Sida Peng},
  \bibinfo{person}{Zhen Xu}, \bibinfo{person}{Hujun Bao}, {and}
  \bibinfo{person}{Xiaowei Zhou}.} \bibinfo{year}{2023}\natexlab{}.
\newblock \showarticletitle{Learning Neural Volumetric Representations of
  Dynamic Humans in Minutes}. In \bibinfo{booktitle}{\emph{Proceedings of the
  IEEE/CVF Conference on Computer Vision and Pattern Recognition (CVPR)}}.
  \bibinfo{pages}{8759--8770}.
\newblock


\bibitem[Google(2024)]%
        {DRACO}
\bibfield{author}{\bibinfo{person}{Google}.} \bibinfo{year}{2024}\natexlab{}.
\newblock \bibinfo{title}{Draco: 3D Graphics Compression}.
\newblock
\newblock
\urldef\tempurl%
\url{https://google.github.io/draco/}
\showURL{%
Retrieved September 20, 2024 from \tempurl}


\bibitem[Gou et~al\mbox{.}(2021)]%
        {DBLP:journals/ijcv/GouYMT21}
\bibfield{author}{\bibinfo{person}{Jianping Gou}, \bibinfo{person}{Baosheng
  Yu}, \bibinfo{person}{Stephen~J. Maybank}, {and} \bibinfo{person}{Dacheng
  Tao}.} \bibinfo{year}{2021}\natexlab{}.
\newblock \showarticletitle{Knowledge Distillation: {A} Survey}.
\newblock \bibinfo{journal}{\emph{Int. J. Comput. Vis.}} \bibinfo{volume}{129},
  \bibinfo{number}{6} (\bibinfo{year}{2021}), \bibinfo{pages}{1789--1819}.
\newblock


\bibitem[Guarda et~al\mbox{.}(2019a)]%
        {DBLP:conf/euvip/GuardaRP19}
\bibfield{author}{\bibinfo{person}{Andr{\'{e}} F.~R. Guarda},
  \bibinfo{person}{Nuno M.~M. Rodrigues}, {and} \bibinfo{person}{Fernando
  Pereira}.} \bibinfo{year}{2019}\natexlab{a}.
\newblock \showarticletitle{Deep Learning-Based Point Cloud Coding: {A}
  Behavior and Performance Study}. In \bibinfo{booktitle}{\emph{8th European
  Workshop on Visual Information Processing, {EUVIP} 2019}}.
  \bibinfo{publisher}{{IEEE}}, \bibinfo{pages}{34--39}.
\newblock


\bibitem[Guarda et~al\mbox{.}(2019b)]%
        {DBLP:conf/pcs/GuardaRP19}
\bibfield{author}{\bibinfo{person}{Andr{\'{e}} F.~R. Guarda},
  \bibinfo{person}{Nuno M.~M. Rodrigues}, {and} \bibinfo{person}{Fernando
  Pereira}.} \bibinfo{year}{2019}\natexlab{b}.
\newblock \showarticletitle{Point Cloud Coding: Adopting a Deep Learning-based
  Approach}. In \bibinfo{booktitle}{\emph{Picture Coding Symposium, {PCS}
  2019}}. \bibinfo{publisher}{{IEEE}}, \bibinfo{pages}{1--5}.
\newblock


\bibitem[Gumhold and Stra{\ss}er(1998)]%
        {DBLP:conf/siggraph/GumholdS98}
\bibfield{author}{\bibinfo{person}{Stefan Gumhold} {and}
  \bibinfo{person}{Wolfgang Stra{\ss}er}.} \bibinfo{year}{1998}\natexlab{}.
\newblock \showarticletitle{Real Time Compression of Triangle Mesh
  Connectivity}. In \bibinfo{booktitle}{\emph{Proceedings of the 25th Annual
  Conference on Computer Graphics and Interactive Techniques, {SIGGRAPH}
  1998}}. \bibinfo{publisher}{{ACM}}, \bibinfo{pages}{133--140}.
\newblock


\bibitem[Guo et~al\mbox{.}(2021)]%
        {DBLP:journals/pami/GuoWHLLB21}
\bibfield{author}{\bibinfo{person}{Yulan Guo}, \bibinfo{person}{Hanyun Wang},
  \bibinfo{person}{Qingyong Hu}, \bibinfo{person}{Hao Liu}, \bibinfo{person}{Li
  Liu}, {and} \bibinfo{person}{Mohammed Bennamoun}.}
  \bibinfo{year}{2021}\natexlab{}.
\newblock \showarticletitle{Deep Learning for 3D Point Clouds: {A} Survey}.
\newblock \bibinfo{journal}{\emph{{IEEE} Trans. Pattern Anal. Mach. Intell.}}
  \bibinfo{volume}{43}, \bibinfo{number}{12} (\bibinfo{year}{2021}),
  \bibinfo{pages}{4338--4364}.
\newblock


\bibitem[Guo et~al\mbox{.}(2023)]%
        {guo2023vmesh}
\bibfield{author}{\bibinfo{person}{Yuan-Chen Guo}, \bibinfo{person}{Yan-Pei
  Cao}, \bibinfo{person}{Chen Wang}, \bibinfo{person}{Yu He},
  \bibinfo{person}{Ying Shan}, \bibinfo{person}{Xiaohu Qie}, {and}
  \bibinfo{person}{Song-Hai Zhang}.} \bibinfo{year}{2023}\natexlab{}.
\newblock \showarticletitle{VMesh: Hybrid Volume-Mesh Representation for
  Efficient View Synthesis}.
\newblock \bibinfo{journal}{\emph{arXiv preprint arXiv:2303.16184}}
  (\bibinfo{year}{2023}).
\newblock


\bibitem[Gurel et~al\mbox{.}(2024)]%
        {gurel2024v2ra}
\bibfield{author}{\bibinfo{person}{Zafer Gurel}, \bibinfo{person}{Alperen~F
  Zengin}, \bibinfo{person}{Ali~C Begen}, \bibinfo{person}{Saba Ahsan},
  \bibinfo{person}{Lukasz Kondrad}, \bibinfo{person}{Kashyap
  Kammachi-Sreedhar}, \bibinfo{person}{Serhan G{\"u}l}, \bibinfo{person}{Gazi
  Illahi}, {and} \bibinfo{person}{Igor~DD Curcio}.}
  \bibinfo{year}{2024}\natexlab{}.
\newblock \showarticletitle{V2RA: a Grid-Based Rate-Adaptation Logic for
  Volumetric Video}. In \bibinfo{booktitle}{\emph{Proceedings of the 16th
  International Workshop on Immersive Mixed and Virtual Environment Systems}}.
  \bibinfo{pages}{50--56}.
\newblock


\bibitem[Guti{\'{e}}rrez et~al\mbox{.}(2022)]%
        {DBLP:journals/tmm/GutierrezPOSCMV22}
\bibfield{author}{\bibinfo{person}{Jes{\'{u}}s Guti{\'{e}}rrez},
  \bibinfo{person}{Pablo P{\'{e}}rez}, \bibinfo{person}{Marta Orduna},
  \bibinfo{person}{Ashutosh Singla}, \bibinfo{person}{Carlos Cort{\'{e}}s},
  \bibinfo{person}{Pramit Mazumdar}, \bibinfo{person}{Irene Viola},
  \bibinfo{person}{Kjell Brunnstr{\"{o}}m}, \bibinfo{person}{Federica
  Battisti}, \bibinfo{person}{Natalia Cieplinska}, \bibinfo{person}{Dawid
  Juszka}, \bibinfo{person}{Lucjan Janowski}, \bibinfo{person}{Mikolaj
  Leszczuk}, \bibinfo{person}{Anthony Adeyemi{-}Ejeye}, \bibinfo{person}{Yaosi
  Hu}, \bibinfo{person}{Zhenzhong Chen}, \bibinfo{person}{Glenn~Van
  Wallendael}, \bibinfo{person}{Peter Lambert}, \bibinfo{person}{C{\'{e}}sar
  D{\'{\i}}az}, \bibinfo{person}{John Hedlund}, \bibinfo{person}{Omar Hamsis},
  \bibinfo{person}{Stephan Fremerey}, \bibinfo{person}{Frank Hofmeyer},
  \bibinfo{person}{Alexander Raake}, \bibinfo{person}{Pablo C{\'{e}}sar},
  \bibinfo{person}{Marco Carli}, {and} \bibinfo{person}{Narciso
  Garc{\'{\i}}a}.} \bibinfo{year}{2022}\natexlab{}.
\newblock \showarticletitle{Subjective Evaluation of Visual Quality and
  Simulator Sickness of Short 360{\textdegree} Videos: {ITU-T} Rec. {P.919}}.
\newblock \bibinfo{journal}{\emph{{IEEE} Trans. Multim.}}  \bibinfo{volume}{24}
  (\bibinfo{year}{2022}), \bibinfo{pages}{3087--3100}.
\newblock


\bibitem[Han(2019)]%
        {DBLP:journals/cm/Han19}
\bibfield{author}{\bibinfo{person}{Bo Han}.} \bibinfo{year}{2019}\natexlab{}.
\newblock \showarticletitle{Mobile Immersive Computing: Research Challenges and
  the Road Ahead}.
\newblock \bibinfo{journal}{\emph{{IEEE} Commun. Mag.}} \bibinfo{volume}{57},
  \bibinfo{number}{10} (\bibinfo{year}{2019}), \bibinfo{pages}{112--118}.
\newblock


\bibitem[Han et~al\mbox{.}(2020)]%
        {DBLP:conf/mobicom/HanLQ20}
\bibfield{author}{\bibinfo{person}{Bo Han}, \bibinfo{person}{Yu Liu}, {and}
  \bibinfo{person}{Feng Qian}.} \bibinfo{year}{2020}\natexlab{}.
\newblock \showarticletitle{ViVo: Visibility-aware Mobile Volumetric Video
  Streaming}. In \bibinfo{booktitle}{\emph{MobiCom '20: The 26th Annual
  International Conference on Mobile Computing and Networking}}.
  \bibinfo{publisher}{{ACM}}, \bibinfo{pages}{11:1--11:13}.
\newblock


\bibitem[Han et~al\mbox{.}(2016)]%
        {DBLP:journals/corr/HanMD15}
\bibfield{author}{\bibinfo{person}{Song Han}, \bibinfo{person}{Huizi Mao},
  {and} \bibinfo{person}{William~J. Dally}.} \bibinfo{year}{2016}\natexlab{}.
\newblock \showarticletitle{Deep Compression: Compressing Deep Neural Network
  with Pruning, Trained Quantization and Huffman Coding}. In
  \bibinfo{booktitle}{\emph{4th International Conference on Learning
  Representations, {ICLR} 2016, Conference Track Proceedings}}.
\newblock


\bibitem[Han et~al\mbox{.}(2018)]%
        {DBLP:journals/tip/HanLVLBHC18}
\bibfield{author}{\bibinfo{person}{Zhizhong Han}, \bibinfo{person}{Zhenbao
  Liu}, \bibinfo{person}{Chi{-}Man Vong}, \bibinfo{person}{Yu{-}Shen Liu},
  \bibinfo{person}{Shuhui Bu}, \bibinfo{person}{Junwei Han}, {and}
  \bibinfo{person}{C.~L.~Philip Chen}.} \bibinfo{year}{2018}\natexlab{}.
\newblock \showarticletitle{Deep Spatiality: Unsupervised Learning of
  Spatially-Enhanced Global and Local 3D Features by Deep Neural Network With
  Coupled Softmax}.
\newblock \bibinfo{journal}{\emph{{IEEE} Trans. Image Process.}}
  \bibinfo{volume}{27}, \bibinfo{number}{6} (\bibinfo{year}{2018}),
  \bibinfo{pages}{3049--3063}.
\newblock


\bibitem[Held and Banks(2008)]%
        {DBLP:conf/apgv/HeldB08}
\bibfield{author}{\bibinfo{person}{Robert~T. Held} {and}
  \bibinfo{person}{Martin~S. Banks}.} \bibinfo{year}{2008}\natexlab{}.
\newblock \showarticletitle{Misperceptions in Stereoscopic Displays: A Vision
  Science Perspective}. In \bibinfo{booktitle}{\emph{Proceedings of the 5th
  Symposium on Applied Perception in Graphics and Visualization, {APGV} 2008}}
  \emph{(\bibinfo{series}{{ACM} International Conference Proceeding Series})}.
  \bibinfo{publisher}{{ACM}}, \bibinfo{pages}{23--32}.
\newblock


\bibitem[Hor{\'{e}} and Ziou(2010)]%
        {DBLP:conf/icpr/HoreZ10}
\bibfield{author}{\bibinfo{person}{Alain Hor{\'{e}}} {and}
  \bibinfo{person}{Djemel Ziou}.} \bibinfo{year}{2010}\natexlab{}.
\newblock \showarticletitle{Image Quality Metrics: {PSNR} vs. {SSIM}}. In
  \bibinfo{booktitle}{\emph{20th International Conference on Pattern
  Recognition, {ICPR} 2010}}. \bibinfo{publisher}{{IEEE} Computer Society},
  \bibinfo{pages}{2366--2369}.
\newblock


\bibitem[Horv{\'{a}}th(2016)]%
        {DBLP:conf/coginfocom/Horvath16b}
\bibfield{author}{\bibinfo{person}{Ildik{\'{o}} Horv{\'{a}}th}.}
  \bibinfo{year}{2016}\natexlab{}.
\newblock \showarticletitle{Innovative Engineering Education in the Cooperative
  {VR} Environment}. In \bibinfo{booktitle}{\emph{7th {IEEE} International
  Conference on Cognitive Infocommunications, CogInfoCom 2016}}.
  \bibinfo{publisher}{{IEEE}}, \bibinfo{pages}{359--364}.
\newblock


\bibitem[Horv{\'a}th(2021)]%
        {horvath2021analysis}
\bibfield{author}{\bibinfo{person}{Ildik{\'o} Horv{\'a}th}.}
  \bibinfo{year}{2021}\natexlab{}.
\newblock \showarticletitle{An Analysis of Personalized Learning Opportunities
  in 3D VR}.
\newblock \bibinfo{journal}{\emph{Frontiers in Computer Science}}
  \bibinfo{volume}{3} (\bibinfo{year}{2021}), \bibinfo{pages}{673826}.
\newblock


\bibitem[Hosseini and Timmerer(2018)]%
        {hosseini2018dynamic}
\bibfield{author}{\bibinfo{person}{Mohammad Hosseini} {and}
  \bibinfo{person}{Christian Timmerer}.} \bibinfo{year}{2018}\natexlab{}.
\newblock \showarticletitle{Dynamic Adaptive Point Cloud Streaming}. In
  \bibinfo{booktitle}{\emph{Proceedings of the 23rd Packet Video Workshop}}.
  \bibinfo{pages}{25--30}.
\newblock


\bibitem[Hu et~al\mbox{.}(2023a)]%
        {fsvvd}
\bibfield{author}{\bibinfo{person}{Kaiyuan Hu}, \bibinfo{person}{Yili Jin},
  \bibinfo{person}{Haowen Yang}, \bibinfo{person}{Junhua Liu}, {and}
  \bibinfo{person}{Fangxin Wang}.} \bibinfo{year}{2023}\natexlab{a}.
\newblock \showarticletitle{{FSVVD:} {A} Dataset of Full Scene Volumetric
  Video}. In \bibinfo{booktitle}{\emph{Proceedings of the 14th Conference on
  {ACM} Multimedia Systems, MMSys 2023}}. \bibinfo{publisher}{{ACM}},
  \bibinfo{pages}{410--415}.
\newblock


\bibitem[Hu et~al\mbox{.}(2023b)]%
        {DBLP:conf/mm/HuYJLCZ023}
\bibfield{author}{\bibinfo{person}{Kaiyuan Hu}, \bibinfo{person}{Haowen Yang},
  \bibinfo{person}{Yili Jin}, \bibinfo{person}{Junhua Liu},
  \bibinfo{person}{Yongting Chen}, \bibinfo{person}{Miao Zhang}, {and}
  \bibinfo{person}{Fangxin Wang}.} \bibinfo{year}{2023}\natexlab{b}.
\newblock \showarticletitle{Understanding User Behavior in Volumetric Video
  Watching: Dataset, Analysis and Prediction}. In
  \bibinfo{booktitle}{\emph{Proceedings of the 31st {ACM} International
  Conference on Multimedia, {MM} 2023}}. \bibinfo{publisher}{{ACM}},
  \bibinfo{pages}{1108--1116}.
\newblock


\bibitem[Hua et~al\mbox{.}(2020)]%
        {hua2020vqa}
\bibfield{author}{\bibinfo{person}{Lei Hua}, \bibinfo{person}{Mei Yu},
  \bibinfo{person}{Gangyi Jiang}, \bibinfo{person}{Zhouyan He}, {and}
  \bibinfo{person}{Yaoya Lin}.} \bibinfo{year}{2020}\natexlab{}.
\newblock \showarticletitle{VQA-CPC: A Novel Visual Quality Assessment Metric
  of Color Point Clouds}. In \bibinfo{booktitle}{\emph{Optoelectronic Imaging
  and Multimedia Technology VII}}, Vol.~\bibinfo{volume}{11550}. SPIE,
  \bibinfo{pages}{244--252}.
\newblock


\bibitem[Huang et~al\mbox{.}(2022)]%
        {huang2022ponder}
\bibfield{author}{\bibinfo{person}{Di Huang}, \bibinfo{person}{Sida Peng},
  \bibinfo{person}{Tong He}, \bibinfo{person}{Xiaowei Zhou}, {and}
  \bibinfo{person}{Wanli Ouyang}.} \bibinfo{year}{2022}\natexlab{}.
\newblock \showarticletitle{Ponder: Point Cloud Pre-training via Neural
  Rendering}.
\newblock \bibinfo{journal}{\emph{arXiv preprint arXiv:2301.00157}}
  (\bibinfo{year}{2022}).
\newblock


\bibitem[Inc.(2023)]%
        {visionpro}
\bibfield{author}{\bibinfo{person}{Apple Inc.}}
  \bibinfo{year}{2023}\natexlab{}.
\newblock \bibinfo{title}{Introducing Apple Vision Pro}.
\newblock
\newblock
\urldef\tempurl%
\url{https://www.apple.com/apple-vision-pro/}
\showURL{%
Retrieved June 14, 2023 from \tempurl}


\bibitem[ITU(2023)]%
        {ITU}
\bibfield{author}{\bibinfo{person}{ITU}.} \bibinfo{year}{2023}\natexlab{}.
\newblock \bibinfo{title}{ITU: Committed to Connecting the World.}
\newblock
\newblock
\urldef\tempurl%
\url{https://www.itu.int/}
\showURL{%
Retrieved June 27, 2023 from \tempurl}


\bibitem[Jacob et~al\mbox{.}(2018)]%
        {DBLP:conf/cvpr/JacobKCZTHAK18}
\bibfield{author}{\bibinfo{person}{Benoit Jacob}, \bibinfo{person}{Skirmantas
  Kligys}, \bibinfo{person}{Bo Chen}, \bibinfo{person}{Menglong Zhu},
  \bibinfo{person}{Matthew Tang}, \bibinfo{person}{Andrew~G. Howard},
  \bibinfo{person}{Hartwig Adam}, {and} \bibinfo{person}{Dmitry Kalenichenko}.}
  \bibinfo{year}{2018}\natexlab{}.
\newblock \showarticletitle{Quantization and Training of Neural Networks for
  Efficient Integer-Arithmetic-Only Inference}. In
  \bibinfo{booktitle}{\emph{2018 {IEEE} Conference on Computer Vision and
  Pattern Recognition, {CVPR} 2018}}. \bibinfo{publisher}{Computer Vision
  Foundation / {IEEE} Computer Society}, \bibinfo{pages}{2704--2713}.
\newblock


\bibitem[Jaderberg et~al\mbox{.}(2014)]%
        {DBLP:conf/bmvc/JaderbergVZ14}
\bibfield{author}{\bibinfo{person}{Max Jaderberg}, \bibinfo{person}{Andrea
  Vedaldi}, {and} \bibinfo{person}{Andrew Zisserman}.}
  \bibinfo{year}{2014}\natexlab{}.
\newblock \showarticletitle{Speeding up Convolutional Neural Networks with Low
  Rank Expansions}. In \bibinfo{booktitle}{\emph{British Machine Vision
  Conference, {BMVC} 2014, Nottingham, UK, September 1-5, 2014}}.
  \bibinfo{publisher}{{BMVA} Press}.
\newblock


\bibitem[Jansen et~al\mbox{.}(2020)]%
        {jansen2020pipeline}
\bibfield{author}{\bibinfo{person}{Jack Jansen}, \bibinfo{person}{Shishir
  Subramanyam}, \bibinfo{person}{Romain Bouqueau}, \bibinfo{person}{Gianluca
  Cernigliaro}, \bibinfo{person}{Marc~Martos Cabr{\'e}},
  \bibinfo{person}{Fernando P{\'e}rez}, {and} \bibinfo{person}{Pablo Cesar}.}
  \bibinfo{year}{2020}\natexlab{}.
\newblock \showarticletitle{A Pipeline for Multiparty Volumetric Video
  Conferencing: Transmission of Point Clouds over Low Latency {DASH}}. In
  \bibinfo{booktitle}{\emph{Proceedings of the 11th ACM Multimedia Systems
  Conference}}. \bibinfo{pages}{341--344}.
\newblock


\bibitem[Jarrold et~al\mbox{.}(2013)]%
        {jarrold2013social}
\bibfield{author}{\bibinfo{person}{William Jarrold}, \bibinfo{person}{Peter
  Mundy}, \bibinfo{person}{Mary Gwaltney}, \bibinfo{person}{Jeremy Bailenson},
  \bibinfo{person}{Naomi Hatt}, \bibinfo{person}{Nancy McIntyre},
  \bibinfo{person}{Kwanguk Kim}, \bibinfo{person}{Marjorie Solomon},
  \bibinfo{person}{Stephanie Novotny}, {and} \bibinfo{person}{Lindsay Swain}.}
  \bibinfo{year}{2013}\natexlab{}.
\newblock \showarticletitle{Social Attention in a Virtual Public Speaking Task
  in Higher Functioning Children with Autism}.
\newblock \bibinfo{journal}{\emph{Autism Research}} \bibinfo{volume}{6},
  \bibinfo{number}{5} (\bibinfo{year}{2013}), \bibinfo{pages}{393--410}.
\newblock


\bibitem[Jin et~al\mbox{.}(2024a)]%
        {tcsvt}
\bibfield{author}{\bibinfo{person}{Yili Jin}, \bibinfo{person}{Xize Duan},
  \bibinfo{person}{Kaiyuan Hu}, \bibinfo{person}{Fangxin Wang}, {and}
  \bibinfo{person}{Xue Liu}.} \bibinfo{year}{2024}\natexlab{a}.
\newblock \showarticletitle{3D Video Conferencing via On-hand Devices}.
\newblock \bibinfo{journal}{\emph{IEEE Transactions on Circuits and Systems for
  Video Technology}} (\bibinfo{year}{2024}).
\newblock


\bibitem[Jin et~al\mbox{.}(2024b)]%
        {HeadsetOff}
\bibfield{author}{\bibinfo{person}{Yili Jin}, \bibinfo{person}{Xize Duan},
  \bibinfo{person}{Fangxin Wang}, {and} \bibinfo{person}{Xue Liu}.}
  \bibinfo{year}{2024}\natexlab{b}.
\newblock \showarticletitle{HeadsetOff: Enabling Photorealistic Video
  Conferencing on Economical {VR} Headsets}. In
  \bibinfo{booktitle}{\emph{Proceedings of the 32st {ACM} International
  Conference on Multimedia, {MM} 2024}}. \bibinfo{publisher}{{ACM}}.
\newblock


\bibitem[Jin et~al\mbox{.}(2024c)]%
        {10634203}
\bibfield{author}{\bibinfo{person}{Yili Jin}, \bibinfo{person}{Junhua Liu},
  \bibinfo{person}{Kaiyuan Hu}, {and} \bibinfo{person}{Fangxin Wang}.}
  \bibinfo{year}{2024}\natexlab{c}.
\newblock \showarticletitle{A Networking Perspective of Volumetric Video
  Service: Architecture, Opportunities and Case Study}.
\newblock \bibinfo{journal}{\emph{IEEE Network}} (\bibinfo{year}{2024}).
\newblock


\bibitem[Jin et~al\mbox{.}(2022)]%
        {DBLP:conf/mm/JinL0C22}
\bibfield{author}{\bibinfo{person}{Yili Jin}, \bibinfo{person}{Junhua Liu},
  \bibinfo{person}{Fangxin Wang}, {and} \bibinfo{person}{Shuguang Cui}.}
  \bibinfo{year}{2022}\natexlab{}.
\newblock \showarticletitle{Where Are You Looking?: {A} Large-Scale Dataset of
  Head and Gaze Behavior for 360-Degree Videos and a Pilot Study}. In
  \bibinfo{booktitle}{\emph{{MM} '22: The 30th {ACM} International Conference
  on Multimedia}}. \bibinfo{publisher}{{ACM}}, \bibinfo{pages}{1025--1034}.
\newblock


\bibitem[Jin et~al\mbox{.}(2023)]%
        {DBLP:journals/iotj/JinLWC23}
\bibfield{author}{\bibinfo{person}{Yili Jin}, \bibinfo{person}{Junhua Liu},
  \bibinfo{person}{Fangxin Wang}, {and} \bibinfo{person}{Shuguang Cui}.}
  \bibinfo{year}{2023}\natexlab{}.
\newblock \showarticletitle{Ebublio: Edge-Assisted Multiuser 360{\textdegree}
  Video Streaming}.
\newblock \bibinfo{journal}{\emph{{IEEE} Internet Things J.}}
  \bibinfo{volume}{10}, \bibinfo{number}{17} (\bibinfo{year}{2023}),
  \bibinfo{pages}{15408--15419}.
\newblock


\bibitem[Jin et~al\mbox{.}(2024d)]%
        {10660498}
\bibfield{author}{\bibinfo{person}{Yili Jin}, \bibinfo{person}{Wenyi Zhang},
  \bibinfo{person}{Zihan Xu}, \bibinfo{person}{Fangxin Wang}, {and}
  \bibinfo{person}{Xue Liu}.} \bibinfo{year}{2024}\natexlab{d}.
\newblock \showarticletitle{Privacy-Preserving Gaze-Assisted Immersive Video
  Streaming}.
\newblock \bibinfo{journal}{\emph{IEEE Transactions on Mobile Computing}}
  (\bibinfo{year}{2024}).
\newblock


\bibitem[Kannisto et~al\mbox{.}(2004)]%
        {DBLP:conf/ict/KannistoVHH04}
\bibfield{author}{\bibinfo{person}{Juha Kannisto}, \bibinfo{person}{Timo
  Vanhatupa}, \bibinfo{person}{Marko H{\"{a}}nnik{\"{a}}inen}, {and}
  \bibinfo{person}{Timo H{\"{a}}m{\"{a}}l{\"{a}}inen}.}
  \bibinfo{year}{2004}\natexlab{}.
\newblock \showarticletitle{Precision Time Protocol Prototype on Wireless
  {LAN}}. In \bibinfo{booktitle}{\emph{Telecommunications and Networking -
  {ICT} 2004, 11th International Conference on Telecommunications,
  Proceedings}} \emph{(\bibinfo{series}{Lecture Notes in Computer Science},
  Vol.~\bibinfo{volume}{3124})}. \bibinfo{publisher}{Springer},
  \bibinfo{pages}{1236--1245}.
\newblock


\bibitem[Karni and Gotsman(2000)]%
        {DBLP:conf/siggraph/KarniG00}
\bibfield{author}{\bibinfo{person}{Zachi Karni} {and} \bibinfo{person}{Craig
  Gotsman}.} \bibinfo{year}{2000}\natexlab{}.
\newblock \showarticletitle{Spectral Compression of Mesh Geometry}. In
  \bibinfo{booktitle}{\emph{Proceedings of the 27th Annual Conference on
  Computer Graphics and Interactive Techniques, {SIGGRAPH} 2000}}.
  \bibinfo{publisher}{{ACM}}, \bibinfo{pages}{279--286}.
\newblock


\bibitem[Kaufman and Bakalash(1988)]%
        {DBLP:journals/cga/KaufmanB88}
\bibfield{author}{\bibinfo{person}{Arie~E. Kaufman} {and}
  \bibinfo{person}{Reuven Bakalash}.} \bibinfo{year}{1988}\natexlab{}.
\newblock \showarticletitle{Memory and Processing Architecture for 3D
  Voxel-based Imagery}.
\newblock \bibinfo{journal}{\emph{{IEEE} Computer Graphics and Applications}}
  \bibinfo{volume}{8}, \bibinfo{number}{6} (\bibinfo{year}{1988}),
  \bibinfo{pages}{10--23}.
\newblock


\bibitem[Knapp(1938)]%
        {knapp1938introduction}
\bibfield{author}{\bibinfo{person}{Arnold Knapp}.}
  \bibinfo{year}{1938}\natexlab{}.
\newblock \showarticletitle{An Introduction to Clinical Perimetry.}
\newblock \bibinfo{journal}{\emph{Archives of Ophthalmology}}
  \bibinfo{volume}{20}, \bibinfo{number}{6} (\bibinfo{year}{1938}),
  \bibinfo{pages}{1116--1117}.
\newblock


\bibitem[Krivoku\'ca et~al\mbox{.}(2018)]%
        {8iVSLF}
\bibfield{author}{\bibinfo{person}{Maja Krivoku\'ca},
  \bibinfo{person}{Philip~A. Chou}, {and} \bibinfo{person}{Patrick Savill}.}
  \bibinfo{year}{2018}\natexlab{}.
\newblock \showarticletitle{8i Voxelized Surface Light Field Dataset}.
\newblock \bibinfo{journal}{\emph{ISO/IEC JTC1/SC29 WG11 (MPEG) input document
  m42914}} (\bibinfo{year}{2018}).
\newblock


\bibitem[Laina et~al\mbox{.}(2016)]%
        {laina2016deeper}
\bibfield{author}{\bibinfo{person}{Iro Laina}, \bibinfo{person}{Christian
  Rupprecht}, \bibinfo{person}{Vasileios Belagiannis},
  \bibinfo{person}{Federico Tombari}, {and} \bibinfo{person}{Nassir Navab}.}
  \bibinfo{year}{2016}\natexlab{}.
\newblock \showarticletitle{Deeper Depth Prediction with Fully Convolutional
  Residual Networks}. In \bibinfo{booktitle}{\emph{2016 Fourth international
  conference on 3D vision (3DV)}}. IEEE, \bibinfo{pages}{239--248}.
\newblock


\bibitem[Laver et~al\mbox{.}(2017)]%
        {laver2017virtual}
\bibfield{author}{\bibinfo{person}{Kate~E Laver}, \bibinfo{person}{Belinda
  Lange}, \bibinfo{person}{Stacey George}, \bibinfo{person}{Judith~E Deutsch},
  \bibinfo{person}{Gustavo Saposnik}, {and} \bibinfo{person}{Maria Crotty}.}
  \bibinfo{year}{2017}\natexlab{}.
\newblock \showarticletitle{Virtual Reality for Stroke Rehabilitation}.
\newblock \bibinfo{journal}{\emph{Cochrane database of systematic reviews}}
  \bibinfo{number}{11} (\bibinfo{year}{2017}).
\newblock


\bibitem[Lavou{\'{e}}(2011)]%
        {DBLP:journals/cgf/Lavoue11}
\bibfield{author}{\bibinfo{person}{Guillaume Lavou{\'{e}}}.}
  \bibinfo{year}{2011}\natexlab{}.
\newblock \showarticletitle{A Multiscale Metric for 3D Mesh Visual Quality
  Assessment}.
\newblock \bibinfo{journal}{\emph{Comput. Graph. Forum}} \bibinfo{volume}{30},
  \bibinfo{number}{5} (\bibinfo{year}{2011}), \bibinfo{pages}{1427--1437}.
\newblock


\bibitem[Lee et~al\mbox{.}(2020)]%
        {DBLP:conf/mobicom/LeeYLCK20}
\bibfield{author}{\bibinfo{person}{Kyungjin Lee}, \bibinfo{person}{Juheon Yi},
  \bibinfo{person}{Youngki Lee}, \bibinfo{person}{Sunghyun Choi}, {and}
  \bibinfo{person}{Young~Min Kim}.} \bibinfo{year}{2020}\natexlab{}.
\newblock \showarticletitle{{GROOT:} A Real-time Streaming System of
  High-fidelity Volumetric Videos}. In \bibinfo{booktitle}{\emph{MobiCom '20:
  The 26th Annual International Conference on Mobile Computing and
  Networking}}. \bibinfo{publisher}{{ACM}}, \bibinfo{pages}{57:1--57:14}.
\newblock


\bibitem[Li et~al\mbox{.}(2023b)]%
        {DBLP:journals/tmm/LiZLHH23}
\bibfield{author}{\bibinfo{person}{Jie Li}, \bibinfo{person}{Cong Zhang},
  \bibinfo{person}{Zhi Liu}, \bibinfo{person}{Richang Hong}, {and}
  \bibinfo{person}{Han Hu}.} \bibinfo{year}{2023}\natexlab{b}.
\newblock \showarticletitle{Optimal Volumetric Video Streaming With Hybrid
  Saliency Based Tiling}.
\newblock \bibinfo{journal}{\emph{{IEEE} Trans. Multim.}}  \bibinfo{volume}{25}
  (\bibinfo{year}{2023}), \bibinfo{pages}{2939--2953}.
\newblock


\bibitem[Li et~al\mbox{.}(2023a)]%
        {li2023compressing}
\bibfield{author}{\bibinfo{person}{Lingzhi Li}, \bibinfo{person}{Zhen Shen},
  \bibinfo{person}{Zhongshu Wang}, \bibinfo{person}{Li Shen}, {and}
  \bibinfo{person}{Liefeng Bo}.} \bibinfo{year}{2023}\natexlab{a}.
\newblock \showarticletitle{Compressing Volumetric Radiance Fields to 1 mb}. In
  \bibinfo{booktitle}{\emph{Proceedings of the IEEE/CVF Conference on Computer
  Vision and Pattern Recognition}}. \bibinfo{pages}{4222--4231}.
\newblock


\bibitem[Li et~al\mbox{.}(2019)]%
        {PU_GAN}
\bibfield{author}{\bibinfo{person}{Ruihui Li}, \bibinfo{person}{Xianzhi Li},
  \bibinfo{person}{Chi{-}Wing Fu}, \bibinfo{person}{Daniel Cohen{-}Or}, {and}
  \bibinfo{person}{Pheng{-}Ann Heng}.} \bibinfo{year}{2019}\natexlab{}.
\newblock \showarticletitle{{PU-GAN:} {A} Point Cloud Upsampling Adversarial
  Network}. In \bibinfo{booktitle}{\emph{2019 {IEEE/CVF} International
  Conference on Computer Vision, {ICCV} 2019}}. \bibinfo{publisher}{{IEEE}},
  \bibinfo{pages}{7202--7211}.
\newblock


\bibitem[Li et~al\mbox{.}(2020)]%
        {DBLP:conf/nips/LiLMKW0K20}
\bibfield{author}{\bibinfo{person}{Xueting Li}, \bibinfo{person}{Sifei Liu},
  \bibinfo{person}{Shalini~De Mello}, \bibinfo{person}{Kihwan Kim},
  \bibinfo{person}{Xiaolong Wang}, \bibinfo{person}{Ming{-}Hsuan Yang}, {and}
  \bibinfo{person}{Jan Kautz}.} \bibinfo{year}{2020}\natexlab{}.
\newblock \showarticletitle{Online Adaptation for Consistent Mesh
  Reconstruction in the Wild}. In \bibinfo{booktitle}{\emph{Advances in Neural
  Information Processing Systems 33: Annual Conference on Neural Information
  Processing Systems 2020, NeurIPS 2020}}.
\newblock


\bibitem[Li and Meng(2022)]%
        {DBLP:conf/atait/LiM22}
\bibfield{author}{\bibinfo{person}{Zhuo Li} {and} \bibinfo{person}{Lin Meng}.}
  \bibinfo{year}{2022}\natexlab{}.
\newblock \showarticletitle{A Survey of Model Pruning for Deep Neural Network}.
  In \bibinfo{booktitle}{\emph{Proceedings of the 4th International Symposium
  on Advanced Technologies and Applications in the Internet of Things {(ATAIT}
  2022)}} \emph{(\bibinfo{series}{{CEUR} Workshop Proceedings},
  Vol.~\bibinfo{volume}{3198})}. \bibinfo{publisher}{CEUR-WS.org},
  \bibinfo{pages}{25--34}.
\newblock


\bibitem[Liang et~al\mbox{.}(2021)]%
        {DBLP:journals/mta/LiangXGBZC21}
\bibfield{author}{\bibinfo{person}{Wei Liang}, \bibinfo{person}{Pengfei Xu},
  \bibinfo{person}{Ling Guo}, \bibinfo{person}{Heng Bai}, \bibinfo{person}{Yang
  Zhou}, {and} \bibinfo{person}{Feng Chen}.} \bibinfo{year}{2021}\natexlab{}.
\newblock \showarticletitle{A Survey of 3D Object Detection}.
\newblock \bibinfo{journal}{\emph{Multim. Tools Appl.}} \bibinfo{volume}{80},
  \bibinfo{number}{19} (\bibinfo{year}{2021}), \bibinfo{pages}{29617--29641}.
\newblock


\bibitem[Liang and Liang(2022)]%
        {DBLP:conf/mir/LiangL22}
\bibfield{author}{\bibinfo{person}{Zujie Liang} {and} \bibinfo{person}{Fan
  Liang}.} \bibinfo{year}{2022}\natexlab{}.
\newblock \showarticletitle{TransPCC: Towards Deep Point Cloud Compression via
  Transformers}. In \bibinfo{booktitle}{\emph{{ICMR} '22: International
  Conference on Multimedia Retrieval}}. \bibinfo{publisher}{{ACM}},
  \bibinfo{pages}{1--5}.
\newblock


\bibitem[Liang et~al\mbox{.}(2024)]%
        {DBLP:journals/tvcg/LiangLDW24}
\bibfield{author}{\bibinfo{person}{Zhicheng Liang}, \bibinfo{person}{Junhua
  Liu}, \bibinfo{person}{Mallesham Dasari}, {and} \bibinfo{person}{Fangxin
  Wang}.} \bibinfo{year}{2024}\natexlab{}.
\newblock \showarticletitle{Fumos: Neural Compression and Progressive
  Refinement for Continuous Point Cloud Video Streaming}.
\newblock \bibinfo{journal}{\emph{{IEEE} Trans. Vis. Comput. Graph.}}
  \bibinfo{volume}{30}, \bibinfo{number}{5} (\bibinfo{year}{2024}),
  \bibinfo{pages}{2849--2859}.
\newblock


\bibitem[Liao et~al\mbox{.}(2023)]%
        {liao2023deep}
\bibfield{author}{\bibinfo{person}{Kang Liao}, \bibinfo{person}{Lang Nie},
  \bibinfo{person}{Shujuan Huang}, \bibinfo{person}{Chunyu Lin},
  \bibinfo{person}{Jing Zhang}, \bibinfo{person}{Yao Zhao},
  \bibinfo{person}{Moncef Gabbouj}, {and} \bibinfo{person}{Dacheng Tao}.}
  \bibinfo{year}{2023}\natexlab{}.
\newblock \showarticletitle{Deep learning for camera calibration and beyond: A
  survey}.
\newblock \bibinfo{journal}{\emph{arXiv preprint arXiv:2303.10559}}
  (\bibinfo{year}{2023}).
\newblock


\bibitem[Ling et~al\mbox{.}(2019)]%
        {DBLP:journals/esticas/LingGGC19}
\bibfield{author}{\bibinfo{person}{Suiyi Ling}, \bibinfo{person}{Jes{\'{u}}s
  Guti{\'{e}}rrez}, \bibinfo{person}{Ke Gu}, {and} \bibinfo{person}{Patrick~Le
  Callet}.} \bibinfo{year}{2019}\natexlab{}.
\newblock \showarticletitle{Prediction of the Influence of Navigation Scan-Path
  on Perceived Quality of Free-Viewpoint Videos}.
\newblock \bibinfo{journal}{\emph{{IEEE} J. Emerg. Sel. Topics Circuits Syst.}}
  \bibinfo{volume}{9}, \bibinfo{number}{1} (\bibinfo{year}{2019}),
  \bibinfo{pages}{204--216}.
\newblock


\bibitem[Liu et~al\mbox{.}(2019)]%
        {DBLP:conf/etra/LiuXDBHJ19}
\bibfield{author}{\bibinfo{person}{Ao Liu}, \bibinfo{person}{Lirong Xia},
  \bibinfo{person}{Andrew~T. Duchowski}, \bibinfo{person}{Reynold Bailey},
  \bibinfo{person}{Kenneth Holmqvist}, {and} \bibinfo{person}{Eakta Jain}.}
  \bibinfo{year}{2019}\natexlab{}.
\newblock \showarticletitle{Differential Privacy for Eye-tracking Data}. In
  \bibinfo{booktitle}{\emph{Proceedings of the 11th {ACM} Symposium on Eye
  Tracking Research {\&} Applications, {ETRA} 2019}}.
  \bibinfo{publisher}{{ACM}}, \bibinfo{pages}{28:1--28:10}.
\newblock


\bibitem[Liu et~al\mbox{.}(2023)]%
        {DBLP:conf/vr/LiuZWJZXC23}
\bibfield{author}{\bibinfo{person}{Junhua Liu}, \bibinfo{person}{Boxiang Zhu},
  \bibinfo{person}{Fangxin Wang}, \bibinfo{person}{Yili Jin},
  \bibinfo{person}{Wenyi Zhang}, \bibinfo{person}{Zihan Xu}, {and}
  \bibinfo{person}{Shuguang Cui}.} \bibinfo{year}{2023}\natexlab{}.
\newblock \showarticletitle{CaV3: Cache-assisted Viewport Adaptive Volumetric
  Video Streaming}. In \bibinfo{booktitle}{\emph{{IEEE} Conference Virtual
  Reality and 3D User Interfaces, {VR} 2023}}. \bibinfo{publisher}{{IEEE}},
  \bibinfo{pages}{173--183}.
\newblock


\bibitem[Liu et~al\mbox{.}(2022)]%
        {DBLP:conf/mobicom/LiuHQNZ22}
\bibfield{author}{\bibinfo{person}{Yu Liu}, \bibinfo{person}{Bo Han},
  \bibinfo{person}{Feng Qian}, \bibinfo{person}{Arvind Narayanan}, {and}
  \bibinfo{person}{Zhi{-}Li Zhang}.} \bibinfo{year}{2022}\natexlab{}.
\newblock \showarticletitle{Vues: Practical Mobile Volumetric Video Streaming
  Through Multiview Transcoding}. In \bibinfo{booktitle}{\emph{{ACM} MobiCom
  '22: The 28th Annual International Conference on Mobile Computing and
  Networking}}. \bibinfo{publisher}{{ACM}}, \bibinfo{pages}{514--527}.
\newblock


\bibitem[Liu et~al\mbox{.}(2021)]%
        {DBLP:journals/network/LiuLCWILJ21}
\bibfield{author}{\bibinfo{person}{Zhi Liu}, \bibinfo{person}{Qiyue Li},
  \bibinfo{person}{Xianfu Chen}, \bibinfo{person}{Celimuge Wu},
  \bibinfo{person}{Susumu Ishihara}, \bibinfo{person}{Jie Li}, {and}
  \bibinfo{person}{Yusheng Ji}.} \bibinfo{year}{2021}\natexlab{}.
\newblock \showarticletitle{Point Cloud Video Streaming: Challenges and
  Solutions}.
\newblock \bibinfo{journal}{\emph{{IEEE} Netw.}} \bibinfo{volume}{35},
  \bibinfo{number}{5} (\bibinfo{year}{2021}), \bibinfo{pages}{202--209}.
\newblock


\bibitem[LLC(2023)]%
        {psvr}
\bibfield{author}{\bibinfo{person}{Sony Interactive~Entertainment LLC}.}
  \bibinfo{year}{2023}\natexlab{}.
\newblock \bibinfo{title}{PlayStation VR: Immerse Yourself in Incredible
  Virtual Reality Games and Experiences.}
\newblock
\newblock
\urldef\tempurl%
\url{https://www.playstation.com/ps-vr}
\showURL{%
Retrieved June 14, 2023 from \tempurl}


\bibitem[Loop et~al\mbox{.}(2021)]%
        {MVUB}
\bibfield{author}{\bibinfo{person}{Charles Loop}, \bibinfo{person}{Qin Cai},
  \bibinfo{person}{Sergio~Orts Escolano}, {and} \bibinfo{person}{Philip~A.
  Chou}.} \bibinfo{year}{2021}\natexlab{}.
\newblock \showarticletitle{JPEG Pleno Database: Microsoft Voxelized Upper
  Bodies - A Voxelized Point Cloud Dataset}.
\newblock \bibinfo{journal}{\emph{ISO/IEC JTC1/SC29 Joint WG11/WG1 (MPEG/JPEG)
  input document m38673/M72012}} (\bibinfo{year}{2021}).
\newblock


\bibitem[Loucks et~al\mbox{.}(2019)]%
        {loucks2019you}
\bibfield{author}{\bibinfo{person}{Laura Loucks}, \bibinfo{person}{Carly
  Yasinski}, \bibinfo{person}{Seth~D Norrholm}, \bibinfo{person}{Jessica
  Maples-Keller}, \bibinfo{person}{Loren Post}, \bibinfo{person}{Liza
  Zwiebach}, \bibinfo{person}{Devika Fiorillo}, \bibinfo{person}{Megan
  Goodlin}, \bibinfo{person}{Tanja Jovanovic}, \bibinfo{person}{Albert~A
  Rizzo}, {et~al\mbox{.}}} \bibinfo{year}{2019}\natexlab{}.
\newblock \showarticletitle{You can do that?!: Feasibility of Virtual Reality
  Exposure Therapy in the Treatment of {PTSD} due to Military Sexual Trauma}.
\newblock \bibinfo{journal}{\emph{Journal of anxiety disorders}}
  \bibinfo{volume}{61} (\bibinfo{year}{2019}), \bibinfo{pages}{55--63}.
\newblock


\bibitem[Love et~al\mbox{.}(2009)]%
        {Love:09}
\bibfield{author}{\bibinfo{person}{Gordon~D. Love}, \bibinfo{person}{David~M.
  Hoffman}, \bibinfo{person}{Philip~J.W. Hands}, \bibinfo{person}{James Gao},
  \bibinfo{person}{Andrew~K. Kirby}, {and} \bibinfo{person}{Martin~S. Banks}.}
  \bibinfo{year}{2009}\natexlab{}.
\newblock \showarticletitle{High-speed Switchable Lens Enables the Development
  of a Volumetric Stereoscopic Display}.
\newblock \bibinfo{journal}{\emph{Opt. Express}} \bibinfo{volume}{17},
  \bibinfo{number}{18} (\bibinfo{year}{2009}), \bibinfo{pages}{15716--15725}.
\newblock


\bibitem[Lucas and Powell(1977)]%
        {lucas1977star}
\bibfield{author}{\bibinfo{person}{George Lucas} {and} \bibinfo{person}{Moray
  Powell}.} \bibinfo{year}{1977}\natexlab{}.
\newblock \bibinfo{booktitle}{\emph{Star wars}}.
\newblock \bibinfo{publisher}{Royal Blind Society of New South Wales.}
\newblock


\bibitem[Makransky and Mayer(2022)]%
        {makransky2022benefits}
\bibfield{author}{\bibinfo{person}{Guido Makransky} {and}
  \bibinfo{person}{Richard~E Mayer}.} \bibinfo{year}{2022}\natexlab{}.
\newblock \showarticletitle{Benefits of Taking a Virtual Field Trip in
  Immersive Virtual Reality: Evidence for the Immersion Principle in Multimedia
  Learning}.
\newblock \bibinfo{journal}{\emph{Educational Psychology Review}}
  \bibinfo{volume}{34}, \bibinfo{number}{3} (\bibinfo{year}{2022}),
  \bibinfo{pages}{1771--1798}.
\newblock


\bibitem[Mallick et~al\mbox{.}(2014)]%
        {6756961}
\bibfield{author}{\bibinfo{person}{Tanwi Mallick},
  \bibinfo{person}{Partha~Pratim Das}, {and} \bibinfo{person}{Arun~Kumar
  Majumdar}.} \bibinfo{year}{2014}\natexlab{}.
\newblock \showarticletitle{Characterizations of Noise in Kinect Depth Images:
  A Review}.
\newblock \bibinfo{journal}{\emph{IEEE Sensors Journal}} \bibinfo{volume}{14},
  \bibinfo{number}{6} (\bibinfo{year}{2014}), \bibinfo{pages}{1731--1740}.
\newblock


\bibitem[Mao et~al\mbox{.}(2017)]%
        {DBLP:journals/comsur/MaoYZHL17}
\bibfield{author}{\bibinfo{person}{Yuyi Mao}, \bibinfo{person}{Changsheng You},
  \bibinfo{person}{Jun Zhang}, \bibinfo{person}{Kaibin Huang}, {and}
  \bibinfo{person}{Khaled~Ben Letaief}.} \bibinfo{year}{2017}\natexlab{}.
\newblock \showarticletitle{A Survey on Mobile Edge Computing: The
  Communication Perspective}.
\newblock \bibinfo{journal}{\emph{{IEEE} Commun. Surv. Tutorials}}
  \bibinfo{volume}{19}, \bibinfo{number}{4} (\bibinfo{year}{2017}),
  \bibinfo{pages}{2322--2358}.
\newblock


\bibitem[Masi{\'{a}} et~al\mbox{.}(2013)]%
        {DBLP:journals/cg/MasiaWDG13}
\bibfield{author}{\bibinfo{person}{Bel{\'{e}}n Masi{\'{a}}},
  \bibinfo{person}{Gordon Wetzstein}, \bibinfo{person}{Piotr Didyk}, {and}
  \bibinfo{person}{Diego Gutierrez}.} \bibinfo{year}{2013}\natexlab{}.
\newblock \showarticletitle{A survey on computational displays: Pushing the
  boundaries of optics, computation, and perception}.
\newblock \bibinfo{journal}{\emph{Comput. Graph.}} \bibinfo{volume}{37},
  \bibinfo{number}{8} (\bibinfo{year}{2013}), \bibinfo{pages}{1012--1038}.
\newblock


\bibitem[Mercier et~al\mbox{.}(2017)]%
        {DBLP:journals/tog/MercierSMZHNL17}
\bibfield{author}{\bibinfo{person}{Olivier Mercier}, \bibinfo{person}{Yusufu
  Sulai}, \bibinfo{person}{Kevin~J. MacKenzie}, \bibinfo{person}{Marina
  Zannoli}, \bibinfo{person}{James Hillis}, \bibinfo{person}{Derek
  Nowrouzezahrai}, {and} \bibinfo{person}{Douglas Lanman}.}
  \bibinfo{year}{2017}\natexlab{}.
\newblock \showarticletitle{Fast Gaze-contingent Optimal Decompositions for
  Multifocal Displays}.
\newblock \bibinfo{journal}{\emph{{ACM} Trans. Graph.}} \bibinfo{volume}{36},
  \bibinfo{number}{6} (\bibinfo{year}{2017}), \bibinfo{pages}{237:1--237:15}.
\newblock


\bibitem[Merkle et~al\mbox{.}(2010)]%
        {DBLP:journals/tce/MerkleMW10}
\bibfield{author}{\bibinfo{person}{Philipp Merkle}, \bibinfo{person}{Karsten
  M{\"{u}}ller}, {and} \bibinfo{person}{Thomas Wiegand}.}
  \bibinfo{year}{2010}\natexlab{}.
\newblock \showarticletitle{3D Video: Acquisition, Coding, and Display}.
\newblock \bibinfo{journal}{\emph{{IEEE} Trans. Consumer Electron.}}
  \bibinfo{volume}{56}, \bibinfo{number}{2} (\bibinfo{year}{2010}),
  \bibinfo{pages}{946--950}.
\newblock


\bibitem[Mertan et~al\mbox{.}(2022)]%
        {mertan2022single}
\bibfield{author}{\bibinfo{person}{Alican Mertan}, \bibinfo{person}{Damien~Jade
  Duff}, {and} \bibinfo{person}{Gozde Unal}.} \bibinfo{year}{2022}\natexlab{}.
\newblock \showarticletitle{Single Image Depth Estimation: An Overview}.
\newblock \bibinfo{journal}{\emph{Digital Signal Processing}}
  \bibinfo{volume}{123} (\bibinfo{year}{2022}), \bibinfo{pages}{103441}.
\newblock


\bibitem[Meynet et~al\mbox{.}(2019)]%
        {DBLP:conf/qomex/MeynetDL19}
\bibfield{author}{\bibinfo{person}{Gabriel Meynet}, \bibinfo{person}{Julie
  Digne}, {and} \bibinfo{person}{Guillaume Lavou{\'{e}}}.}
  \bibinfo{year}{2019}\natexlab{}.
\newblock \showarticletitle{{PC-MSDM:} {A} Quality Metric for 3D Point Clouds}.
  In \bibinfo{booktitle}{\emph{11th International Conference on Quality of
  Multimedia Experience QoMEX 2019}}. \bibinfo{publisher}{{IEEE}},
  \bibinfo{pages}{1--3}.
\newblock


\bibitem[Meynet et~al\mbox{.}(2020)]%
        {DBLP:conf/qomex/MeynetNDL20}
\bibfield{author}{\bibinfo{person}{Gabriel Meynet}, \bibinfo{person}{Yana
  Nehm{\'{e}}}, \bibinfo{person}{Julie Digne}, {and} \bibinfo{person}{Guillaume
  Lavou{\'{e}}}.} \bibinfo{year}{2020}\natexlab{}.
\newblock \showarticletitle{{PCQM:} {A} Full-Reference Quality Metric for
  Colored 3D Point Clouds}. In \bibinfo{booktitle}{\emph{Twelfth International
  Conference on Quality of Multimedia Experience, QoMEX 2020}}.
  \bibinfo{publisher}{{IEEE}}, \bibinfo{pages}{1--6}.
\newblock


\bibitem[Microsoft(2023)]%
        {Azure_Kinect}
\bibfield{author}{\bibinfo{person}{Microsoft}.}
  \bibinfo{year}{2023}\natexlab{}.
\newblock \bibinfo{title}{About Azure Kinect DK}.
\newblock
\newblock
\urldef\tempurl%
\url{https://learn.microsoft.com/en-us/azure/kinect-dk/about-azure-kinect-dk}
\showURL{%
Retrieved July 12, 2023 from \tempurl}


\bibitem[Mieloch et~al\mbox{.}(2022)]%
        {DBLP:journals/tcsv/MielochGMJJRS22}
\bibfield{author}{\bibinfo{person}{Dawid Mieloch}, \bibinfo{person}{Patrick
  Garus}, \bibinfo{person}{Marta Milovanovic}, \bibinfo{person}{Jo{\"{e}}l
  Jung}, \bibinfo{person}{Jun~Young Jeong},
  \bibinfo{person}{Smitha~Lingadahalli Ravi}, {and} \bibinfo{person}{Basel
  Salahieh}.} \bibinfo{year}{2022}\natexlab{}.
\newblock \showarticletitle{Overview and Efficiency of Decoder-Side Depth
  Estimation in {MPEG} Immersive Video}.
\newblock \bibinfo{journal}{\emph{{IEEE} Trans. Circuits Syst. Video Technol.}}
  \bibinfo{volume}{32}, \bibinfo{number}{9} (\bibinfo{year}{2022}),
  \bibinfo{pages}{6360--6374}.
\newblock


\bibitem[Mildenhall et~al\mbox{.}(2022)]%
        {DBLP:journals/cacm/MildenhallSTBRN22}
\bibfield{author}{\bibinfo{person}{Ben Mildenhall}, \bibinfo{person}{Pratul~P.
  Srinivasan}, \bibinfo{person}{Matthew Tancik}, \bibinfo{person}{Jonathan~T.
  Barron}, \bibinfo{person}{Ravi Ramamoorthi}, {and} \bibinfo{person}{Ren Ng}.}
  \bibinfo{year}{2022}\natexlab{}.
\newblock \showarticletitle{NeRF: Representing Scenes as Neural Radiance Fields
  for View Synthesis}.
\newblock \bibinfo{journal}{\emph{Commun. {ACM}}} \bibinfo{volume}{65},
  \bibinfo{number}{1} (\bibinfo{year}{2022}), \bibinfo{pages}{99--106}.
\newblock


\bibitem[Miller et~al\mbox{.}(2020)]%
        {miller2020personal}
\bibfield{author}{\bibinfo{person}{Mark~Roman Miller},
  \bibinfo{person}{Fernanda Herrera}, \bibinfo{person}{Hanseul Jun},
  \bibinfo{person}{James~A Landay}, {and} \bibinfo{person}{Jeremy~N
  Bailenson}.} \bibinfo{year}{2020}\natexlab{}.
\newblock \showarticletitle{Personal Identifiability of User Tracking Data
  during Observation of 360-degree VR Video}.
\newblock \bibinfo{journal}{\emph{Scientific Reports}} \bibinfo{volume}{10},
  \bibinfo{number}{1} (\bibinfo{year}{2020}), \bibinfo{pages}{17404}.
\newblock


\bibitem[Ming et~al\mbox{.}(2021)]%
        {DBLP:journals/ijon/MingMFY21}
\bibfield{author}{\bibinfo{person}{Yue Ming}, \bibinfo{person}{Xuyang Meng},
  \bibinfo{person}{Chunxiao Fan}, {and} \bibinfo{person}{Hui Yu}.}
  \bibinfo{year}{2021}\natexlab{}.
\newblock \showarticletitle{Deep Learning for Monocular Depth Estimation: {A}
  review}.
\newblock \bibinfo{journal}{\emph{Neurocomputing}}  \bibinfo{volume}{438}
  (\bibinfo{year}{2021}), \bibinfo{pages}{14--33}.
\newblock


\bibitem[MPEG(2023a)]%
        {MPEG-PCC}
\bibfield{author}{\bibinfo{person}{MPEG}.} \bibinfo{year}{2023}\natexlab{a}.
\newblock \bibinfo{title}{MPEG Point Cloud Compression}.
\newblock
\newblock
\urldef\tempurl%
\url{https://mpeg-pcc.org/}
\showURL{%
Retrieved June 27, 2023 from \tempurl}


\bibitem[MPEG(2023b)]%
        {MPEG}
\bibfield{author}{\bibinfo{person}{MPEG}.} \bibinfo{year}{2023}\natexlab{b}.
\newblock \bibinfo{title}{MPEG: The Moving Picture Experts Group}.
\newblock
\newblock
\urldef\tempurl%
\url{https://mpeg.chiariglione.org/}
\showURL{%
Retrieved June 27, 2023 from \tempurl}


\bibitem[Murthy et~al\mbox{.}(2019)]%
        {DBLP:conf/bigdatasec/MurthyBRR19}
\bibfield{author}{\bibinfo{person}{Suntherasvaran Murthy},
  \bibinfo{person}{Asmidar~Abu Bakar}, \bibinfo{person}{Fiza~Abdul Rahim},
  {and} \bibinfo{person}{Ramona Ramli}.} \bibinfo{year}{2019}\natexlab{}.
\newblock \showarticletitle{A Comparative Study of Data Anonymization
  Techniques}. In \bibinfo{booktitle}{\emph{5th {IEEE} International Conference
  on Big Data Security on Cloud, {IEEE} International Conference on High
  Performance and Smart Computing, and {IEEE} International Conference on
  Intelligent Data and Security, BigDataSecurity/HPSC/IDS 2019}}.
  \bibinfo{publisher}{{IEEE}}, \bibinfo{pages}{306--309}.
\newblock


\bibitem[Nadenau et~al\mbox{.}(2003)]%
        {DBLP:journals/tip/NadenauRK03}
\bibfield{author}{\bibinfo{person}{Marcus~J. Nadenau}, \bibinfo{person}{Julien
  Reichel}, {and} \bibinfo{person}{Murat Kunt}.}
  \bibinfo{year}{2003}\natexlab{}.
\newblock \showarticletitle{Wavelet-based Color Image Compression: Exploiting
  the Contrast Sensitivity Function}.
\newblock \bibinfo{journal}{\emph{{IEEE} Trans. Image Process.}}
  \bibinfo{volume}{12}, \bibinfo{number}{1} (\bibinfo{year}{2003}),
  \bibinfo{pages}{58--70}.
\newblock


\bibitem[Newcombe et~al\mbox{.}(2011)]%
        {DBLP:conf/ismar/NewcombeIHMKDKSHF11}
\bibfield{author}{\bibinfo{person}{Richard~A. Newcombe},
  \bibinfo{person}{Shahram Izadi}, \bibinfo{person}{Otmar Hilliges},
  \bibinfo{person}{David Molyneaux}, \bibinfo{person}{David Kim},
  \bibinfo{person}{Andrew~J. Davison}, \bibinfo{person}{Pushmeet Kohli},
  \bibinfo{person}{Jamie Shotton}, \bibinfo{person}{Steve Hodges}, {and}
  \bibinfo{person}{Andrew~W. Fitzgibbon}.} \bibinfo{year}{2011}\natexlab{}.
\newblock \showarticletitle{KinectFusion: Real-time Dense Surface Mapping and
  Tracking}. In \bibinfo{booktitle}{\emph{10th {IEEE} International Symposium
  on Mixed and Augmented Reality, {ISMAR} 2011}}. \bibinfo{publisher}{{IEEE}
  Computer Society}, \bibinfo{pages}{127--136}.
\newblock


\bibitem[Orts-Escolano et~al\mbox{.}(2016)]%
        {orts2016holoportation}
\bibfield{author}{\bibinfo{person}{Sergio Orts-Escolano},
  \bibinfo{person}{Christoph Rhemann}, \bibinfo{person}{Sean Fanello},
  \bibinfo{person}{Wayne Chang}, \bibinfo{person}{Adarsh Kowdle},
  \bibinfo{person}{Yury Degtyarev}, \bibinfo{person}{David Kim},
  \bibinfo{person}{Philip~L Davidson}, \bibinfo{person}{Sameh Khamis},
  \bibinfo{person}{Mingsong Dou}, {et~al\mbox{.}}}
  \bibinfo{year}{2016}\natexlab{}.
\newblock \showarticletitle{Holoportation: Virtual 3d Teleportation in
  Real-time}. In \bibinfo{booktitle}{\emph{Proceedings of the 29th annual
  symposium on user interface software and technology}}.
  \bibinfo{pages}{741--754}.
\newblock


\bibitem[Owen(1998)]%
        {DBLP:conf/imr/Owen98}
\bibfield{author}{\bibinfo{person}{Steven~J. Owen}.}
  \bibinfo{year}{1998}\natexlab{}.
\newblock \showarticletitle{A Survey of Unstructured Mesh Generation
  Technology}. In \bibinfo{booktitle}{\emph{Proceedings of the 7th
  International Meshing Roundtable, {IMR} 1998}}. \bibinfo{pages}{239--267}.
\newblock


\bibitem[Pag{\'e}s et~al\mbox{.}(2021)]%
        {Volograms}
\bibfield{author}{\bibinfo{person}{Rafael Pag{\'e}s}, \bibinfo{person}{Emin
  Zerman}, \bibinfo{person}{Konstantinos Amplianitis}, \bibinfo{person}{Jan
  Ond{\v{r}}ej}, {and} \bibinfo{person}{Aljosa Smolic}.}
  \bibinfo{year}{2021}\natexlab{}.
\newblock \showarticletitle{Volograms \& {V-SENSE} {V}olumetric {V}ideo
  {D}ataset}.
\newblock \bibinfo{journal}{\emph{ISO/IEC JTC1/SC29/WG07 MPEG2021/m56767}}
  (\bibinfo{year}{2021}).
\newblock


\bibitem[Palmer(1999)]%
        {palmer1999vision}
\bibfield{author}{\bibinfo{person}{Stephen~E Palmer}.}
  \bibinfo{year}{1999}\natexlab{}.
\newblock \bibinfo{booktitle}{\emph{Vision Science: Photons to Phenomenology}}.
\newblock \bibinfo{publisher}{MIT press}.
\newblock


\bibitem[Pan et~al\mbox{.}(2005)]%
        {DBLP:journals/tmm/PanCB05}
\bibfield{author}{\bibinfo{person}{Yixin Pan}, \bibinfo{person}{Irene Cheng},
  {and} \bibinfo{person}{Anup Basu}.} \bibinfo{year}{2005}\natexlab{}.
\newblock \showarticletitle{Quality Metric for Approximating Subjective
  Evaluation of 3-D Objects}.
\newblock \bibinfo{journal}{\emph{{IEEE} Trans. Multim.}} \bibinfo{volume}{7},
  \bibinfo{number}{2} (\bibinfo{year}{2005}), \bibinfo{pages}{269--279}.
\newblock


\bibitem[Papagiannakis et~al\mbox{.}(2018)]%
        {DBLP:conf/siggrapha/PapagiannakisLK18}
\bibfield{author}{\bibinfo{person}{George Papagiannakis},
  \bibinfo{person}{Nikos Lydatakis}, \bibinfo{person}{Steve Kateros},
  \bibinfo{person}{Stelios Georgiou}, {and} \bibinfo{person}{Paul Zikas}.}
  \bibinfo{year}{2018}\natexlab{}.
\newblock \showarticletitle{Transforming Medical Education and Training with
  {VR} using {M.A.G.E.S}}. In \bibinfo{booktitle}{\emph{{SIGGRAPH} Asia 2018
  Posters}}. \bibinfo{publisher}{{ACM}}, \bibinfo{pages}{83:1--83:2}.
\newblock


\bibitem[Park et~al\mbox{.}(2019)]%
        {park2019rate}
\bibfield{author}{\bibinfo{person}{Jounsup Park}, \bibinfo{person}{Philip~A
  Chou}, {and} \bibinfo{person}{Jenq-Neng Hwang}.}
  \bibinfo{year}{2019}\natexlab{}.
\newblock \showarticletitle{Rate-utility Optimized Streaming of Volumetric
  Media for Augmented Reality}.
\newblock \bibinfo{journal}{\emph{IEEE Journal on Emerging and Selected Topics
  in Circuits and Systems}} \bibinfo{volume}{9}, \bibinfo{number}{1}
  (\bibinfo{year}{2019}), \bibinfo{pages}{149--162}.
\newblock


\bibitem[Pastuszak and Abramowski(2016)]%
        {DBLP:journals/tcsv/PastuszakA16}
\bibfield{author}{\bibinfo{person}{Grzegorz Pastuszak} {and}
  \bibinfo{person}{Andrzej Abramowski}.} \bibinfo{year}{2016}\natexlab{}.
\newblock \showarticletitle{Algorithm and Architecture Design of the
  {H.265/HEVC} Intra Encoder}.
\newblock \bibinfo{journal}{\emph{{IEEE} Trans. Circuits Syst. Video Technol.}}
  \bibinfo{volume}{26}, \bibinfo{number}{1} (\bibinfo{year}{2016}),
  \bibinfo{pages}{210--222}.
\newblock


\bibitem[Paudyal et~al\mbox{.}(2021)]%
        {DBLP:journals/tbc/PaudyalBCGC21}
\bibfield{author}{\bibinfo{person}{Pradip Paudyal}, \bibinfo{person}{Federica
  Battisti}, \bibinfo{person}{Patrick~Le Callet}, \bibinfo{person}{Jes{\'{u}}s
  Guti{\'{e}}rrez}, {and} \bibinfo{person}{Marco Carli}.}
  \bibinfo{year}{2021}\natexlab{}.
\newblock \showarticletitle{Perceptual Quality of Light Field Images and Impact
  of Visualization Techniques}.
\newblock \bibinfo{journal}{\emph{{IEEE} Trans. Broadcast.}}
  \bibinfo{volume}{67}, \bibinfo{number}{2} (\bibinfo{year}{2021}),
  \bibinfo{pages}{395--408}.
\newblock


\bibitem[Perkis et~al\mbox{.}(2020)]%
        {Qualinet1}
\bibfield{author}{\bibinfo{person}{Andrew Perkis}, \bibinfo{person}{Christian
  Timmerer}, \bibinfo{person}{Sabina Barakovic},
  \bibinfo{person}{Jasmina~Barakovic Husic}, \bibinfo{person}{S{\o}ren Bech},
  \bibinfo{person}{Sebastian Bosse}, \bibinfo{person}{Jean Botev},
  \bibinfo{person}{Kjell Brunnstr{\"{o}}m},
  \bibinfo{person}{Lu{\'{\i}}s~Alberto da Silva~Cruz},
  \bibinfo{person}{Katrien~De Moor}, \bibinfo{person}{Andrea de Polo~Saibanti},
  \bibinfo{person}{Wouter Durnez}, \bibinfo{person}{Sebastian Egger{-}Lampl},
  \bibinfo{person}{Ulrich Engelke}, \bibinfo{person}{Tiago~H. Falk},
  \bibinfo{person}{Asim Hameed}, \bibinfo{person}{Andrew Hines},
  \bibinfo{person}{Tanja Kojic}, \bibinfo{person}{Dragan Kukolj},
  \bibinfo{person}{Eirini Liotou}, \bibinfo{person}{Dragorad Milovanovic},
  \bibinfo{person}{Sebastian M{\"{o}}ller}, \bibinfo{person}{Niall Murray},
  \bibinfo{person}{Babak Naderi}, \bibinfo{person}{Manuela Pereira},
  \bibinfo{person}{Stuart~W. Perry}, \bibinfo{person}{Ant{\'{o}}nio M.~G.
  Pinheiro}, \bibinfo{person}{Andres~Pinilla Palacios},
  \bibinfo{person}{Alexander Raake}, \bibinfo{person}{Sarvesh~Rajesh Agrawal},
  \bibinfo{person}{Ulrich Reiter}, \bibinfo{person}{Rafael Rodrigues},
  \bibinfo{person}{Raimund Schatz}, \bibinfo{person}{Peter Schelkens},
  \bibinfo{person}{Steven Schmidt}, \bibinfo{person}{Saeed~Shafiee Sabet},
  \bibinfo{person}{Ashutosh Singla}, \bibinfo{person}{Lea Skorin{-}Kapov},
  \bibinfo{person}{Mirko Suznjevic}, \bibinfo{person}{Stefan Uhrig},
  \bibinfo{person}{Sara Vlahovic}, \bibinfo{person}{Jan{-}Niklas
  Voigt{-}Antons}, {and} \bibinfo{person}{Saman Zadtootaghaj}.}
  \bibinfo{year}{2020}\natexlab{}.
\newblock \bibinfo{booktitle}{\emph{QUALINET White Paper on Definitions of
  Immersive Media Experience (IMEx)}}.
\newblock \bibinfo{type}{{T}echnical {R}eport}. \bibinfo{institution}{{Qualinet
  (www.qualinet.eu)}}.
\newblock


\bibitem[Pfister et~al\mbox{.}(2000)]%
        {pfister2000surfels}
\bibfield{author}{\bibinfo{person}{Hanspeter Pfister},
  \bibinfo{person}{Matthias Zwicker}, \bibinfo{person}{Jeroen Van~Baar}, {and}
  \bibinfo{person}{Markus Gross}.} \bibinfo{year}{2000}\natexlab{}.
\newblock \showarticletitle{Surfels: Surface Elements as Rendering Primitives}.
  In \bibinfo{booktitle}{\emph{Proceedings of the 27th annual conference on
  Computer graphics and interactive techniques}}. \bibinfo{pages}{335--342}.
\newblock


\bibitem[Postolache et~al\mbox{.}(2021)]%
        {DBLP:journals/jsac/PostolacheHAGGK21}
\bibfield{author}{\bibinfo{person}{Octavian Postolache},
  \bibinfo{person}{D.~Jude Hemanth}, \bibinfo{person}{Ricardo Alexandre},
  \bibinfo{person}{Deepak Gupta}, \bibinfo{person}{Oana Geman}, {and}
  \bibinfo{person}{Ashish Khanna}.} \bibinfo{year}{2021}\natexlab{}.
\newblock \showarticletitle{Remote Monitoring of Physical Rehabilitation of
  Stroke Patients Using IoT and Virtual Reality}.
\newblock \bibinfo{journal}{\emph{{IEEE} J. Sel. Areas Commun.}}
  \bibinfo{volume}{39}, \bibinfo{number}{2} (\bibinfo{year}{2021}),
  \bibinfo{pages}{562--573}.
\newblock


\bibitem[Qi et~al\mbox{.}(2017)]%
        {PointNet++}
\bibfield{author}{\bibinfo{person}{Charles~Ruizhongtai Qi}, \bibinfo{person}{Li
  Yi}, \bibinfo{person}{Hao Su}, {and} \bibinfo{person}{Leonidas~J. Guibas}.}
  \bibinfo{year}{2017}\natexlab{}.
\newblock \showarticletitle{PointNet++: Deep Hierarchical Feature Learning on
  Point Sets in a Metric Space}. In \bibinfo{booktitle}{\emph{Advances in
  Neural Information Processing Systems 30: Annual Conference on Neural
  Information Processing Systems 2017}}. \bibinfo{pages}{5099--5108}.
\newblock


\bibitem[Qian et~al\mbox{.}(2019)]%
        {DBLP:conf/wmcsa/QianHPG19}
\bibfield{author}{\bibinfo{person}{Feng Qian}, \bibinfo{person}{Bo Han},
  \bibinfo{person}{Jarrell Pair}, {and} \bibinfo{person}{Vijay
  Gopalakrishnan}.} \bibinfo{year}{2019}\natexlab{}.
\newblock \showarticletitle{Toward Practical Volumetric Video Streaming on
  Commodity Smartphones}. In \bibinfo{booktitle}{\emph{Proceedings of the 20th
  International Workshop on Mobile Computing Systems and Applications,
  HotMobile 2019}}. \bibinfo{publisher}{{ACM}}, \bibinfo{pages}{135--140}.
\newblock


\bibitem[Quach et~al\mbox{.}(2019)]%
        {DBLP:conf/icip/QuachVD19}
\bibfield{author}{\bibinfo{person}{Maurice Quach}, \bibinfo{person}{Giuseppe
  Valenzise}, {and} \bibinfo{person}{Fr{\'{e}}d{\'{e}}ric Dufaux}.}
  \bibinfo{year}{2019}\natexlab{}.
\newblock \showarticletitle{Learning Convolutional Transforms for Lossy Point
  Cloud Geometry Compression}. In \bibinfo{booktitle}{\emph{2019 {IEEE}
  International Conference on Image Processing, {ICIP} 2019}}.
  \bibinfo{publisher}{{IEEE}}, \bibinfo{pages}{4320--4324}.
\newblock


\bibitem[Quach et~al\mbox{.}(2020)]%
        {DBLP:conf/mmsp/QuachVD20}
\bibfield{author}{\bibinfo{person}{Maurice Quach}, \bibinfo{person}{Giuseppe
  Valenzise}, {and} \bibinfo{person}{Fr{\'{e}}d{\'{e}}ric Dufaux}.}
  \bibinfo{year}{2020}\natexlab{}.
\newblock \showarticletitle{Improved Deep Point Cloud Geometry Compression}. In
  \bibinfo{booktitle}{\emph{22nd {IEEE} International Workshop on Multimedia
  Signal Processing, {MMSP} 2020}}. \bibinfo{publisher}{{IEEE}},
  \bibinfo{pages}{1--6}.
\newblock


\bibitem[Rakotosaona et~al\mbox{.}(2023)]%
        {rakotosaona2023nerfmeshing}
\bibfield{author}{\bibinfo{person}{Marie-Julie Rakotosaona},
  \bibinfo{person}{Fabian Manhardt}, \bibinfo{person}{Diego~Martin Arroyo},
  \bibinfo{person}{Michael Niemeyer}, \bibinfo{person}{Abhijit Kundu}, {and}
  \bibinfo{person}{Federico Tombari}.} \bibinfo{year}{2023}\natexlab{}.
\newblock \showarticletitle{NeRFMeshing: Distilling Neural Radiance Fields into
  Geometrically-Accurate 3D Meshes}.
\newblock \bibinfo{journal}{\emph{arXiv preprint arXiv:2303.09431}}
  (\bibinfo{year}{2023}).
\newblock


\bibitem[Ratcliffe et~al\mbox{.}(2021)]%
        {DBLP:conf/chi/RatcliffeSBTF21}
\bibfield{author}{\bibinfo{person}{Jack Ratcliffe}, \bibinfo{person}{Francesco
  Soave}, \bibinfo{person}{Nick Bryan{-}Kinns}, \bibinfo{person}{Laurissa
  Tokarchuk}, {and} \bibinfo{person}{Ildar Farkhatdinov}.}
  \bibinfo{year}{2021}\natexlab{}.
\newblock \showarticletitle{Extended Reality {(XR)} Remote Research: a Survey
  of Drawbacks and Opportunities}. In \bibinfo{booktitle}{\emph{{CHI} '21:
  {CHI} Conference on Human Factors in Computing Systems}}.
  \bibinfo{publisher}{{ACM}}, \bibinfo{pages}{527:1--527:13}.
\newblock


\bibitem[Reimat et~al\mbox{.}(2021)]%
        {cwipc}
\bibfield{author}{\bibinfo{person}{Ignacio Reimat}, \bibinfo{person}{Evangelos
  Alexiou}, \bibinfo{person}{Jack Jansen}, \bibinfo{person}{Irene Viola},
  \bibinfo{person}{Shishir Subramanyam}, {and} \bibinfo{person}{Pablo Cesar}.}
  \bibinfo{year}{2021}\natexlab{}.
\newblock \showarticletitle{{CWIPC-SXR}: Point Cloud Dynamic Human Dataset for
  Social {XR}}. In \bibinfo{booktitle}{\emph{MMSys '21: 12th {ACM} Multimedia
  Systems Conference}}. \bibinfo{publisher}{{ACM}}, \bibinfo{pages}{300--306}.
\newblock


\bibitem[Rejaie et~al\mbox{.}(2000)]%
        {DBLP:journals/jsac/RejaieHE00}
\bibfield{author}{\bibinfo{person}{Reza Rejaie}, \bibinfo{person}{Mark
  Handley}, {and} \bibinfo{person}{Deborah Estrin}.}
  \bibinfo{year}{2000}\natexlab{}.
\newblock \showarticletitle{Layered Quality Adaptation for Internet video
  streaming}.
\newblock \bibinfo{journal}{\emph{{IEEE} J. Sel. Areas Commun.}}
  \bibinfo{volume}{18}, \bibinfo{number}{12} (\bibinfo{year}{2000}),
  \bibinfo{pages}{2530--2543}.
\newblock


\bibitem[Remondino and Fraser(2006)]%
        {camera_calibration_ETH}
\bibfield{author}{\bibinfo{person}{Fabio Remondino} {and}
  \bibinfo{person}{Clive Fraser}.} \bibinfo{year}{2006}\natexlab{}.
\newblock \showarticletitle{Digital Camera Calibration Methods. Considerations
  and Comparisons}.
\newblock \bibinfo{journal}{\emph{International Archives of the Photogrammetry,
  Remote Sensing and Spatial Information Sciences}} \bibinfo{volume}{XXXVI},
  \bibinfo{number}{5}, \bibinfo{pages}{266 -- 272}.
\newblock


\bibitem[Riegler and Koltun(2020)]%
        {riegler2020free}
\bibfield{author}{\bibinfo{person}{Gernot Riegler} {and}
  \bibinfo{person}{Vladlen Koltun}.} \bibinfo{year}{2020}\natexlab{}.
\newblock \showarticletitle{Free View Synthesis}. In
  \bibinfo{booktitle}{\emph{Computer Vision--ECCV 2020: 16th European
  Conference, Proceedings, Part XIX 16}}. Springer, \bibinfo{pages}{623--640}.
\newblock


\bibitem[Rossignac(1999)]%
        {DBLP:journals/tvcg/Rossignac99}
\bibfield{author}{\bibinfo{person}{Jarek Rossignac}.}
  \bibinfo{year}{1999}\natexlab{}.
\newblock \showarticletitle{Edgebreaker: Connectivity Compression for Triangle
  Meshes}.
\newblock \bibinfo{journal}{\emph{{IEEE} Trans. Vis. Comput. Graph.}}
  \bibinfo{volume}{5}, \bibinfo{number}{1} (\bibinfo{year}{1999}),
  \bibinfo{pages}{47--61}.
\newblock


\bibitem[Rossignac(2001)]%
        {DBLP:conf/smi/Rossignac01}
\bibfield{author}{\bibinfo{person}{Jarek Rossignac}.}
  \bibinfo{year}{2001}\natexlab{}.
\newblock \showarticletitle{3D Compression Made Simple: Edgebreaker with
  Zip{\&}Wrap on a Corner-Table}. In \bibinfo{booktitle}{\emph{2001
  International Conference on Shape Modeling and Applications}}.
  \bibinfo{publisher}{{IEEE} Computer Society}, \bibinfo{pages}{278}.
\newblock


\bibitem[Rusu and Cousins(2011)]%
        {DBLP:conf/icra/RusuC11}
\bibfield{author}{\bibinfo{person}{Radu~Bogdan Rusu} {and}
  \bibinfo{person}{Steve Cousins}.} \bibinfo{year}{2011}\natexlab{}.
\newblock \showarticletitle{3D is here: Point Cloud Library {(PCL)}}. In
  \bibinfo{booktitle}{\emph{{IEEE} International Conference on Robotics and
  Automation, {ICRA} 2011}}. \bibinfo{publisher}{{IEEE}}.
\newblock


\bibitem[Sampsell(1994)]%
        {sampsell1994digital}
\bibfield{author}{\bibinfo{person}{Jeffrey~B Sampsell}.}
  \bibinfo{year}{1994}\natexlab{}.
\newblock \showarticletitle{Digital Micromirror Device and Its Application to
  Projection Displays}.
\newblock \bibinfo{journal}{\emph{Journal of Vacuum Science \& Technology B:
  Microelectronics and Nanometer Structures Processing, Measurement, and
  Phenomena}} \bibinfo{volume}{12}, \bibinfo{number}{6} (\bibinfo{year}{1994}),
  \bibinfo{pages}{3242--3246}.
\newblock


\bibitem[Sandri et~al\mbox{.}(2019)]%
        {DBLP:journals/tip/SandriQC19}
\bibfield{author}{\bibinfo{person}{Gustavo Sandri}, \bibinfo{person}{Ricardo~L.
  de Queiroz}, {and} \bibinfo{person}{Philip~A. Chou}.}
  \bibinfo{year}{2019}\natexlab{}.
\newblock \showarticletitle{Compression of Plenoptic Point Clouds}.
\newblock \bibinfo{journal}{\emph{{IEEE} Trans. Image Process.}}
  \bibinfo{volume}{28}, \bibinfo{number}{3} (\bibinfo{year}{2019}),
  \bibinfo{pages}{1419--1427}.
\newblock


\bibitem[Schnabel and Klein(2006)]%
        {DBLP:conf/spbg/SchnabelK06}
\bibfield{author}{\bibinfo{person}{Ruwen Schnabel} {and}
  \bibinfo{person}{Reinhard Klein}.} \bibinfo{year}{2006}\natexlab{}.
\newblock \showarticletitle{Octree-based Point-Cloud Compression}. In
  \bibinfo{booktitle}{\emph{3rd Symposium on Point Based Graphics, PBG@SIGGRAPH
  2006}}. \bibinfo{publisher}{Eurographics Association},
  \bibinfo{pages}{111--120}.
\newblock


\bibitem[Sch{\"{o}}nberger and Frahm(2016)]%
        {DBLP:conf/cvpr/SchonbergerF16}
\bibfield{author}{\bibinfo{person}{Johannes~L. Sch{\"{o}}nberger} {and}
  \bibinfo{person}{Jan{-}Michael Frahm}.} \bibinfo{year}{2016}\natexlab{}.
\newblock \showarticletitle{Structure-from-Motion Revisited}. In
  \bibinfo{booktitle}{\emph{2016 {IEEE} Conference on Computer Vision and
  Pattern Recognition, {CVPR} 2016}}. \bibinfo{publisher}{{IEEE} Computer
  Society}, \bibinfo{pages}{4104--4113}.
\newblock


\bibitem[Schwarz et~al\mbox{.}(2019)]%
        {DBLP:journals/esticas/SchwarzPBBCCCKL19}
\bibfield{author}{\bibinfo{person}{Sebastian Schwarz}, \bibinfo{person}{Marius
  Preda}, \bibinfo{person}{Vittorio Baroncini}, \bibinfo{person}{Madhukar
  Budagavi}, \bibinfo{person}{Pablo C{\'{e}}sar}, \bibinfo{person}{Philip~A.
  Chou}, \bibinfo{person}{Robert~A. Cohen}, \bibinfo{person}{Maja Krivokuca},
  \bibinfo{person}{Sebastien Lasserre}, \bibinfo{person}{Zhu Li},
  \bibinfo{person}{Joan Llach}, \bibinfo{person}{Khaled Mammou},
  \bibinfo{person}{Rufael Mekuria}, \bibinfo{person}{Ohji Nakagami},
  \bibinfo{person}{Ernestasia Siahaan}, \bibinfo{person}{Ali~J. Tabatabai},
  \bibinfo{person}{Alexis~M. Tourapis}, {and} \bibinfo{person}{Vladyslav
  Zakharchenko}.} \bibinfo{year}{2019}\natexlab{}.
\newblock \showarticletitle{Emerging {MPEG} Standards for Point Cloud
  Compression}.
\newblock \bibinfo{journal}{\emph{{IEEE} J. Emerg. Sel. Topics Circuits Syst.}}
  \bibinfo{volume}{9}, \bibinfo{number}{1} (\bibinfo{year}{2019}),
  \bibinfo{pages}{133--148}.
\newblock


\bibitem[Scott et~al\mbox{.}(1982)]%
        {scott1982blade}
\bibfield{author}{\bibinfo{person}{Ridley Scott}, \bibinfo{person}{Harrison
  Ford}, \bibinfo{person}{Rutger Hauer}, \bibinfo{person}{Sean Young},
  \bibinfo{person}{Hampton Fancher}, {and} \bibinfo{person}{Vangelis}.}
  \bibinfo{year}{1982}\natexlab{}.
\newblock \bibinfo{booktitle}{\emph{Blade Runner}}.
\newblock \bibinfo{publisher}{Warner Home Video Los Angeles}.
\newblock


\bibitem[Serafin et~al\mbox{.}(2018)]%
        {DBLP:journals/cga/SerafinGENN18}
\bibfield{author}{\bibinfo{person}{Stefania Serafin}, \bibinfo{person}{Michele
  Geronazzo}, \bibinfo{person}{Cumhur Erkut}, \bibinfo{person}{Niels~C.
  Nilsson}, {and} \bibinfo{person}{Rolf Nordahl}.}
  \bibinfo{year}{2018}\natexlab{}.
\newblock \showarticletitle{Sonic Interactions in Virtual Reality: State of the
  Art, Current Challenges, and Future Directions}.
\newblock \bibinfo{journal}{\emph{{IEEE} Computer Graphics and Applications}}
  \bibinfo{volume}{38}, \bibinfo{number}{2} (\bibinfo{year}{2018}),
  \bibinfo{pages}{31--43}.
\newblock


\bibitem[Shi et~al\mbox{.}(2023)]%
        {DBLP:conf/mmsys/ShiVDO23}
\bibfield{author}{\bibinfo{person}{Yuang Shi}, \bibinfo{person}{Pranav
  Venkatram}, \bibinfo{person}{Yifan Ding}, {and} \bibinfo{person}{Wei~Tsang
  Ooi}.} \bibinfo{year}{2023}\natexlab{}.
\newblock \showarticletitle{Enabling Low Bit-Rate {MPEG} V-PCC-encoded
  Volumetric Video Streaming with 3D Sub-sampling}. In
  \bibinfo{booktitle}{\emph{Proceedings of the 14th Conference on {ACM}
  Multimedia Systems, MMSys 2023}}. \bibinfo{publisher}{{ACM}},
  \bibinfo{pages}{108--118}.
\newblock


\bibitem[Shrestha et~al\mbox{.}(2007)]%
        {DBLP:conf/mm/ShresthaBW07}
\bibfield{author}{\bibinfo{person}{Prarthana Shrestha}, \bibinfo{person}{Mauro
  Barbieri}, {and} \bibinfo{person}{Hans Weda}.}
  \bibinfo{year}{2007}\natexlab{}.
\newblock \showarticletitle{Synchronization of Multi-camera Video Recordings
  based on Audio}. In \bibinfo{booktitle}{\emph{Proceedings of the 15th
  International Conference on Multimedia 2007}}. \bibinfo{publisher}{{ACM}},
  \bibinfo{pages}{545--548}.
\newblock


\bibitem[Smid and Branstad(1988)]%
        {DBLP:journals/pieee/SmidB88}
\bibfield{author}{\bibinfo{person}{Miles~E. Smid} {and}
  \bibinfo{person}{Dennis~K. Branstad}.} \bibinfo{year}{1988}\natexlab{}.
\newblock \showarticletitle{Data Encryption Standard: Past and Future}.
\newblock \bibinfo{journal}{\emph{Proc. {IEEE}}} \bibinfo{volume}{76},
  \bibinfo{number}{5} (\bibinfo{year}{1988}), \bibinfo{pages}{550--559}.
\newblock


\bibitem[SMPTE(2023)]%
        {SMPTE}
\bibfield{author}{\bibinfo{person}{SMPTE}.} \bibinfo{year}{2023}\natexlab{}.
\newblock \bibinfo{title}{SMPTE: The Home of Media Professionals,
  Technologists, and Engineers.}
\newblock
\newblock
\urldef\tempurl%
\url{https://www.smpte.org/}
\showURL{%
Retrieved June 27, 2023 from \tempurl}


\bibitem[Sommer et~al\mbox{.}(2010)]%
        {DBLP:journals/comsur/SommerGFKMSS10}
\bibfield{author}{\bibinfo{person}{J{\"{o}}rg Sommer},
  \bibinfo{person}{Sebastian Gunreben}, \bibinfo{person}{F. Feller},
  \bibinfo{person}{Martin K{\"{o}}hn}, \bibinfo{person}{Ahlem Mifdaoui},
  \bibinfo{person}{Detlef Sass}, {and} \bibinfo{person}{Joachim Scharf}.}
  \bibinfo{year}{2010}\natexlab{}.
\newblock \showarticletitle{Ethernet - {A} Survey on its Fields of
  Application}.
\newblock \bibinfo{journal}{\emph{{IEEE} Commun. Surv. Tutorials}}
  \bibinfo{volume}{12}, \bibinfo{number}{2} (\bibinfo{year}{2010}),
  \bibinfo{pages}{263--284}.
\newblock


\bibitem[Sorkine et~al\mbox{.}(2004)]%
        {DBLP:conf/sgp/SorkineCLARS04}
\bibfield{author}{\bibinfo{person}{Olga Sorkine}, \bibinfo{person}{Daniel
  Cohen{-}Or}, \bibinfo{person}{Yaron Lipman}, \bibinfo{person}{Marc Alexa},
  \bibinfo{person}{Christian R{\"{o}}ssl}, {and} \bibinfo{person}{Hans{-}Peter
  Seidel}.} \bibinfo{year}{2004}\natexlab{}.
\newblock \showarticletitle{Laplacian Surface Editing}. In
  \bibinfo{booktitle}{\emph{Second Eurographics Symposium on Geometry
  Processing}} \emph{(\bibinfo{series}{{ACM} International Conference
  Proceeding Series}, Vol.~\bibinfo{volume}{71})}.
  \bibinfo{publisher}{Eurographics Association}, \bibinfo{pages}{175--184}.
\newblock


\bibitem[Speicher et~al\mbox{.}(2019)]%
        {DBLP:conf/chi/SpeicherHN19}
\bibfield{author}{\bibinfo{person}{Maximilian Speicher},
  \bibinfo{person}{Brian~D. Hall}, {and} \bibinfo{person}{Michael Nebeling}.}
  \bibinfo{year}{2019}\natexlab{}.
\newblock \showarticletitle{What is Mixed Reality?}. In
  \bibinfo{booktitle}{\emph{Proceedings of the 2019 {CHI} Conference on Human
  Factors in Computing Systems, {CHI} 2019}}. \bibinfo{publisher}{{ACM}},
  \bibinfo{pages}{537}.
\newblock


\bibitem[Sterzentsenko et~al\mbox{.}(2018)]%
        {sterzentsenko2018low}
\bibfield{author}{\bibinfo{person}{Vladimiros Sterzentsenko},
  \bibinfo{person}{Antonis Karakottas}, \bibinfo{person}{Alexandros
  Papachristou}, \bibinfo{person}{Nikolaos Zioulis},
  \bibinfo{person}{Alexandros Doumanoglou}, \bibinfo{person}{Dimitrios
  Zarpalas}, {and} \bibinfo{person}{Petros Daras}.}
  \bibinfo{year}{2018}\natexlab{}.
\newblock \showarticletitle{A Low-cost, Flexible and Portable Volumetric
  Capturing System}. In \bibinfo{booktitle}{\emph{2018 14th International
  Conference on Signal-Image Technology \& Internet-Based Systems (SITIS)}}.
  IEEE, \bibinfo{pages}{200--207}.
\newblock


\bibitem[Subramanyam et~al\mbox{.}(2020)]%
        {DBLP:conf/vr/SubramanyamLVC20}
\bibfield{author}{\bibinfo{person}{Shishir Subramanyam}, \bibinfo{person}{Jie
  Li}, \bibinfo{person}{Irene Viola}, {and} \bibinfo{person}{Pablo
  C{\'{e}}sar}.} \bibinfo{year}{2020}\natexlab{}.
\newblock \showarticletitle{Comparing the Quality of Highly Realistic Digital
  Humans in 3DoF and 6DoF: {A} Volumetric Video Case Study}. In
  \bibinfo{booktitle}{\emph{{IEEE} Conference on Virtual Reality and 3D User
  Interfaces, {VR} 2010}}. \bibinfo{publisher}{{IEEE}},
  \bibinfo{pages}{127--136}.
\newblock


\bibitem[Sun et~al\mbox{.}(2023)]%
        {DBLP:conf/mmsys/SunHSOHH23}
\bibfield{author}{\bibinfo{person}{Yuan{-}Chun Sun}, \bibinfo{person}{I{-}Chun
  Huang}, \bibinfo{person}{Yuang Shi}, \bibinfo{person}{Wei~Tsang Ooi},
  \bibinfo{person}{Chun{-}Ying Huang}, {and} \bibinfo{person}{Cheng{-}Hsin
  Hsu}.} \bibinfo{year}{2023}\natexlab{}.
\newblock \showarticletitle{A Dynamic 3D Point Cloud Dataset for Immersive
  Applications}. In \bibinfo{booktitle}{\emph{Proceedings of the 14th
  Conference on {ACM} Multimedia Systems, MMSys 2023}}.
  \bibinfo{publisher}{{ACM}}, \bibinfo{pages}{376--383}.
\newblock


\bibitem[Tan et~al\mbox{.}(2018)]%
        {DBLP:conf/cvpr/Tan0LX18}
\bibfield{author}{\bibinfo{person}{Qingyang Tan}, \bibinfo{person}{Lin Gao},
  \bibinfo{person}{Yu{-}Kun Lai}, {and} \bibinfo{person}{Shihong Xia}.}
  \bibinfo{year}{2018}\natexlab{}.
\newblock \showarticletitle{Variational Autoencoders for Deforming 3D Mesh
  Models}. In \bibinfo{booktitle}{\emph{2018 {IEEE} Conference on Computer
  Vision and Pattern Recognition, {CVPR} 2018}}. \bibinfo{publisher}{Computer
  Vision Foundation / {IEEE} Computer Society}, \bibinfo{pages}{5841--5850}.
\newblock


\bibitem[Tang et~al\mbox{.}(2022)]%
        {DBLP:conf/nips/TangC0Z22}
\bibfield{author}{\bibinfo{person}{Jiaxiang Tang}, \bibinfo{person}{Xiaokang
  Chen}, \bibinfo{person}{Jingbo Wang}, {and} \bibinfo{person}{Gang Zeng}.}
  \bibinfo{year}{2022}\natexlab{}.
\newblock \showarticletitle{Compressible-composable NeRF via Rank-residual
  Decomposition}. In \bibinfo{booktitle}{\emph{NeurIPS}}.
\newblock


\bibitem[Tarnanas et~al\mbox{.}(2013)]%
        {tarnanas2013ecological}
\bibfield{author}{\bibinfo{person}{Ioannis Tarnanas}, \bibinfo{person}{Winfried
  Schlee}, \bibinfo{person}{Magda Tsolaki}, \bibinfo{person}{Ren{\'e}
  M{\"u}ri}, \bibinfo{person}{Urs Mosimann}, \bibinfo{person}{Tobias Nef},
  {et~al\mbox{.}}} \bibinfo{year}{2013}\natexlab{}.
\newblock \showarticletitle{Ecological Validity of Virtual Reality Daily Living
  Activities Screening for Early Dementia: Longitudinal Study}.
\newblock \bibinfo{journal}{\emph{JMIR serious games}} \bibinfo{volume}{1},
  \bibinfo{number}{1} (\bibinfo{year}{2013}), \bibinfo{pages}{e2778}.
\newblock


\bibitem[Tian et~al\mbox{.}(2017)]%
        {DBLP:conf/icip/TianOFCV17}
\bibfield{author}{\bibinfo{person}{Dong Tian}, \bibinfo{person}{Hideaki
  Ochimizu}, \bibinfo{person}{Chen Feng}, \bibinfo{person}{Robert~A. Cohen},
  {and} \bibinfo{person}{Anthony Vetro}.} \bibinfo{year}{2017}\natexlab{}.
\newblock \showarticletitle{Geometric Distortion Metrics for Point Cloud
  Compression}. In \bibinfo{booktitle}{\emph{2017 {IEEE} International
  Conference on Image Processing, {ICIP} 2017}}. \bibinfo{publisher}{{IEEE}},
  \bibinfo{pages}{3460--3464}.
\newblock


\bibitem[Torkhani et~al\mbox{.}(2015)]%
        {DBLP:journals/spic/TorkhaniWC15}
\bibfield{author}{\bibinfo{person}{Fakhri Torkhani}, \bibinfo{person}{Kai
  Wang}, {and} \bibinfo{person}{Jean{-}Marc Chassery}.}
  \bibinfo{year}{2015}\natexlab{}.
\newblock \showarticletitle{Perceptual Quality Assessment of 3D Dynamic Meshes:
  Subjective and Objective Studies}.
\newblock \bibinfo{journal}{\emph{Signal Process. Image Commun.}}
  \bibinfo{volume}{31} (\bibinfo{year}{2015}), \bibinfo{pages}{185--204}.
\newblock


\bibitem[Torrence(2006)]%
        {DBLP:conf/siggraph/Torrence06a}
\bibfield{author}{\bibinfo{person}{Ann Torrence}.}
  \bibinfo{year}{2006}\natexlab{}.
\newblock \showarticletitle{Martin Newell's Original Teapot}. In
  \bibinfo{booktitle}{\emph{International Conference on Computer Graphics and
  Interactive Techniques, {SIGGRAPH} 2006, Teapot}}.
  \bibinfo{publisher}{{ACM}}, \bibinfo{pages}{29}.
\newblock


\bibitem[Touma and Gotsman(1998)]%
        {DBLP:conf/graphicsinterface/ToumaG98}
\bibfield{author}{\bibinfo{person}{Costa Touma} {and} \bibinfo{person}{Craig
  Gotsman}.} \bibinfo{year}{1998}\natexlab{}.
\newblock \showarticletitle{Triangle Mesh Compression}. In
  \bibinfo{booktitle}{\emph{Proceedings of the Graphics Interface 1998
  Conference}}. \bibinfo{publisher}{Canadian Human-Computer Communications
  Society}, \bibinfo{pages}{26--34}.
\newblock


\bibitem[Troje(2023)]%
        {troje2023zoom}
\bibfield{author}{\bibinfo{person}{Nikolaus~F Troje}.}
  \bibinfo{year}{2023}\natexlab{}.
\newblock \showarticletitle{Zoom Disrupts Eye Contact Behaviour: Problems and
  Solutions}.
\newblock \bibinfo{journal}{\emph{Trends in Cognitive Sciences}}
  (\bibinfo{year}{2023}).
\newblock


\bibitem[Turk and Levoy(1994)]%
        {DBLP:conf/siggraph/TurkL94}
\bibfield{author}{\bibinfo{person}{Greg Turk} {and} \bibinfo{person}{Marc
  Levoy}.} \bibinfo{year}{1994}\natexlab{}.
\newblock \showarticletitle{Zippered Polygon Meshes from Range Images}. In
  \bibinfo{booktitle}{\emph{Proceedings of the 21th Annual Conference on
  Computer Graphics and Interactive Techniques, {SIGGRAPH} 1994}}.
  \bibinfo{publisher}{{ACM}}, \bibinfo{pages}{311--318}.
\newblock


\bibitem[van~der Hooft et~al\mbox{.}(2023)]%
        {DBLP:journals/comsur/HooftAVSSST23}
\bibfield{author}{\bibinfo{person}{Jeroen van~der Hooft}, \bibinfo{person}{Hadi
  Amirpour}, \bibinfo{person}{Maria~Torres Vega}, \bibinfo{person}{Yago
  Sanchez}, \bibinfo{person}{Raimund Schatz}, \bibinfo{person}{Thomas Schierl},
  {and} \bibinfo{person}{Christian Timmerer}.} \bibinfo{year}{2023}\natexlab{}.
\newblock \showarticletitle{A Tutorial on Immersive Video Delivery: From
  Omnidirectional Video to Holography}.
\newblock \bibinfo{journal}{\emph{{IEEE} Commun. Surv. Tutorials}}
  \bibinfo{volume}{25}, \bibinfo{number}{2} (\bibinfo{year}{2023}),
  \bibinfo{pages}{1336--1375}.
\newblock


\bibitem[van~der Hooft et~al\mbox{.}(2020)]%
        {DBLP:conf/qomex/HooftVTBTS20}
\bibfield{author}{\bibinfo{person}{Jeroen van~der Hooft},
  \bibinfo{person}{Maria~Torres Vega}, \bibinfo{person}{Christian Timmerer},
  \bibinfo{person}{Ali~C. Begen}, \bibinfo{person}{Filip~De Turck}, {and}
  \bibinfo{person}{Raimund Schatz}.} \bibinfo{year}{2020}\natexlab{}.
\newblock \showarticletitle{Objective and Subjective QoE Evaluation for
  Adaptive Point Cloud Streaming}. In \bibinfo{booktitle}{\emph{Twelfth
  International Conference on Quality of Multimedia Experience, QoMEX 2020}}.
  \bibinfo{publisher}{{IEEE}}, \bibinfo{pages}{1--6}.
\newblock


\bibitem[van~der Hooft et~al\mbox{.}(2019)]%
        {DBLP:conf/mm/HooftWTTH19}
\bibfield{author}{\bibinfo{person}{Jeroen van~der Hooft}, \bibinfo{person}{Tim
  Wauters}, \bibinfo{person}{Filip~De Turck}, \bibinfo{person}{Christian
  Timmerer}, {and} \bibinfo{person}{Hermann Hellwagner}.}
  \bibinfo{year}{2019}\natexlab{}.
\newblock \showarticletitle{Towards 6DoF {HTTP} Adaptive Streaming Through
  Point Cloud Compression}. In \bibinfo{booktitle}{\emph{Proceedings of the
  27th {ACM} International Conference on Multimedia, {MM} 2019}}.
  \bibinfo{publisher}{{ACM}}, \bibinfo{pages}{2405--2413}.
\newblock


\bibitem[V{\'{a}}sa and Skala(2011)]%
        {DBLP:journals/tvcg/VasaS11}
\bibfield{author}{\bibinfo{person}{Libor V{\'{a}}sa} {and}
  \bibinfo{person}{V{\'{a}}clav Skala}.} \bibinfo{year}{2011}\natexlab{}.
\newblock \showarticletitle{A Perception Correlated Comparison Method for
  Dynamic Meshes}.
\newblock \bibinfo{journal}{\emph{{IEEE} Trans. Vis. Comput. Graph.}}
  \bibinfo{volume}{17}, \bibinfo{number}{2} (\bibinfo{year}{2011}),
  \bibinfo{pages}{220--230}.
\newblock


\bibitem[Vijayanarasimhan et~al\mbox{.}(2017)]%
        {vijayanarasimhan2017sfm}
\bibfield{author}{\bibinfo{person}{Sudheendra Vijayanarasimhan},
  \bibinfo{person}{Susanna Ricco}, \bibinfo{person}{Cordelia Schmid},
  \bibinfo{person}{Rahul Sukthankar}, {and} \bibinfo{person}{Katerina
  Fragkiadaki}.} \bibinfo{year}{2017}\natexlab{}.
\newblock \showarticletitle{Sfm-net: Learning of Structure and Motion from
  Video}.
\newblock \bibinfo{journal}{\emph{arXiv preprint arXiv:1704.07804}}
  (\bibinfo{year}{2017}).
\newblock


\bibitem[Viola and Cesar(2023)]%
        {IVT15}
\bibfield{author}{\bibinfo{person}{Irene Viola} {and} \bibinfo{person}{Pablo
  Cesar}.} \bibinfo{year}{2023}\natexlab{}.
\newblock \showarticletitle{Volumetric Video Streaming: Current Approaches and
  Implementations}.
\newblock In \bibinfo{booktitle}{\emph{Immersive Video Technologies}}.
  \bibinfo{publisher}{Elsevier}, \bibinfo{pages}{425--443}.
\newblock


\bibitem[Viola et~al\mbox{.}(2022)]%
        {DBLP:conf/vr/SubramanyamLVC202}
\bibfield{author}{\bibinfo{person}{Irene Viola}, \bibinfo{person}{Shishir
  Subramanyam}, \bibinfo{person}{Jie Li}, {and} \bibinfo{person}{Pablo Cesar}.}
  \bibinfo{year}{2022}\natexlab{}.
\newblock \showarticletitle{On the Impact of VR Assessment on the Quality of
  Experience of Highly Realistic Digital Humans: A Volumetric Video Case
  Study}.
\newblock \bibinfo{journal}{\emph{Quality and User Experience}}
  \bibinfo{volume}{7} (\bibinfo{year}{2022}).
\newblock


\bibitem[VQEG(2024)]%
        {VQEG}
\bibfield{author}{\bibinfo{person}{VQEG}.} \bibinfo{year}{2024}\natexlab{}.
\newblock \bibinfo{title}{VQEG Brings International Experts Together.}
\newblock
\newblock
\urldef\tempurl%
\url{https://vqeg.org/}
\showURL{%
Retrieved June 21, 2024 from \tempurl}


\bibitem[Wang et~al\mbox{.}(2020)]%
        {wang2020holographic}
\bibfield{author}{\bibinfo{person}{Di Wang}, \bibinfo{person}{Chao Liu},
  \bibinfo{person}{Chuan Shen}, \bibinfo{person}{Yan Xing}, {and}
  \bibinfo{person}{Qiong-Hua Wang}.} \bibinfo{year}{2020}\natexlab{}.
\newblock \showarticletitle{Holographic Capture and Projection System of Real
  Object based on Tunable Zoom Lens}.
\newblock \bibinfo{journal}{\emph{PhotoniX}}  \bibinfo{volume}{1}
  (\bibinfo{year}{2020}), \bibinfo{pages}{1--15}.
\newblock


\bibitem[Wang et~al\mbox{.}(2023)]%
        {DBLP:journals/comsur/WangSZXLLS23}
\bibfield{author}{\bibinfo{person}{Yuntao Wang}, \bibinfo{person}{Zhou Su},
  \bibinfo{person}{Ning Zhang}, \bibinfo{person}{Rui Xing},
  \bibinfo{person}{Dongxiao Liu}, \bibinfo{person}{Tom~H. Luan}, {and}
  \bibinfo{person}{Xuemin Shen}.} \bibinfo{year}{2023}\natexlab{}.
\newblock \showarticletitle{A Survey on Metaverse: Fundamentals, Security, and
  Privacy}.
\newblock \bibinfo{journal}{\emph{{IEEE} Commun. Surv. Tutorials}}
  \bibinfo{volume}{25}, \bibinfo{number}{1} (\bibinfo{year}{2023}),
  \bibinfo{pages}{319--352}.
\newblock


\bibitem[Wang et~al\mbox{.}(2019)]%
        {MPU}
\bibfield{author}{\bibinfo{person}{Yifan Wang}, \bibinfo{person}{Shihao Wu},
  \bibinfo{person}{Hui Huang}, \bibinfo{person}{Daniel Cohen{-}Or}, {and}
  \bibinfo{person}{Olga Sorkine{-}Hornung}.} \bibinfo{year}{2019}\natexlab{}.
\newblock \showarticletitle{Patch-Based Progressive 3D Point Set Upsampling}.
  In \bibinfo{booktitle}{\emph{{IEEE} Conference on Computer Vision and Pattern
  Recognition, {CVPR} 2019}}. \bibinfo{publisher}{Computer Vision Foundation /
  {IEEE}}, \bibinfo{pages}{5958--5967}.
\newblock


\bibitem[Wang et~al\mbox{.}(2022)]%
        {DBLP:conf/nossdav/WangLH0CQ22}
\bibfield{author}{\bibinfo{person}{Zelong Wang}, \bibinfo{person}{Zhenxiao
  Luo}, \bibinfo{person}{Miao Hu}, \bibinfo{person}{Di Wu},
  \bibinfo{person}{Youlong Cao}, {and} \bibinfo{person}{Yi Qin}.}
  \bibinfo{year}{2022}\natexlab{}.
\newblock \showarticletitle{Revisiting Super-resolution for Internet Video
  Streaming}. In \bibinfo{booktitle}{\emph{Proceedings of the 32nd {ACM}
  Workshop on Network and Operating Systems Support for Digital Audio and
  Video, {NOSSDAV} 2022}}. \bibinfo{publisher}{{ACM}}, \bibinfo{pages}{8--14}.
\newblock


\bibitem[Wei et~al\mbox{.}(2020)]%
        {wei2020deepsfm}
\bibfield{author}{\bibinfo{person}{Xingkui Wei}, \bibinfo{person}{Yinda Zhang},
  \bibinfo{person}{Zhuwen Li}, \bibinfo{person}{Yanwei Fu}, {and}
  \bibinfo{person}{Xiangyang Xue}.} \bibinfo{year}{2020}\natexlab{}.
\newblock \showarticletitle{Deepsfm: Structure from Motion via Deep Bundle
  Adjustment}. In \bibinfo{booktitle}{\emph{Computer Vision--ECCV 2020: 16th
  European Conference, Glasgow, UK, August 23--28, 2020, Proceedings, Part I
  16}}. Springer, \bibinfo{pages}{230--247}.
\newblock


\bibitem[Werner et~al\mbox{.}(2009)]%
        {werner2009use}
\bibfield{author}{\bibinfo{person}{Perla Werner}, \bibinfo{person}{Sarit
  Rabinowitz}, \bibinfo{person}{Evelyne Klinger}, \bibinfo{person}{Amos~D
  Korczyn}, {and} \bibinfo{person}{Naomi Josman}.}
  \bibinfo{year}{2009}\natexlab{}.
\newblock \showarticletitle{Use of the Virtual Action Planning Supermarket for
  the Diagnosis of Mild Cognitive Impairment}.
\newblock \bibinfo{journal}{\emph{Dementia and geriatric cognitive disorders}}
  \bibinfo{volume}{27}, \bibinfo{number}{4} (\bibinfo{year}{2009}),
  \bibinfo{pages}{301--309}.
\newblock


\bibitem[Wimbauer et~al\mbox{.}(2023)]%
        {Wimbauer_2023_CVPR}
\bibfield{author}{\bibinfo{person}{Felix Wimbauer}, \bibinfo{person}{Nan Yang},
  \bibinfo{person}{Christian Rupprecht}, {and} \bibinfo{person}{Daniel
  Cremers}.} \bibinfo{year}{2023}\natexlab{}.
\newblock \showarticletitle{Behind the Scenes: Density Fields for Single View
  Reconstruction}. In \bibinfo{booktitle}{\emph{Proceedings of the IEEE/CVF
  Conference on Computer Vision and Pattern Recognition (CVPR)}}.
  \bibinfo{pages}{9076--9086}.
\newblock


\bibitem[Winkler(2006)]%
        {winkler2006perceptual}
\bibfield{author}{\bibinfo{person}{Stefan Winkler}.}
  \bibinfo{year}{2006}\natexlab{}.
\newblock \showarticletitle{Perceptual Video Quality Metrics — A Review}.
\newblock In \bibinfo{booktitle}{\emph{Digital Video Image Quality and
  Perceptual Coding} (\bibinfo{edition}{1st} ed.)}. \bibinfo{publisher}{CRC
  Press}, \bibinfo{pages}{155--176}.
\newblock


\bibitem[Wu et~al\mbox{.}(2020)]%
        {9412653}
\bibfield{author}{\bibinfo{person}{Shan Wu}, \bibinfo{person}{Amnir Hadachi},
  \bibinfo{person}{Damien Vivet}, {and} \bibinfo{person}{Yadu Prabhakar}.}
  \bibinfo{year}{2020}\natexlab{}.
\newblock \showarticletitle{NetCalib: {A} Novel Approach for LiDAR-Camera
  Auto-calibration Based on Deep Learning}. In \bibinfo{booktitle}{\emph{25th
  International Conference on Pattern Recognition, {ICPR} 2020}}.
  \bibinfo{publisher}{{IEEE}}, \bibinfo{pages}{6648--6655}.
\newblock


\bibitem[Wu et~al\mbox{.}(2024a)]%
        {ijcai2024p920}
\bibfield{author}{\bibinfo{person}{Yixuan Wu}, \bibinfo{person}{Kaiyuan Hu},
  \bibinfo{person}{Danny~Z. Chen}, {and} \bibinfo{person}{Jian Wu}.}
  \bibinfo{year}{2024}\natexlab{a}.
\newblock \showarticletitle{AI-Enhanced Virtual Reality in Medicine: A
  Comprehensive Survey}. In \bibinfo{booktitle}{\emph{Proceedings of the
  Thirty-Third International Joint Conference on Artificial Intelligence,
  {IJCAI-24}}}. \bibinfo{publisher}{International Joint Conferences on
  Artificial Intelligence Organization}, \bibinfo{pages}{8326--8334}.
\newblock
\newblock
\shownote{Survey Track}.


\bibitem[Wu et~al\mbox{.}(2024b)]%
        {hu2024teleor}
\bibfield{author}{\bibinfo{person}{Yixuan Wu}, \bibinfo{person}{Kaiyuan Hu},
  \bibinfo{person}{Qian Shao}, \bibinfo{person}{Jintai Chen},
  \bibinfo{person}{Danny~Z Chen}, {and} \bibinfo{person}{Jian Wu}.}
  \bibinfo{year}{2024}\natexlab{b}.
\newblock \showarticletitle{TeleOR: Real-time Telemedicine System for
  Full-Scene Operating Room}.
\newblock \bibinfo{journal}{\emph{arXiv preprint arXiv:2407.19763}}
  (\bibinfo{year}{2024}).
\newblock


\bibitem[Wu et~al\mbox{.}(2015)]%
        {DBLP:conf/cvpr/WuSKYZTX15}
\bibfield{author}{\bibinfo{person}{Zhirong Wu}, \bibinfo{person}{Shuran Song},
  \bibinfo{person}{Aditya Khosla}, \bibinfo{person}{Fisher Yu},
  \bibinfo{person}{Linguang Zhang}, \bibinfo{person}{Xiaoou Tang}, {and}
  \bibinfo{person}{Jianxiong Xiao}.} \bibinfo{year}{2015}\natexlab{}.
\newblock \showarticletitle{3D ShapeNets: {A} Deep Representation for
  Volumetric Shapes}. In \bibinfo{booktitle}{\emph{{IEEE} Conference on
  Computer Vision and Pattern Recognition, {CVPR} 2015}}.
  \bibinfo{publisher}{{IEEE} Computer Society}, \bibinfo{pages}{1912--1920}.
\newblock


\bibitem[Xiao et~al\mbox{.}(2018)]%
        {DBLP:journals/tog/XiaoKFCL18}
\bibfield{author}{\bibinfo{person}{Lei Xiao}, \bibinfo{person}{Anton
  Kaplanyan}, \bibinfo{person}{Alexander Fix}, \bibinfo{person}{Matthew
  Chapman}, {and} \bibinfo{person}{Douglas Lanman}.}
  \bibinfo{year}{2018}\natexlab{}.
\newblock \showarticletitle{DeepFocus: Learned Image Synthesis for
  Computational Displays}.
\newblock \bibinfo{journal}{\emph{{ACM} Trans. Graph.}} \bibinfo{volume}{37},
  \bibinfo{number}{6} (\bibinfo{year}{2018}), \bibinfo{pages}{200}.
\newblock


\bibitem[Xie et~al\mbox{.}(2022)]%
        {DBLP:journals/cgf/XieTSLYKTTSS22}
\bibfield{author}{\bibinfo{person}{Yiheng Xie}, \bibinfo{person}{Towaki
  Takikawa}, \bibinfo{person}{Shunsuke Saito}, \bibinfo{person}{Or Litany},
  \bibinfo{person}{Shiqin Yan}, \bibinfo{person}{Numair Khan},
  \bibinfo{person}{Federico Tombari}, \bibinfo{person}{James Tompkin},
  \bibinfo{person}{Vincent Sitzmann}, {and} \bibinfo{person}{Srinath Sridhar}.}
  \bibinfo{year}{2022}\natexlab{}.
\newblock \showarticletitle{Neural Fields in Visual Computing and Beyond}.
\newblock \bibinfo{journal}{\emph{Comput. Graph. Forum}} \bibinfo{volume}{41},
  \bibinfo{number}{2} (\bibinfo{year}{2022}), \bibinfo{pages}{641--676}.
\newblock


\bibitem[Xu et~al\mbox{.}(2021a)]%
        {DBLP:conf/dcc/XuFGMJZW21}
\bibfield{author}{\bibinfo{person}{Jiacheng Xu}, \bibinfo{person}{Zhijun Fang},
  \bibinfo{person}{Yongbin Gao}, \bibinfo{person}{Siwei Ma},
  \bibinfo{person}{Yaochu Jin}, \bibinfo{person}{Heng Zhou}, {and}
  \bibinfo{person}{Anjie Wang}.} \bibinfo{year}{2021}\natexlab{a}.
\newblock \showarticletitle{Point {AE-DCGAN:} {A} Deep Learning Model for 3D
  Point Cloud Lossy Geometry Compression}. In \bibinfo{booktitle}{\emph{31st
  Data Compression Conference, {DCC} 2021}}. \bibinfo{publisher}{{IEEE}},
  \bibinfo{pages}{379}.
\newblock


\bibitem[Xu et~al\mbox{.}(2022a)]%
        {xu2022point}
\bibfield{author}{\bibinfo{person}{Qiangeng Xu}, \bibinfo{person}{Zexiang Xu},
  \bibinfo{person}{Julien Philip}, \bibinfo{person}{Sai Bi},
  \bibinfo{person}{Zhixin Shu}, \bibinfo{person}{Kalyan Sunkavalli}, {and}
  \bibinfo{person}{Ulrich Neumann}.} \bibinfo{year}{2022}\natexlab{a}.
\newblock \showarticletitle{Point-nerf: Point-based Neural Radiance Fields}. In
  \bibinfo{booktitle}{\emph{Proceedings of the IEEE/CVF Conference on Computer
  Vision and Pattern Recognition}}. \bibinfo{pages}{5438--5448}.
\newblock


\bibitem[Xu et~al\mbox{.}(2018)]%
        {xu2018monoperfcap}
\bibfield{author}{\bibinfo{person}{Weipeng Xu}, \bibinfo{person}{Avishek
  Chatterjee}, \bibinfo{person}{Michael Zollh{\"o}fer}, \bibinfo{person}{Helge
  Rhodin}, \bibinfo{person}{Dushyant Mehta}, \bibinfo{person}{Hans-Peter
  Seidel}, {and} \bibinfo{person}{Christian Theobalt}.}
  \bibinfo{year}{2018}\natexlab{}.
\newblock \showarticletitle{Monoperfcap: Human Performance Capture from
  Monocular Video}.
\newblock \bibinfo{journal}{\emph{ACM Transactions on Graphics (ToG)}}
  \bibinfo{volume}{37}, \bibinfo{number}{2} (\bibinfo{year}{2018}),
  \bibinfo{pages}{1--15}.
\newblock


\bibitem[Xu et~al\mbox{.}(2017)]%
        {Owlii}
\bibfield{author}{\bibinfo{person}{Yi Xu}, \bibinfo{person}{Yao Lu}, {and}
  \bibinfo{person}{Ziyu Wen}.} \bibinfo{year}{2017}\natexlab{}.
\newblock \showarticletitle{Owlii Dynamic Human Mesh Sequence Dataset}.
\newblock \bibinfo{journal}{\emph{ISO/IEC JTC1/SC29/WG11 m41658, 120th MPEG
  Meeting}} (\bibinfo{year}{2017}).
\newblock


\bibitem[Xu et~al\mbox{.}(2021b)]%
        {xu2021voxel}
\bibfield{author}{\bibinfo{person}{Yusheng Xu}, \bibinfo{person}{Xiaohua Tong},
  {and} \bibinfo{person}{Uwe Stilla}.} \bibinfo{year}{2021}\natexlab{b}.
\newblock \showarticletitle{Voxel-based Representation of 3D Point Clouds:
  Methods, Applications, and its Potential Use in the Construction Industry}.
\newblock \bibinfo{journal}{\emph{Automation in Construction}}
  \bibinfo{volume}{126} (\bibinfo{year}{2021}), \bibinfo{pages}{103675}.
\newblock


\bibitem[Xu et~al\mbox{.}(2022b)]%
        {DBLP:journals/tbc/XuYYH22}
\bibfield{author}{\bibinfo{person}{Yiling Xu}, \bibinfo{person}{Qi Yang},
  \bibinfo{person}{Le Yang}, {and} \bibinfo{person}{Jenq{-}Neng Hwang}.}
  \bibinfo{year}{2022}\natexlab{b}.
\newblock \showarticletitle{{EPES:} Point Cloud Quality Modeling Using Elastic
  Potential Energy Similarity}.
\newblock \bibinfo{journal}{\emph{{IEEE} Trans. Broadcast.}}
  \bibinfo{volume}{68}, \bibinfo{number}{1} (\bibinfo{year}{2022}),
  \bibinfo{pages}{33--42}.
\newblock


\bibitem[Yang et~al\mbox{.}(2024)]%
        {yang2024depth}
\bibfield{author}{\bibinfo{person}{Lihe Yang}, \bibinfo{person}{Bingyi Kang},
  \bibinfo{person}{Zilong Huang}, \bibinfo{person}{Xiaogang Xu},
  \bibinfo{person}{Jiashi Feng}, {and} \bibinfo{person}{Hengshuang Zhao}.}
  \bibinfo{year}{2024}\natexlab{}.
\newblock \showarticletitle{Depth Anything: Unleashing the Power of Large-scale
  Unlabeled Data}. In \bibinfo{booktitle}{\emph{Proceedings of the IEEE/CVF
  Conference on Computer Vision and Pattern Recognition}}.
  \bibinfo{pages}{10371--10381}.
\newblock


\bibitem[Yang et~al\mbox{.}(2022)]%
        {DBLP:journals/pami/YangMXLS22}
\bibfield{author}{\bibinfo{person}{Qi Yang}, \bibinfo{person}{Zhan Ma},
  \bibinfo{person}{Yiling Xu}, \bibinfo{person}{Zhu Li}, {and}
  \bibinfo{person}{Jun Sun}.} \bibinfo{year}{2022}\natexlab{}.
\newblock \showarticletitle{Inferring Point Cloud Quality via Graph
  Similarity}.
\newblock \bibinfo{journal}{\emph{{IEEE} Trans. Pattern Anal. Mach. Intell.}}
  \bibinfo{volume}{44}, \bibinfo{number}{6} (\bibinfo{year}{2022}),
  \bibinfo{pages}{3015--3029}.
\newblock


\bibitem[Yaqoob et~al\mbox{.}(2020)]%
        {DBLP:journals/comsur/YaqoobBM20}
\bibfield{author}{\bibinfo{person}{Abid Yaqoob}, \bibinfo{person}{Ting Bi},
  {and} \bibinfo{person}{Gabriel{-}Miro Muntean}.}
  \bibinfo{year}{2020}\natexlab{}.
\newblock \showarticletitle{A Survey on Adaptive 360{\textdegree} Video
  Streaming: Solutions, Challenges and Opportunities}.
\newblock \bibinfo{journal}{\emph{{IEEE} Commun. Surv. Tutorials}}
  \bibinfo{volume}{22}, \bibinfo{number}{4} (\bibinfo{year}{2020}),
  \bibinfo{pages}{2801--2838}.
\newblock


\bibitem[Yu et~al\mbox{.}(2021)]%
        {DBLP:conf/iccv/YuLT0NK21}
\bibfield{author}{\bibinfo{person}{Alex Yu}, \bibinfo{person}{Ruilong Li},
  \bibinfo{person}{Matthew Tancik}, \bibinfo{person}{Hao Li},
  \bibinfo{person}{Ren Ng}, {and} \bibinfo{person}{Angjoo Kanazawa}.}
  \bibinfo{year}{2021}\natexlab{}.
\newblock \showarticletitle{PlenOctrees for Real-time Rendering of Neural
  Radiance Fields}. In \bibinfo{booktitle}{\emph{2021 {IEEE/CVF} International
  Conference on Computer Vision, {ICCV} 2021}}. \bibinfo{publisher}{{IEEE}},
  \bibinfo{pages}{5732--5741}.
\newblock


\bibitem[Yuan et~al\mbox{.}(2021)]%
        {DBLP:journals/mta/YuanAL21}
\bibfield{author}{\bibinfo{person}{Juefei Yuan}, \bibinfo{person}{Hameed
  Abdul{-}Rashid}, {and} \bibinfo{person}{Bo Li}.}
  \bibinfo{year}{2021}\natexlab{}.
\newblock \showarticletitle{A Survey of Recent 3D Scene Analysis and Processing
  Methods}.
\newblock \bibinfo{journal}{\emph{Multim. Tools Appl.}} \bibinfo{volume}{80},
  \bibinfo{number}{13} (\bibinfo{year}{2021}), \bibinfo{pages}{19491--19511}.
\newblock


\bibitem[Zeisl and Pollefeys(2016)]%
        {DBLP:conf/icra/ZeislP16}
\bibfield{author}{\bibinfo{person}{Bernhard Zeisl} {and} \bibinfo{person}{Marc
  Pollefeys}.} \bibinfo{year}{2016}\natexlab{}.
\newblock \showarticletitle{Structure-based Auto-calibration of {RGB-D}
  Sensors}. In \bibinfo{booktitle}{\emph{2016 {IEEE} International Conference
  on Robotics and Automation, {ICRA} 2016}}. \bibinfo{publisher}{{IEEE}},
  \bibinfo{pages}{5076--5083}.
\newblock


\bibitem[Zerman et~al\mbox{.}(2020)]%
        {DBLP:conf/qomex/ZermanOGS20}
\bibfield{author}{\bibinfo{person}{Emin Zerman}, \bibinfo{person}{Cagri
  Ozcinar}, \bibinfo{person}{Pan Gao}, {and} \bibinfo{person}{Aljosa Smolic}.}
  \bibinfo{year}{2020}\natexlab{}.
\newblock \showarticletitle{Textured Mesh vs Coloured Point Cloud: {A}
  Subjective Study for Volumetric Video Compression}. In
  \bibinfo{booktitle}{\emph{Twelfth International Conference on Quality of
  Multimedia Experience, QoMEX 2020}}. \bibinfo{publisher}{{IEEE}},
  \bibinfo{pages}{1--6}.
\newblock


\bibitem[Zhai and Min(2020)]%
        {DBLP:journals/chinaf/ZhaiM20}
\bibfield{author}{\bibinfo{person}{Guangtao Zhai} {and}
  \bibinfo{person}{Xiongkuo Min}.} \bibinfo{year}{2020}\natexlab{}.
\newblock \showarticletitle{Perceptual Image Quality Assessment: A Survey}.
\newblock \bibinfo{journal}{\emph{Sci. China Inf. Sci.}} \bibinfo{volume}{63},
  \bibinfo{number}{11} (\bibinfo{year}{2020}).
\newblock


\bibitem[Zhang et~al\mbox{.}(2021a)]%
        {DBLP:conf/wmcsa/ZhangW0021}
\bibfield{author}{\bibinfo{person}{Anlan Zhang}, \bibinfo{person}{Chendong
  Wang}, \bibinfo{person}{Bo Han}, {and} \bibinfo{person}{Feng Qian}.}
  \bibinfo{year}{2021}\natexlab{a}.
\newblock \showarticletitle{Efficient Volumetric Video Streaming Through Super
  Resolution}. In \bibinfo{booktitle}{\emph{HotMobile '21: The 22nd
  International Workshop on Mobile Computing Systems and Applications}}.
  \bibinfo{publisher}{{ACM}}, \bibinfo{pages}{106--111}.
\newblock


\bibitem[Zhang et~al\mbox{.}(2022)]%
        {DBLP:conf/nsdi/ZhangW0022}
\bibfield{author}{\bibinfo{person}{Anlan Zhang}, \bibinfo{person}{Chendong
  Wang}, \bibinfo{person}{Bo Han}, {and} \bibinfo{person}{Feng Qian}.}
  \bibinfo{year}{2022}\natexlab{}.
\newblock \showarticletitle{YuZu: Neural-Enhanced Volumetric Video Streaming}.
  In \bibinfo{booktitle}{\emph{19th {USENIX} Symposium on Networked Systems
  Design and Implementation, {NSDI} 2022}}. \bibinfo{publisher}{{USENIX}
  Association}, \bibinfo{pages}{137--154}.
\newblock


\bibitem[Zhang et~al\mbox{.}(2021b)]%
        {DBLP:conf/mm/ZhangYX21}
\bibfield{author}{\bibinfo{person}{Yujie Zhang}, \bibinfo{person}{Qi Yang},
  {and} \bibinfo{person}{Yiling Xu}.} \bibinfo{year}{2021}\natexlab{b}.
\newblock \showarticletitle{MS-GraphSIM: Inferring Point Cloud Quality via
  Multiscale Graph Similarity}. In \bibinfo{booktitle}{\emph{{MM} '21: {ACM}
  Multimedia Conference}}. \bibinfo{publisher}{{ACM}},
  \bibinfo{pages}{1230--1238}.
\newblock


\bibitem[Zhang(2000)]%
        {camera_calibration_ZZY}
\bibfield{author}{\bibinfo{person}{Zhengyou Zhang}.}
  \bibinfo{year}{2000}\natexlab{}.
\newblock \showarticletitle{A Flexible New Technique for Camera Calibration}.
\newblock \bibinfo{journal}{\emph{{IEEE} Trans. Pattern Anal. Mach. Intell.}}
  \bibinfo{volume}{22}, \bibinfo{number}{11} (\bibinfo{year}{2000}),
  \bibinfo{pages}{1330--1334}.
\newblock


\bibitem[Zhang(2004)]%
        {principal_points}
\bibfield{author}{\bibinfo{person}{Zhengyou Zhang}.}
  \bibinfo{year}{2004}\natexlab{}.
\newblock \showarticletitle{Camera Calibration with One-Dimensional Objects}.
\newblock \bibinfo{journal}{\emph{{IEEE} Trans. Pattern Anal. Mach. Intell.}}
  \bibinfo{volume}{26}, \bibinfo{number}{7} (\bibinfo{year}{2004}),
  \bibinfo{pages}{892--899}.
\newblock


\bibitem[Zhong et~al\mbox{.}(2021)]%
        {DBLP:journals/tog/ZhongJYHWM21}
\bibfield{author}{\bibinfo{person}{Fangcheng Zhong}, \bibinfo{person}{Akshay
  Jindal}, \bibinfo{person}{Ali~{\"{O}}zg{\"{u}}r Y{\"{o}}ntem},
  \bibinfo{person}{Param Hanji}, \bibinfo{person}{Simon~J. Watt}, {and}
  \bibinfo{person}{Rafal~K. Mantiuk}.} \bibinfo{year}{2021}\natexlab{}.
\newblock \showarticletitle{Reproducing Reality with a High-dynamic-range
  Multi-focal Stereo Display}.
\newblock \bibinfo{journal}{\emph{{ACM} Trans. Graph.}} \bibinfo{volume}{40},
  \bibinfo{number}{6} (\bibinfo{year}{2021}), \bibinfo{pages}{241:1--241:14}.
\newblock


\bibitem[Zhong et~al\mbox{.}(2015)]%
        {DBLP:conf/iwcmc/ZhongKCFA15}
\bibfield{author}{\bibinfo{person}{Zhenzhe Zhong}, \bibinfo{person}{Parag
  Kulkarni}, \bibinfo{person}{Fengming Cao}, \bibinfo{person}{Zhong Fan}, {and}
  \bibinfo{person}{Simon Armour}.} \bibinfo{year}{2015}\natexlab{}.
\newblock \showarticletitle{Issues and Challenges in Dense WiFi Networks}. In
  \bibinfo{booktitle}{\emph{International Wireless Communications and Mobile
  Computing Conference, {IWCMC} 2015}}. \bibinfo{publisher}{{IEEE}},
  \bibinfo{pages}{947--951}.
\newblock


\bibitem[Zhou et~al\mbox{.}(2021)]%
        {zhou2021review}
\bibfield{author}{\bibinfo{person}{Shuyao Zhou}, \bibinfo{person}{Tianqian
  Zhu}, \bibinfo{person}{Kanle Shi}, \bibinfo{person}{Yazi Li},
  \bibinfo{person}{Wen Zheng}, {and} \bibinfo{person}{Junhai Yong}.}
  \bibinfo{year}{2021}\natexlab{}.
\newblock \showarticletitle{Review of Light Field Technologies}.
\newblock \bibinfo{journal}{\emph{Visual Computing for Industry, Biomedicine,
  and Art}} \bibinfo{volume}{4}, \bibinfo{number}{1} (\bibinfo{year}{2021}),
  \bibinfo{pages}{29}.
\newblock


\bibitem[Zhu et~al\mbox{.}(2023)]%
        {10.1109/TPAMI.2023.3330935}
\bibfield{author}{\bibinfo{person}{Wentao Zhu}, \bibinfo{person}{Xiaoxuan Ma},
  \bibinfo{person}{Dongwoo Ro}, \bibinfo{person}{Hai Ci},
  \bibinfo{person}{Jinlu Zhang}, \bibinfo{person}{Jiaxin Shi},
  \bibinfo{person}{Feng Gao}, \bibinfo{person}{Qi Tian}, {and}
  \bibinfo{person}{Yizhou Wang}.} \bibinfo{year}{2023}\natexlab{}.
\newblock \showarticletitle{Human Motion Generation: A Survey}.
\newblock \bibinfo{journal}{\emph{IEEE Trans. Pattern Anal. Mach. Intell.}}
  \bibinfo{volume}{46}, \bibinfo{number}{4} (\bibinfo{year}{2023}),
  \bibinfo{pages}{2430–2449}.
\newblock
\showISSN{0162-8828}


\end{thebibliography}

\appendix

\end{document}